\documentclass[useAMS, usenatbib, usegraphicx]{mn2e}
\usepackage{dcolumn}
\usepackage{bm}
\usepackage{amssymb}
\usepackage{color}

\newcommand{\m}{\phantom{-}}
\newcommand{\z}{\phantom{0}}
\newcommand{\zz}{\phantom{00}}

\newcommand{\lowentry}[1]{\smash{\lower 1.5 ex \hbox{#1}}}
\newcommand{\highentry}[1]{\smash{\raise 1.5 ex \hbox{#1}}}

\newcommand{\cnn}{$ \mathrm{CN}_\mathrm{nipr} $}
\newcommand{\cni}{$ \mathrm{CN}_\mathrm{ipr} $}

\definecolor{darkgreen}{rgb}{0,0.6,0}
\definecolor{cyan}{rgb}{0,0.8,0.8}

\begin{document}

\title[Phase-transition-induced collapse of neutron stars]
      {Relativistic simulations of the phase-transition-induced
       collapse of neutron stars}

\author[E.~B.~Abdikamalov, H.~Dimmelmeier, L.~Rezzolla and J.~C.~Miller]
        {Ernazar~B.~Abdikamalov$^{1,2}$\thanks{E-mail:abdik@sissa.it},
        Harald~Dimmelmeier$^{3,4}$,
        Luciano~Rezzolla$^{5,6,7}$ and
        \newauthor John~C.~Miller$^{1,8}$ \vspace*{0.2 em} \\
        $^1$International School of Advanced Studies (SISSA) and INFN,
        Via Beirut 2\,--\,4, I-34014 Trieste, Italy \\
        $^2$Institute of Nuclear Physics, Uzbekistan Academy of
        Sciences, Ulughbek, Uzbekistan \\
        $^3$Department of Physics, Aristotle University of
        Thessaloniki, GR-54124 Thessaloniki, Greece \\
        $^4$Max-Planck-Institut f\"ur Astrophysik,
        Karl-Schwarzschild-Strasse~1, D-85741 Garching, Germany \\
        $^5$Max-Planck-Institut f\"ur Gravitationsphysik,
        Albert-Einstein-Institut, Am M\"uhlenberg 1, D-14476 Golm, Germany \\
        $^6$Department of Physics and Astronomy, Louisiana State University, Baton
        Rouge, LA 70803, USA \\
        $^7$INFN, Department of Physics, University of Trieste, Trieste, Italy \\
        $^8$Department of Physics (Astrophysics), University of
        Oxford, Keble Road, Oxford OX1 3RH, UK}

\date{Accepted $<$date$>$. Received $<$date$>$; in original form $<$date$>$}

\pagerange{\pageref{firstpage}--\pageref{lastpage}} \pubyear{2009}

\label{firstpage}

\maketitle


\begin{abstract}
  An increase in the central density of a neutron star may trigger a
  phase transition from hadronic matter to deconfined quark matter in
  the core, causing it to collapse to a more compact hybrid-star
  configuration. We present a study of this, building on previous work
  by \citet{lin_06_a}. We follow them in considering a supersonic phase
  transition and using a simplified equation of state, but our
  calculations are general relativistic (using 2D simulations in the
  conformally flat approximation) as compared with their 3D Newtonian
  treatment. We also improved the treatment of the initial phase
  transformation, avoiding the introduction of artificial convection. As
  before, we find that the emitted gravitational-wave spectrum is
  dominated by the fundamental quasi-radial and quadrupolar pulsation
  modes but the strain amplitudes are much smaller than suggested
  previously, which is disappointing for the detection
  prospects. However, we see significantly smaller damping and observe a
  nonlinear mode resonance which substantially enhances the emission in
  some cases. We explain the damping mechanisms operating, giving a
  different view from the previous work. Finally, we discuss the
  detectability of the gravitational waves, showing that the
  signal-to-noise ratio for current or second generation interferometers
  could be high enough to detect such events in our Galaxy, although
  third generation detectors would be needed to observe them out to the
  Virgo cluster, which would be necessary for having a reasonable event
  rate.
\end{abstract}

\begin{keywords}
  hydrodynamics -- relativity -- methods: numerical -- stars: neutron --
  stars: pulsations -- stars: phase-transition -- stars: rotation
\end{keywords}


\section{Introduction}
\label{section:introduction}

The existence of compact stellar objects partially or totally 
consisting of matter in a deconfined quark phase was already predicted 
long ago \citep{bodmer_71_a, itoh_70_a, witten_84_a}. Such stars are 
thought likely to originate as a result of the conversion of purely 
hadronic matter in the interior of a neutron star (NS) into a 
deconfined quark matter phase when the density exceeds a certain 
threshold \citep[for a review see, e.g.][]{weber_99_a, 
glendenning_02_a}. We focus here on cases where the conversion occurs 
only in the core of the NS, while the outer parts remain unchanged. 
The theory of dense nuclear matter predicts that such compact stars 
[hereafter we will use the term ``hybrid quark star'' (HQS) when 
referring to these objects] would generally be more compact than the 
progenitor standard NS, and their equilibrium radii would be smaller 
by up to 20\%. The potential energy $ W $ released by the phase 
transition is expected to be of order
\begin{displaymath}
  \label{eqq:energy_release}
  \Delta W \sim \frac{M^2}{R} \frac{\Delta R}{R} \sim
  10^{52} \mathrm{\ erg},
  \label{eq:DW}
\end{displaymath}
where $ R $ and $ M $ are the typical radius and mass of the NS, 
respectively, and $ \Delta R $ is the decrease in radius.

A first-order phase transition is expected to be the most interesting
case as far as the dynamics and structure of the star are
concerned. Such a transition would proceed by the conversion of
initially metastable hadronic matter in the core into the new
deconfined quark phase \citep{zdunik_07_a, zdunik_08_a}. The
metastable phase could be formed as the central density of the NS
increases due to mass accretion, spin-down or cooling. This could
happen soon after the birth of the NS in a supernova or it could occur
for an old NS accreting from a binary companion. The first of these
channels is likely to give the higher event rate but the second is
also interesting. In the widely accepted scenario for the formation of
millisecond pulsars, where an old NS is spun up by accretion from a
binary companion, the amount of spin-up would be directly related to
the amount of mass accreted \citep{burderi_99_a}, meaning that there
would be a population of rapidly rotating NSs with rather high mass
(and hence high central densities). Recent observational data seems to
confirm that a significant proportion of millisecond pulsars do indeed
tend to be high-mass objects \citep{freire_08_a} which would then be
candidates for undergoing (or having undergone) a
phase-transition-induced collapse of the type which we are discussing
here. While some significant proportion of the potential energy
release given by Eq.~(\ref{eq:DW}) would probably go into neutrino
emission, a significant proportion might also go into pulsations of
the newly-formed HQS, if the conversion process to the new phase is
rapid enough, and this could be an interesting source of gravitational
waves \citep[GWs; see][]{marranghello_02_a, miniutti_03_a,
  lin_06_a}. If detected, these GW signals, and in particular the
identification of quasi-normal mode frequencies in their spectrum,
could help to constrain the properties of matter at the high densities
encountered here. For non-rotating cold NSs with various compositions,
the related theory of asteroseismology has already been formulated in
recent years \citep{andersson_98_a, kokkotas_01_a, benhar_04_a}.

A problem for making any detailed studies of phase-transition-induced 
collapse is that the description of the physics of the phase 
transition remains very uncertain and controversial \citep[for a 
recent review see][]{horvath_07_a}. \citet{drago_07_a} discussed 
possible modes of burning of hadronic matter into quark matter in the 
framework of relativistic hydrodynamics using a microphysical equation 
of state (EoS). They found that the conversion process always 
corresponds to a deflagration and never to a detonation. They also 
argued that hydrodynamical instabilities can develop at the burning 
front. They estimated the corresponding increase in the propagation 
velocity of the phase transition and noticed that, although the 
increase is significant, it is not sufficient to transform the 
deflagration into a detonation in essentially all realistic scenarios. 
On the other hand, \citet{bhattacharyya_06_a} considered the 
transition as being a two-step process, in which the hadronic matter 
is first converted to two-flavour $ u $ and $ d $ quark matter, which 
then subsequently transforms into strange quark matter ($ u $, $ d $ 
and $ s $ quarks) in a second step. They used the relativistic 
hydrodynamic equations to calculate the propagation velocity of the 
first front and found that, in this first stage, a detonation wave 
develops in the hadronic matter. After this front passes through, 
leaving behind two-flavour matter, a second front is generated which 
transforms the two-flavour matter to three-flavour matter via weak 
interactions. The timescale for the second conversion is $ \sim 100 
\mathrm{\ s} $, while that for the first step is about $ 1 \mathrm{\ 
ms} $.

Against this background, \citeauthor{lin_06_a} (\citeyear{lin_06_a};
hereafter referred to as LCCS) carried out a first study of GW
emission by a phase-transition-induced collapse of a rotating NS to a
HQS, using a very simplified model for the microphysics and treating
the phase transition as occurring instantaneously. On the basis of 3D
calculations using Newtonian gravity and hydrodynamics, they obtained
waveforms of the emitted GWs for several collapse models, and found
that the typical predicted dimensionless GW strain $ h $ ranged from
$ 3 $ to $ 15 \times 10^{-23} $ for a source at a distance of
$ 10 \mathrm{\ Mpc} $. The corresponding energy $ E_\mathrm{gw} $
carried away by the GWs was found to be between $ 0.3 $ and
$ 2.8 \times 10^{51} \mathrm{\ ergs} $. They also determined the modes
of stellar pulsation excited by the collapse and showed that the
spectrum of the emitted GWs was dominated by the fundamental
quasi-radial and quadrupolar pulsation modes of the final star. They
suggested that the damping of the stellar pulsations observed in their
calculations was due to the production of shock waves leading to the
development of differential rotation, which proceeds on a timescale of
about $ 5 \mathrm{\ ms} $ for typical collapse models.

The study by LCCS treated the conversion process as being 
instantaneous, and this was mimicked by replacing, at the initial 
time, the EoS describing the hadronic nuclear matter in the core (with 
which the initial equilibrium model had been built) by one describing 
a central zone of deconfined quarks surrounded by a region of mixed 
phase. The material outward of this remained in the original hadronic 
phase. In a first-order phase transition that proceeds via a 
detonation, the conversion front propagates supersonically with 
respect to the matter ahead of it. The sound speed in the stellar 
interior typically has a value of the order of $ 0.3 $ to $ 0.5\,c$. 
Assuming that the quark-matter core has a radius of about $ R \sim 5 
\mathrm{\ km} $, a rough estimate for the timescale of a supersonic 
conversion gives $ \tau \sim R / v \sim 0.05 \mathrm{\ ms} $. Clearly, 
this value is not much smaller than the dynamical timescale of a NS 
and so one does not know, a priori, the impact that the properties of 
the conversion would have for the subsequent stellar dynamics taking 
place on similar timescales. Therefore, one should first check how the 
dynamics of the star after the phase transition would depend on the 
finite propagation velocity of the front, and we do this here.

LCCS represented the hadronic matter by means of a polytropic EoS, 
initially having an adiabatic index $ \gamma = 2 $. When the phase 
transition was triggered (by changing the EoS in the central regions, 
reducing the pressure support), they also replaced the original 
hadronic EoS by a softer ideal-gas type of EoS (with $ \gamma $ then 
ranging from $ 1.95 $ to $ 1.75 $ depending on the model), which 
artificially lowered the pressure \emph{outside} the deconfined quark 
matter core as well [see their Eq.~(44)]. The lowering of the 
adiabatic index also in the outer regions leads to an increased 
release of gravitational binding energy and is hard to motivate on 
physical grounds.

Although the study by LCCS was an important step forward, it was 
clearly using an extremely simplified model for the physics of the NS 
matter and of the phase transition (as well as not including emission 
processes other than GW emission) and one must question how closely it 
represents such processes occurring in real NSs. Since the full 
problem is complex and involves some input physics which is very 
incompletely known at present, the only way to make progress at this 
stage is, indeed, to work in terms of simplified models but making 
improvements where possible. The strategy of this paper is to take the 
work of LCCS as a starting point and then to take some steps forward 
in the direction of including further aspects. In particular, the 
effects of general relativity (GR) certainly play an important role in 
such studies: for instance, the total rest mass of a typical Newtonian 
NS may be almost twice as large as that of its counterpart in GR with 
the same central density and the same EoS. The impact of GR is even 
more pronounced if the star is near to the maximum mass limit. 
Therefore, one expects the Newtonian and GR descriptions of NS 
collapse to differ significantly, thus necessitating a proper 
treatment of the GR effects.

In this study we extend the previous work of LCCS in a number of ways: 
(i) we take into account the effects of GR; (ii) we modify the EoS of 
the stellar matter only in regions which undergo the phase transition 
and take care to avoid introducing spurious convection; (iii) we 
consider a larger set of physical models for different values of the 
stellar parameters and different properties of the stellar matter; 
(iv) we study the effect of introducing a finite timescale for the 
initial phase transformation which destabilises the core rather than 
treating it as occurring instantaneously. Our model for the HQS is 
based on that of LCCS. The numerical hydrodynamics simulations are 
here performed in axisymmetry using the conformal flatness 
approximation to GR. While our model remains extremely simplified, we 
believe that these modifications do represent valid and significant 
steps forward. We feel, however, that we should caution against making 
further elaborate extensions of the hydrodynamics (e.g.\ to 3D GR 
hydrodynamics or GR magneto-hydrodynamics) until such time as one has the 
possibility of including a greatly improved treatment of the 
microphysics and emission processes so that the models come closer to 
those of real NSs.

This article is organized as follows: in 
Section~\ref{section:numerics} we summarize the numerical methods 
used, while in Section~\ref{section:models} we introduce the models 
investigated. In Section~\ref{subsection:comparison} we validate our 
code by performing a comparison with the Newtonian models of LCCS. In 
Section~\ref{section:gr_collapse_models} we discuss the results of our 
simulations of GR models and their dependence on the various 
parameters, and in Section~\ref{section:conclusion} we conclude with a 
summary. In Appendix~\ref{appendix:conversion_velocity} we examine the 
impact on the collapse dynamics of considering a finite timescale for 
the initial phase transformation, and in 
Appendix~\ref{appendix:damping_times} we present our method for 
determining the damping times for the emitted gravitational radiation 
waveforms.

Unless otherwise noted, we choose geometrical units for all physical 
quantities by setting the speed of light and the gravitational 
constant to one, $ c = G = 1 $. Latin indices run from 1 to 3, Greek 
indices from 0 to 3.


\section{Numerical methods}
\label{section:numerics}

We construct our initial rotating NS equilibrium models using a 
variant of the self-consistent field method described by 
\citet{komatsu_89_a} (KEH hereafter), as implemented in the code 
\textsc{RNS} \citep{stergioulas_95_a}. This code solves the GR 
hydro-stationary equations for rotating matter distributions whose 
pressure obeys an EoS given by a polytropic relation [see 
Eq.~(\ref{eq:polytrope}) below]. The resulting equilibrium models, 
which we choose to be rotating uniformly, are taken as initial data 
for the evolution code.

The time-dependent numerical simulations were performed with the 
\textsc{CoCoNuT} code developed by \citet{dimmelmeier_02_a, 
dimmelmeier_02_b} with a metric solver based on spectral methods as 
described in \citet{dimmelmeier_05_a}. The code solves the GR field 
equations for a curved spacetime in the 3\,+\,1 split under the 
assumption of the conformal flatness condition (CFC) for the 
three-metric. The hydrodynamics equations are formulated in 
conservation form, and are evolved with high-resolution 
shock-capturing schemes.

In the following subsections, we summarise the mathematical 
formulation of the metric and hydrodynamic equations, and the 
numerical methods used for solving them.


\subsection{Metric equations}
\label{subsection:metric_equations}

We adopt the ADM 3\,+\,1 formalism of \citet{arnowitt_62_a} to foliate 
a spacetime endowed with a four-metric $ g_{\mu\nu} $ into a set of 
non-intersecting spacelike hypersurfaces. The line element is then 
given by
\begin{equation}
  ds^2 = g_{\mu\nu} \, dx^\mu \! dx^\nu = - \alpha^2 dt^2 +
  \gamma_{ij} (dx^i + \beta^i dt) (dx^j + \beta^j dt), ~
  \label{eq:line_element}
\end{equation}
where $ \alpha $ is the lapse function, $ \beta^i $ is the spacelike 
shift three-vector, and $ \gamma_{ij} $ is the spatial three-metric.

In the 3\,+\,1 formalism, the Einstein equations are split into 
evolution equations for the three-metric $ \gamma_{ij} $ and the 
extrinsic curvature $ K_{ij} $, and constraint equations (the 
Hamiltonian and momentum constraints) which must be fulfilled at every 
spacelike hypersurface.

The fluid is generally specified by means of the rest-mass density $ 
\rho $, the four-velocity $ u^\mu $, and the pressure $ P $, with the 
specific enthalpy defined as $ h = 1 + \epsilon + P / \rho $, where $ 
\epsilon $ is the specific internal energy. The three-velocity of the 
fluid as measured by an Eulerian observer is given by $ v^i = u^i / 
(\alpha u^0) + \beta^i / \alpha $, and the Lorentz factor $ W = \alpha 
u^0 $ satisfies the relation $ W = 1 / \sqrt{1 - v_i v^i} $.

Based on the ideas of \citet{isenberg_78_a} and \citet{wilson_96_a}, 
and as done in the work of \citet{dimmelmeier_02_a, dimmelmeier_02_b}, 
we approximate the general metric $ g_{\mu\nu} $ by replacing the 
spatial three-metric $ \gamma_{ij} $ with the conformally flat 
three-metric
\begin{equation}
  \gamma_{ij} = \phi^4 \hat{\gamma}_{ij},
  \label{eq:cfc_metric}
\end{equation}
where $ \hat{\gamma}_{ij} $ is the flat metric and $ \phi $ is a 
conformal factor. In this CFC approximation, the ADM equations for the 
spacetime metric reduce to a set of five coupled elliptic nonlinear 
equations for the metric components,
\begin{equation}
  \setlength{\arraycolsep}{0.14 em}
  \begin{array}{rcl}
    \hat{\Delta} \phi & = & \displaystyle - 2 \pi \phi^5
    \left( \rho h W^2 - P \right) - \phi^5 \frac{K_{ij} K^{ij}}{8},
    \\ [1.2 em]
    \hat{\Delta} (\alpha \phi) & = & \displaystyle 2 \pi \alpha \phi^5
    \left( \rho h (3 W^2 - 2) + 5 P \right) +
    \alpha \phi^5 \frac{7 K_{ij} K^{ij}}{8},
    \\ [1.2 em]
    \hat{\Delta} \beta^i & = & \displaystyle 16 \pi \alpha \phi^4
    \rho h W^2 v^i + 2 \phi^{10} K^{ij} \hat{\nabla}_{\!j}
    \! \left( \alpha \phi^{-6} \right) -
    \frac{1}{3} \hat{\nabla}^i \hat{\nabla}_{\!k} \beta^k\!,
    \!\!\!\!\!\!\!\!\!\!\!\!\!\!\!\!\!\!\!\!\!
  \end{array}
  \label{eq:cfc_metric_equations}
\end{equation}
where the maximal slicing condition, $ K^i_i = 0 $, is imposed. Here 
$ \hat{\nabla}_{\!i} $ and $ \hat{\Delta} $ are the flat space Nabla 
and Laplace operators, respectively. For the extrinsic curvature we 
have the expression
\begin{equation}
  K_{ij} = \frac {1}{2 \alpha}
  \left( \! \nabla_{\!i} \beta_j + \nabla_{\!j} \beta_i -
  \frac{2}{3} \gamma_{ij} \nabla_{\!k} \beta^k \! \right),
  \label{eq:definition_of_extrinsic_curvature}
\end{equation}
which closes the system~(\ref{eq:cfc_metric_equations}).

We rewrite the above metric equations in a mathematically equivalent 
form by introducing an auxiliary vector field $ W^i $ and obtain
\begin{equation}
  \setlength{\arraycolsep}{0.14 em}
  \begin{array}{rcl}
    \hat{\Delta} \phi & = & \displaystyle - 2 \pi \phi^5
    \left( \rho h W^2 - P \right) -
    \phi^{-7} \frac{\hat{K}_{ij} \hat{K}^{ij}}{8},
    \\ [1.2 em]
    \hat{\Delta} (\alpha \phi) & = & \displaystyle 2 \pi \alpha \phi^5
    \left( \rho h (3 W^2 - 2) + 5 P \right) +
    \alpha \phi^{-7} \frac{7 \hat{K}_{ij} \hat{K}^{ij}}{8},
    \\ [1.2 em]
    \hat{\Delta} \beta^i & = & \displaystyle 2 \hat{\nabla}_{\!j} \!
    \left( 2 \alpha \phi^{-6} \hat{K}^{ij} \right) -
    \frac{1}{3} \hat{\nabla}^i \hat{\nabla}_{\!k} \beta^k\!,
    \\ [1.2 em]
    \hat{\Delta} W^i & = & \displaystyle 8 \pi \phi^{10} \rho h W^2 v^i -
    \frac{1}{3} \hat{\nabla}^i \hat{\nabla}_{\!k} W^k\!,
  \end{array}
  \label{eq:cfc_metric_equations_new}
\end{equation}
where the flat space extrinsic curvature is given by
\begin{equation}
  \hat{K}_{ij} = \hat{\nabla}_{\!i} W_j + \hat{\nabla}_{\!j} W_i -
  \frac{2}{3} \hat{\gamma}_{ij} \hat{\nabla}_{\!k} W^k
  \label{eq:definition_of_flat_extrinsic_curvature}
\end{equation}
and relates to the regular extrinsic curvature as $ \hat{K}_{ij} = 
\phi^2 K_{ij} $ and $ \hat{K}^{ij} = \phi^{10} K^{ij} $. The 
advantages of this new formulation of the metric equations will be 
discussed in a future publication.

Note that the metric equations do not contain explicit time 
derivatives, and thus the metric is calculated by a fully constrained 
approach, at the cost of neglecting some evolutionary degrees of 
freedom in the spacetime metric (e.g.\ dynamical GW degrees of 
freedom).

The accuracy of the CFC approximation for isolated compact stellar
objects has been tested in various works, both in the context of
stellar core collapse and for equilibrium models of NSs \citep[for a
detailed comparison between the CFC approximation of GR and full GR,
see][and references therein]{ott_07_a}. In particular,
\citet{dimmelmeier_06_b} compared collapse simulations of rotating
massive stellar cores in the CFC approximation with the full GR
calculations of \citet{shibata_05_a}. Although such massive stellar
cores are rather unmotivated astrophysically, they are nice toy models
where the spacetime dynamics during collapse is violent, similar to
what is expected in the case of the collapse of NSs to HQSs. For
example, some models almost collapse to black holes, with the value of
the lapse function reaching $ 0.29 $. The comparsion by
\citet{dimmelmeier_06_b} reveals very good agreement between the CFC
and full GR calculations. We thus conclude that the spacetime of
rapidly rotating NS models (whether uniformly or differentially
rotating) is still very well approximated by the CFC
metric~(\ref{eq:cfc_metric}). The accuracy of the approximation is
expected to degrade only in extreme cases, such as a rapidly rotating
black hole, a self-gravitating thin disc or a compact binary system.


\subsection{General relativistic hydrodynamics}
\label{subsection:gr_hydrodynamics}

The hydrodynamic evolution of a standard relativistic perfect fluid is 
determined by the local conservation equations
\begin{equation}
  \nabla_{\!\mu} J^{\mu} = 0, \qquad \nabla_{\!\mu} T^{\mu \nu} = 0,
  \label{eq:gr_equations_of_motion}
\end{equation}
where $ J^{\mu} = \rho u^{\mu} $ is the rest-mass current, and 
$\nabla_{\!\mu}$ denotes the covariant derivative with respect to the 
four-metric $ g_{\mu \nu} $. Following \citet{banyuls_97_a}, we 
introduce a set of conserved variables in terms of the primitive 
(physical) variables $ (\rho, v_i, \epsilon) $:
\begin{displaymath}
  D = \rho W,
  \qquad
  S_i = \rho h W^2 v_i,
  \qquad
  \tau = \rho h W^2 - P - D.
\end{displaymath}
Using these, the local conservation 
laws~(\ref{eq:gr_equations_of_motion}) can be written as a 
first-order, flux-conservative hyperbolic system of equations,
\begin{equation}
  \frac{\partial \sqrt{\gamma} \bm{U}}{\partial t} +
  \frac{\partial \sqrt{- g} \bm{F}^i}{\partial x^i} =
  \sqrt{- g} \bm{S},
  \label{eq:hydro_conservation_equation}
\end{equation}
with the state vector, flux vector, and source vector being
\begin{equation}
  \setlength{\arraycolsep}{0.14 em}
  \begin{array}{rcl}
  \bm{U} & = & [D, S_j, \tau], \\ [1.0 em]
  \bm{F}^i & = & \displaystyle
  \left[ D \hat{v}^i, S_j \hat{v}^i + \delta^i_j P,
  \tau \hat{v}^i + P v^i \right], \\ [1.0 em]
  \bm{S} & = & \displaystyle
  \Bigl[ 0, \frac{1}{2} T^{\mu \nu}
  \frac{\partial g_{\mu \nu}}{\partial x^j},
  T^{00} \!\left(\!\! K_{ij} \beta^i \beta^j - \beta^j
  \frac{\partial \alpha}{\partial x^j} \!\!\right) + \\ [1.0 em]
  & & \displaystyle \qquad \qquad \qquad \;\:\:
  T^{0j} \!\left(\!\! 2 K_{ij} \beta^i -
  \frac{\partial \alpha}{\partial x^j} \!\!\right)\! +
  T^{ij} K_{ij} \Bigr],
  \end{array}
  \label{eq:hydro_conservation_equation_constituents}
\end{equation}
respectively\footnote{Note that here we use an analytically 
  equivalent reformulation of the energy source term as compared with 
  the one presented in e.g.\ \citet{dimmelmeier_02_a}.}. Here $ 
\hat{v}^i = v^i - \beta^i / \alpha $, and $ \sqrt{-g} = \alpha 
\sqrt{\gamma} $, with $ g = \det (g_{\mu \nu}) $ and $ \gamma = \det 
(\gamma_{ij}) $; the $ {\it \Gamma}^\lambda_{\mu \nu} $ are the 
Christoffel symbols associated with $ g_{\mu \nu} $.

The system of hydrodynamic 
equations~(\ref{eq:hydro_conservation_equation}) is closed by an EoS, 
which relates the pressure to some thermodynamically independent 
quantities; in our case $ P = P (\rho, \epsilon) $.


\subsection{Numerical methods for solving the metric and hydrodynamics
  equations}
\label{subsection:numerical_methods}

The hydrodynamic solver performs the numerical time integration of the 
system of conservation 
equations~(\ref{eq:hydro_conservation_equation}) using a 
high-resolution shock-capturing (HRSC) scheme on a finite-difference 
grid. In (upwind) HRSC methods, a Riemann problem has to be solved at 
each cell interface, which requires the reconstruction of the 
primitive variables $ (\rho, v^i, \epsilon) $ at these interfaces. We 
use the PPM method for making the reconstruction, which yields 
third-order accuracy in space for smooth flows and away from extrema. 
The numerical fluxes are computed by means of Marquina's approximate 
flux formula \citep{donat_98_a}. The time update of the state vector $ 
\bm{U} $ is done using the method of lines in combination with a 
Runge--Kutta scheme having second-order accuracy in time. Once the 
state vector has been updated in time, the primitive variables are 
recovered using an iterative Newton--Raphson method. To numerically 
solve the elliptic CFC metric 
equations~(\ref{eq:cfc_metric_equations}) we utilise an iterative 
nonlinear solver based on spectral methods. The combination of HRSC 
methods for the hydrodynamics and spectral methods for the metric 
equations within a multidimensional numerical code has been presented 
in detail in \citet{dimmelmeier_05_a}.

The \texttt{CoCoNuT} code uses Eulerian spherical polar coordinates $ 
\{r, \theta\} $; for the models discussed in this work we are assuming 
axisymmetry with respect to the rotation axis and also equatorial 
symmetry. The finite-difference grid consists of 200 radial grid 
points and 40 angular grid points, which are equidistantly spaced. A 
small part of the grid, which initially corresponds to 60 radial 
gridpoints, covers an artificial low-density atmosphere, extending 
beyond the stellar surface, whose rest-mass density is $ 10^{-17} $ of 
the initial central rest-mass density of the NS.

Since the calculation of the spacetime metric is computationally 
expensive, it is updated only once every 25 hydrodynamic timesteps 
during the evolution and is extrapolated in between. The suitability 
of this procedure has been tested and discussed in detail in 
\citet{dimmelmeier_02_a}. We also note that tests with different grid 
resolutions were performed to check that the regular grid resolution 
specified above is appropriate for our simulations. A check on the 
relative violation in the conservation of total rest mass, ADM mass 
and total angular momentum showed that each of these quantities is 
conserved to within 1\% during the entire evolution time.


\subsection{Gravitational waves}
\label{subsection:gravitational_waves}

The GWs emitted by the collapsing NS are computed using the Newtonian
quadrupole formula in its first time-integrated form \citep[the
  first-moment of momentum density formulation as described in detail
  in][]{dimmelmeier_02_b} in the variant of \citet{shibata_04_a}. This
yields the quadrupole wave amplitude $ A^\mathrm{E2}_{20} $ as the
lowest-order term in a multipole expansion of the radiation field into
pure-spin tensor harmonics \citep{thorne_80_a}. The wave amplitude is
related to the dimensionless GW strain $ h $ in the equatorial plane
by \citep{dimmelmeier_02_b}
\begin{equation}
  h = \frac{1}{8} \sqrt{\frac{15}{\pi}} \frac{A^\mathrm{E2}_{20}}{r} =
  8.8524 \times 10^{-21}
  \left(\! \frac{A^\mathrm{E2}_{20}}{10^3 \mathrm{\ cm}} \!\right)
  \left(\! \frac{10 \mathrm{\ kpc}}{r} \!\right)\!,
\end{equation}
where $ r $ is the distance from the emitting source.

We point out that although the quadrupole formula is not gauge 
invariant and is only strictly valid in the Newtonian slow-motion 
limit, for GWs emitted by pulsations of rotating NSs it gives results 
that agree very well in phase and to about $ 10 \% \mbox{\,--\,} 20\% 
$ in amplitude with more sophisticated methods \citep{shibata_03_a, 
nagar_07_a}.


\section{Stellar models and treatment of the phase transition}
\label{section:models}


\subsection{Initial neutron star model}
\label{subsection:initial_model}

Following LCCS, we compute the initial equilibrium NS model before the phase
transition using a polytropic EoS,
\begin{equation}
  P = K \rho^\gamma,
  \label{eq:polytrope}
\end{equation}
where $ K $ and $ \gamma $ are constants. We choose $ \gamma = 2 $ 
and $ K = 100 $ (in units where $ M_\odot = 1 $) for all of the GR 
initial models considered in the present study. On the initial 
timeslice, we also need to specify the specific internal energy $ 
\epsilon $. For a polytropic EoS, the thermodynamically consistent $ 
\epsilon $ is given by
\begin{equation}
  \epsilon = \frac{K}{\gamma - 1} \rho^{\gamma - 1}.
\end{equation}


\subsection{Hybrid quark star model}
\label{subsection:quark_star_model}

Due to the complexity of the fundamental theory of strong 
interactions, all theoretical studies of quark matter in compact stars 
are based on phenomenological models. The MIT bag model EoS, which has 
been used extensively for this \citep[for a review see, 
e.g.][]{weber_99_a, glendenning_02_a}, is based on the following 
assumptions: (i) quarks appear in colour neutral clusters confined to 
a finite region of space, the volume of which is limited by the 
negative pressure of the QCD vacuum; (ii) within this region, 
interactions between the quarks are weak and can be treated using 
low-order perturbation theory in the coupling constant. These two 
assumptions allow the two main features of QCD to be modelled, namely 
colour confinement and asymptotic freedom. The parameters of the bag 
model are the bag constant $ B $, the masses of the quarks and the 
running coupling constant $ \alpha_s $, whose value at the scale 
relevant for typical quark chemical potentials is $ \alpha_s \in [0.4, 
0.6] $ \citep{benhar_07_a}.

At the moment there is no general consensus about the value of $ B $. 
Fits to the spectrum of light hadrons give $ B^{1/4} \approx 145 
\mathrm{\ MeV} $ \citep{degrand_75_a}, while the adjustment of $ B $ 
with hadronic structure functions suggests $ B^{1/4} \sim 170 
\mathrm{\ MeV} $ \citep{steffens_95_a}. On the other hand, lattice QCD 
calculations predict values up to $ B^{1/4} \sim 190 \mathrm{\ MeV} $ 
\citep{satz_82_a}. Hereafter, and following LCCS, we take $ B^{1/4} = 
170 \mathrm{\ MeV} $.

The masses of the $ u $ and $ d $ quarks are of the order of a few MeV 
\citep{hagiwara_02_a} and can therefore be mostly neglected, whereas 
the mass of the $ s $ quark is much larger, its value being in the 
range $ m_s \in [80, 155] \mathrm{\ MeV} $. Nevertheless, including 
this mass for the $ s $ quark, rather than taking it to be massless, 
would decrease the pressure by only a few percent \citep{alcock_88_a}. 
We do not expect that this would change our results qualitatively, and 
so we neglect it in our study. We also neglect the residual 
interaction between the quarks and approximate their temperature as 
being zero. The EoS of the MIT bag model for massless and 
non-interacting quarks at zero temperature is given by
\begin{equation}
  P_\mathrm{q} = \frac{1}{3}(e - 4 B),
  \label{eq:bag_eos} 
\end{equation} 
where $ e $ is the total energy density. 

A fundamental problem appears to arise in using the hydrodynamic 
equations of Section~\ref{section:numerics} to describe the quark 
medium since the quarks are being treated as having zero rest mass. 
The quantity $ e $ represents only the internal energy density of the 
quarks and contains no rest-mass contribution. Rest mass appears as a 
fluid quantity, within this picture, only when the quarks become 
confined. Nevertheless, the continuity equation [the first equation of 
the system of the hydrodynamic 
equations~(\ref{eq:hydro_conservation_equation})] remains well-defined 
if thought of in terms of baryon number (which \emph{is} defined for 
the quark medium) rather than in terms of rest mass. In order to have 
a unified treatment for the quark matter and the hadronic matter 
(which is necessary since we have transformation between the two), one 
can define a quantity $ \rho = n_\mathrm{b} m_\mathrm{b} $ in the 
quark medium (where $n_\mathrm{b}$ is the baryon number density and 
$m_\mathrm{b}$ is the rest mass per baryon in the hadronic medium) and 
then formally split the internal energy density of the quark medium 
into two parts, writing $e = \rho+\rho\epsilon$ as usual but bearing 
in mind that the first term on the right hand side represents just a 
part of the internal energy density in the quark phase. If one does 
this, it is easy to show that the treatment of the hydrodynamics goes 
through unchanged in a consistent way, using this $ \rho $ and $ 
\epsilon $.

For the normal hadron matter, during the evolution we use an ideal-gas 
type of EoS
\begin{equation}
  P_\mathrm{h} = (\gamma - 1) \rho \epsilon.
  \label{eq:ideal_gas_eos}
\end{equation}
However, in contrast to LCCS, we do not reduce the adiabatic index $ 
\gamma $ from its initial value of $ 2 $, because we see no physical 
mechanism which could be responsible for a global reduction of 
pressure in the \emph{hadronic} matter phase. Consequently, in our GR 
models the collapse is caused solely by the pressure change due to the 
introduction of \emph{quark} matter in the core of the NS.

As first shown by \citet{glendenning_91_a, glendenning_92_a}, if the 
surface tension between the phases is not too large, relaxing the 
condition of local electrical neutrality would allow for the 
possibility of having coexistence between the two phases within a 
certain range of densities. In a region where this applied, one would 
then have many inter-mixed microscopic zones of the lower-density 
hadronic matter and of the higher-density quark matter. Each zone would
have a net electric charge but with the mixture being electrically 
neutral on average and with the volume fraction occupied by the 
higher-density phase growing from 0 at the lower-density boundary of 
the mixed phase $ \rho_\mathrm{hm} $ up to 1 at the upper-density 
boundary $ \rho_\mathrm{qm} $.

The value of the lower threshold density $ \rho_\mathrm{hm} $, above 
which free quarks start to appear, is rather uncertain. From simple 
geometrical considerations, nucleons should begin to touch at $ \rho 
\sim (4 \pi r_\mathrm{nuc}^2 / 3) $, which for the characteristic 
nucleon radius $ r_\mathrm{nuc} \sim 1 \mathrm{\ fm} $ gives a few 
times nuclear saturation density, $ \rho_\mathrm{nuc} = 2.7 \times 
10^{14} \mathrm{\ g\ cm}^{-3} $ \citep{glendenning_89_a}. For 
densities above this, one expects that the boundaries of particles 
like $ p $, $ n $, $ \Sigma^- $, $ \Lambda $ and $ K^- $ would 
dissolve and that quarks would start to populate free states outside 
the hadrons \citep{weber_99_a}. The value of the upper threshold 
density $ \rho_\mathrm{qm} $, marking the boundary between the mixed 
phase and the pure deconfined quark matter phase, is also uncertain 
(and is model dependent) but it is very probably in the range $ 
4\mbox{\,--\,}10 \, \rho_\mathrm{nuc} $.

According to this picture, a hybrid star would then be composed of 
either two or three parts: (i) a \textit{pure hadronic} matter phase 
for $ \rho < \rho_\mathrm{hm} $, (ii) a \textit{mixed} phase of the 
confined-deconfined matter for $ \rho_\mathrm{hm} < \rho < 
\rho_\mathrm{qm} $, and (iii) a \textit{pure quark} matter phase for 
$ \rho > \rho_\mathrm{qm} $ (this might or might not be present in 
practice, depending on the maximum density reached).

In this paper, we follow LCCS in adopting the picture outlined above 
and we also follow their prescription in formulating the EoS for the 
HQS matter:
\begin{equation}
  P = \left\{
    \begin{array}{ll}
      P_\mathrm{h} & \quad
      \mathrm{for\ } \rho < \rho_\mathrm{hm},
      \\
      \alpha P_\mathrm{q} + (1 - \alpha) P_\mathrm{h} & \quad
      \mathrm{for\ } \rho_\mathrm{hm} \le \rho \le \rho_\mathrm{qm},
      \\
      P_\mathrm{q} & \quad
      \mathrm{for\ } \rho_\mathrm{qm} < \rho,
    \end{array}
  \right.
  \label{eq:mixed_eos}
\end{equation}
where
\begin{equation}
  \alpha = 1 - \left( \frac{\rho_\mathrm{qm} - \rho}
  {\rho_\mathrm{qm} - \rho_\mathrm{hm}} \right)^\delta
  \label{eq:alpha}
\end{equation}
is a factor quantifying the contribution of each of the components of 
the mixed phase to the total pressure. As stated above, we take $ 
P_\mathrm{q} $ to be given by the MIT bag model~(\ref{eq:bag_eos}), 
while $ P_\mathrm{h} $ is calculated using the ideal-gas 
EoS~(\ref{eq:ideal_gas_eos}). The parameter $ \delta $ is introduced 
in order to control the quark matter contribution to the pressure in 
the mixed phase: with larger $ \delta $, the contribution from $ 
P_\mathrm{q} $ increases. For $ \delta = 1 $ we recover the EoS of 
LCCS. We again emphasize that, in contrast to LCCS, we do not reduce 
the effective adiabatic index of the nuclear matter in our GR models, 
but rather keep it at its initial value $ \gamma = 2 $ during the 
evolution.

\begin{table}
  \centering
  \caption{Summary of the set of initial models: $ p $ is the rotation
    period of the NS, $ M_0 $ is the total rest mass, $ M $ is the
    gravitational mass, $ T $ is the rotational mass energy, $ W $ is
    the gravitational binding energy, $ \rho_\mathrm{c,i} $ is the
    central rest-mass density and $ r_\mathrm{e} / r_\mathrm{p} $ is the ratio
    of the equatorial and polar radii. Note that the initial models
    A5, B4 and C3 are identical. Model N is one of the Newtonian
    initial models used by LCCS. }
  \label{tab:initial_models}
  \begin{tabular}{@{}lccc@{~~}c@{~~}cc@{}}
    \hline \\ [-1 em]
    Model &
    $ p $ &
    $ M_0 $ &
    $ M $ &
    $ \rho_\mathrm{c,i} $ &
    $ r_\mathrm{e} / r_\mathrm{p} $ &
    $ T / |W| $ \\
    &
    [ms] &
    [$ M_\odot $] &
    [$ M_\odot $] &
    [$ 10^{14} \mathrm{\ g\ cm}^{-3} $] & &
    [\%] \\
    \hline \\ [- 1 em]
    A1 & 1.00 & 1.98 & 1.81 &  11.25 & 0.635 & 8.44 \\
    A2 & 1.20 & 1.85 & 1.70 &  11.25 & 0.785 & 5.30 \\
    A3 & 1.40 & 1.80 & 1.65 &  11.25 & 0.847 & 3.75 \\
    A4 & 1.60 & 1.77 & 1.62 &  11.25 & 0.885 & 2.80 \\
    A5 & 1.80 & 1.75 & 1.60 &  11.25 & 0.910 & 2.18 \\
    A6 & 2.00 & 1.73 & 1.59 &  11.25 & 0.928 & 1.74 \\
    A7 & 2.99 & 1.70 & 1.57 &  11.25 & 0.968 & 0.76 \\
    A8 & 5.98 & 1.69 & 1.55 &  11.25 & 0.992 & 0.50 \\ [0.5 em]
    B1 & 1.30 & 1.75 & 1.62 & \z8.42 & 0.746 & 6.32 \\
    B2 & 1.40 & 1.75 & 1.61 & \z9.48 & 0.815 & 4.57 \\
    B3 & 1.60 & 1.75 & 1.61 &  10.63 & 0.878 & 2.98 \\
    B4 & 1.80 & 1.75 & 1.60 &  11.25 & 0.910 & 2.18 \\
    B5 & 2.00 & 1.75 & 1.60 &  11.80 & 0.931 & 1.66 \\
    B6 & 2.98 & 1.75 & 1.60 &  12.86 & 0.972 & 0.66 \\ [0.5 em]
    C1 & 1.80 & 1.65 & 1.53 & \z8.74 & 0.882 & 2.89 \\
    C2 & 1.80 & 1.70 & 1.57 & \z9.88 & 0.897 & 2.51 \\
    C3 & 1.80 & 1.75 & 1.60 &  11.25 & 0.910 & 2.18 \\
    C4 & 1.80 & 1.80 & 1.65 &  13.70 & 0.927 & 1.75 \\ [0.5 em]
    N  & 1.20 & 2.20 & 1.89 & \z9.34 & 0.695 & 7.71 \\
    \hline
  \end{tabular}
\end{table}

For our GR models we define the transition density $ \rho_\mathrm{hm} 
$ from the pure hadronic matter phase to the mixed phase to be where $ 
P_\mathrm{q} $ vanishes, similarly to LCCS (although they were 
identifying energy density and rest-mass density within their 
Newtonian regime). This corresponds to $ \rho_\mathrm{hm} = 6.97 
\times 10^{14} \mathrm{\ g\ cm}^{-3} = 2.58 \, \rho_\mathrm{nuc} $ for 
$ B^{1/4} = 170 \mathrm{\ MeV} $ and $ \rho_\mathrm{nuc} = 2.7 \times 
10^{14} \mathrm{\ g\ cm}^{-3} $. Following LCCS, we set $ 
\rho_\mathrm{qm} = 9 \, \rho_\mathrm{nuc} $ in our simulations 
(corresponding to $ 24.3 \times 10^{14} \mathrm{\ g\ cm}^{-3} $ for 
our value of $ \rho_\mathrm{nuc} $), but this is just a rough estimate 
to give a working value. When we make direct comparisons with the 
Newtonian models of LCCS in Section~\ref{subsection:comparison}, we 
use their values $ \rho_\mathrm{hm} = 7.24 \times 10^{14} \mathrm{\ g\ 
cm}^{-3} $ and $ \rho_\mathrm{qm} = 25.2 \times 10^{14} \mathrm{\ g\ 
cm}^{-3} $, which slightly differ from our standard values.

We want to stress that this treatment of the EoS via 
Eqs.~(\ref{eq:mixed_eos}, \ref{eq:alpha}) is extremely rough as a 
representation of matter in the mixed phase experiencing the phase 
transition, especially when one bears in mind the behaviour of fluid 
elements undergoing successive compression and decompression and 
changing the proportions of the phases. In particular, it neglects any 
possible effect from local heating and from the creation and 
subsequent emission of neutrinos during the collapse and the 
subsequent bounces. However, it does have the advantage of being 
simple and parametrisable. By changing the values of the free 
parameters (e.g.\ $ \delta $, $ \rho_\mathrm{hm} $ or $ 
\rho_\mathrm{qm} $) it allows us to modify easily the properties of 
the EoS.


\subsection{Parameter space}
\label{subsection:parameter_space}

\begin{table*}
  \centering
  \caption{Summary of the set of collapse models: $ \gamma $ is the
    adiabatic index of the hadronic matter during the evolution, $
    \delta $ is the EoS parameter that specifies the contribution of
    the quark matter pressure in the mixed phase, $ \rho_\mathrm{c,b}
    $ is the value of the central rest-mass density at bounce, $
    |h|_\mathrm{max} $ is the maximum value of the GW signal strain
    during the evolution, $ E_\mathrm{gw} $ is the energy emitted in
    GWs (during a total emission time of $ t_\mathrm{f} = 50 \mathrm{\ 
    ms} $), $ f_F $ and $ f_{{}^{2\!}f} $ are the frequencies of the 
    fundamental quasi-radial and quadrupolar modes, respectively, 
    while $ \tau_F $ and $ \tau_{{}^{\,2\!}f} $ are the damping times 
    of those modes in the GW signal. In addition, we give the phases $ 
    \phi_1 $ and $ \phi_2 $ of the $ F $-mode and the $ {}^{2\!}f 
    $-mode and the relative amplitude $ A_1 / A_2 $ as obtained from a 
    fit to the GW signal according to Eq.~(\ref{eq:curve_fit}). Note 
    that the collapse models CA5, CB4 and CC3 are identical, as are 
    models CB2 and CD2, and models CB5 and CD7. During the 
    contraction, models CB6, CC4 and CD9 form an apparent horizon and 
    become black holes. The phase-transition-induced collapse models 
    CN1, CN2 and CN3, as well as the test collapse models \cnn\ and 
    \cni\ are computed with a Newtonian treatment. Where no values 
    are given for the damping times, this signifies that the model 
    either collapses to form a black hole or that no unambiguous 
    damping could be diagnosed in the GW signal (mostly due to mode 
    resonance; see Section~\ref{subsection:mode_resonance}).}
  \label{tab:collapse_models}
  \begin{tabular}{@{}lcccc@{~~}c@{~~}cccccccc@{}}
    \hline \\ [-1 em]
    Model &
    Initial &
    $ \gamma $ &
    $ \delta $ &
    $ \rho_\mathrm{c,b} $ &
    $ |h|_\mathrm{max} $ &
    $ E_\mathrm{gw} $ &
    $ f_F $ &
    $ f_{{}^{\,2\!}f} $ &
    $ \tau_F $ &
    $ \tau_{{}^{\,2\!}f} $ &
    $ \phi_1 $ &
    $ \phi_2 $ &
    $ A_1 / A_2 $ \\
    &
    \highentry{model} & & &
    \highentry{[$ 10^{14} \mathrm{\ g\ cm}^{-3} $]} &
    $ \displaystyle \left[ \!\!\!
      \begin{array}{c}
        10^{-23} \\ [-0.2 em]
        \mathrm{\ at\ 10\ Mpc}
      \end{array}
    \! \right] $ &
    \highentry{[$ 10^{-4} M_\odot c^2 $]} &
    \highentry{[kHz]} &
    \highentry{[kHz]} &
    \highentry{[ms]} &
    \highentry{[ms]} &
    \highentry{[rad]} &
    \highentry{[rad]} \\ [0.8 em]
    \hline \\ [-1 em]
    CA1  & A1 & 2.00 &  2  &  15.81 & \z1.45 & 0.04 & 0.87 & 2.01 & \z40 & \z12 & $ \m0.06 $ & $  -3.07 $ & 0.93 \\
    CA2  & A2 & 2.00 &  2  &  16.04 & \z1.33 & 0.29 & 0.99 & 2.08 & \z49 &  130 & $  -0.48 $ & $ \m3.24 $ & 0.40 \\
    CA3  & A3 & 2.00 &  2  &  15.81 & \z1.90 & 0.59 & 1.05 & 2.08 &  --- & \z18 & $  -3.91 $ & $  -6.44 $ & 0.15 \\
    CA4  & A4 & 2.00 &  2  &  15.92 & \z1.10 & 0.17 & 1.07 & 2.08 &  319 & \z67 & $  -0.15 $ & $ \m0.30 $ & 0.64 \\
    CA5  & A5 & 2.00 &  2  &  15.99 & \z0.62 & 0.11 & 1.09 & 2.06 &  418 &  --- & $ \m0.09 $ & $ \m3.00 $ & 1.44 \\
    CA6  & A6 & 2.00 &  2  &  16.02 & \z0.52 & 0.11 & 1.12 & 2.04 &  270 & \z51 & $  -0.15 $ & $ \m1.50 $ & 1.78 \\
    CA7  & A7 & 2.00 &  2  &  16.10 & \z0.25 & 0.05 & 1.16 & 2.02 &  711 &  --- & $  -0.46 $ & $ \m3.44 $ & 3.01 \\
    CA8  & A8 & 2.00 &  2  &  16.07 & \z0.09 & 0.01 & 1.19 & 2.00 &  --- &  --- & $ \m5.98 $ & $ \m6.35 $ & 1.69 \\ [0.5 em]
    CB1  & B1 & 2.00 &  2  & \z9.52 & \z0.42 & 0.01 & 1.05 & 1.78 & \z99 & \z37 & $ \m0.06 $ & $ \m3.17 $ & 0.57 \\
    CB2  & B2 & 2.00 &  2  &  11.70 & \z0.69 & 0.05 & 1.06 & 1.90 &  133 & \z44 & $ \m0.04 $ & $ \m3.06 $ & 0.71 \\
    CB3  & B3 & 2.00 &  2  &  14.36 & \z0.74 & 0.09 & 1.10 & 2.02 &  196 & \z59 & $ \m0.10 $ & $  -3.11 $ & 0.87 \\
    CB4  & B4 & 2.00 &  2  &  15.99 & \z0.62 & 0.11 & 1.10 & 2.06 &  418 &  --- & $ \m0.09 $ & $ \m3.00 $ & 0.64 \\
    CB5  & B5 & 2.00 &  2  &  17.92 & \z2.41 & 1.16 & 1.08 & 2.10 &  687 & \z76 & $ \m6.16 $ & $ \m6.02 $ & 0.06 \\
    CB6  & B6 & 2.00 &  2  &    --- &    --- &  --- &  --- &  --- &  --- &  --- &   \zz---   &   \zz---   &  --- \\ [0.5 em]
    CC1  & C1 & 2.00 &  2  &  10.31 & \z0.36 & 0.01 & 1.14 & 1.76 &  143 & \z53 & $ \m0.13 $ & $ \m3.21 $ & 0.77 \\
    CC2  & C2 & 2.00 &  2  &  12.71 & \z0.51 & 0.04 & 1.13 & 1.90 & \z71 & \z47 & $  -0.03 $ & $ \m3.27 $ & 0.86 \\
    CC3  & C3 & 2.00 &  2  &  15.99 & \z0.62 & 0.11 & 1.10 & 2.06 &  418 &  --- & $ \m0.09 $ & $ \m3.00 $ & 0.64 \\
    CC4  & C4 & 2.00 &  2  &    --- &    --- &  --- &  --- &  --- &  --- &  --- &   \zz---   &   \zz---   &  --- \\ [0.5 em]
    CD1  & B2 & 2.00 &  1  &  10.47 & \z0.40 & 0.01 & 1.15 & 1.86 &  150 & \z36 & $ \m0.00 $ & $ \m3.17 $ & 0.69 \\
    CD2  & B2 & 2.00 &  2  &  11.70 & \z0.69 & 0.05 & 1.07 & 1.90 &  133 & \z44 & $ \m0.04 $ & $ \m3.06 $ & 0.71 \\
    CD3  & B2 & 2.00 &  3  &  12.93 & \z1.07 & 0.08 & 1.04 & 1.94 & \z54 & \z39 & $ \m0.15 $ & $ \m2.96 $ & 0.78 \\
    CD4  & B2 & 2.00 &  4  &  14.16 & \z1.47 & 0.14 & 1.10 & 1.96 & \z19 & \z48 & $ \m0.35 $ & $  -1.42 $ & 2.32 \\
    CD5  & B2 & 2.00 &  5  &  15.35 & \z1.84 & 0.15 & 1.14 & 1.97 & \zz8 & \z44 & $ \m0.04 $ & $ \m3.17 $ & 0.80 \\
    CD6  & B5 & 2.00 &  1  &  14.41 & \z0.31 & 0.04 & 1.13 & 2.04 &  248 & \z62 & $ \m0.00 $ & $  -3.18 $ & 1.04 \\
    CD7  & B5 & 2.00 &  2  &  17.92 & \z2.41 & 1.16 & 1.09 & 2.10 &  687 & \z76 & $ \m6.16 $ & $ \m6.02 $ & 0.06 \\
    CD8  & B5 & 2.00 &  3  &  24.80 & \z1.14 & 0.24 & 1.02 & 2.15 &  --- & \z34 & $  -0.61 $ & $ \m3.41 $ & 0.46 \\
    CD9  & B5 & 2.00 &  4  &    --- &    --- &  --- &  --- &  --- &  --- &  --- &   \zz---   &   \zz---   &  --- \\ [0.5 em]
    CN1  & N  & 1.95 &  1  &  10.37 & \z4.41 & 0.88 & 2.68 & 1.98 & \z73 & \z80 & $ \m3.17 $ & $ \m0.00 $ & 1.12 \\
    CN2  & N  & 1.85 &  1  &  12.31 &  15.38 & 4.10 & 2.88 & 2.14 & \zz9 & \z20 & $ \m2.95 $ & $  -0.08 $ & 1.03 \\
    CN3  & N  & 1.75 &  1  &  14.19 &  21.22 & 5.70 & 3.18 & 2.34 & \zz1 & \z15 & $ \m2.85 $ & $  -0.38 $ & 0.98 \\
    \cnn & N  & 1.90 & --- &  12.32 &  10.11 & 4.96 & 2.67 & 2.04 & \z71 & \z67 & $ \m3.32 $ & $ \m0.06 $ & 1.10 \\
    \cni & N  & 2.00 & --- &  11.82 & \z7.63 & 2.86 & 2.83 & 2.04 & \z86 & \z67 & $ \m2.97 $ & $  -0.01 $ & 1.05 \\
    \hline
  \end{tabular}
\end{table*}

The properties of our models for the phase-transition-induced collapse 
depend on a number of free parameters including the initial rotation 
period $ p $, the total stellar rest mass $ M_0 $, the pressure 
contribution of the quark component in the mixed phase (controlled by 
$ \delta $), etc. In order to study how the collapse dynamics depends 
on these quantities, we performed simulations for various sequences of 
models where one parameter was held fixed. For instance, in order to 
investigate the impact of rotation we used the models A1 to A8 and B1 
to B6 (see Table~\ref{tab:initial_models}). The models of sequence A 
have a fixed central rest-mass density $ \rho_\mathrm{c,i} = 11.25 
\times 10^{15} \mathrm{\ g\ cm}^{-3} $ and a varying rotation period 
in the range from $ p = 1.00 $ to $ 6.18 \mathrm{\ ms} $. The ones of 
sequence B have a fixed rest mass $ M_0 = 1.75 \, M_\odot $ and 
rotation periods from $ p = 1.30 $ to $ 2.89 \mathrm{\ ms} $. The 
consequences of a variation in the rest mass of the initial NS are 
explored in the sequence of models C1 to C4, where the rotation period 
is held fixed at $ p = 1.80 \mathrm{\ ms} $ and the rest mass takes 
values from $ M_0 = 1.65 $ to $ 1.80 \, M_\odot $.

These initial models were then evolved as sequences CA, CB, and CC 
with fixed EoS parameters, choosing $ \delta = 2 $ in each case. For 
the CD sequence, however, we used different values for $ \delta $ 
(from $ 1 $ to $ 5 $) in order to assess how varying the quark 
contribution to the EoS in the mixed phase influences the dynamics. In 
this sequence, the models CD1 to CD5 use the initial model A2, while 
CD6 to CD9 are based on the initial model A5.

In addition, in order to validate our code and to discuss the results 
obtained by LCCS in more detail, we simulated some of their models 
with a Newtonian version of the \texttt{CoCoNuT} code. Our Newtonian 
models CN1, CN2 and CN3 (all based on the equilibrium model N) are 
identical to their models G1.95, R and G1.75, with the adiabatic index 
$ \gamma $ of the EoS being reduced everywhere as they had done. For 
these models we used exactly the same EoS as in LCCS, which differs 
from our regular EoS in various ways (see 
Section~\ref{subsection:quark_star_model}). The models labelled \cnn\ 
and \cni\ are Newtonian test models whose properties are discussed in 
detail in Section~\ref{subsection:convection_newtonian}.

In Table~\ref{tab:collapse_models} we list important quantities 
obtained in the simulations of the collapse models which are discussed 
in the following sections. Note that the total evolution time for all 
models is $ t_\mathrm{f} = 50 \mathrm{\ ms} $. At that point we simply 
terminate the evolution, which however could be extended to much 
longer times given the long-term stability of our code.


\section{Comparison with Newtonian models}
\label{subsection:comparison}

Before discussing the results for our GR models, we present here our 
simulations of three of the Newtonian models also studied by LCCS. We 
begin, however, with a description of the qualitative features of 
phase-transition-induced collapse of a rotating NS to a HQS which is 
relevant also for the later GR models.


\subsection{Collapse dynamics and gravitational radiation waveform}
\label{subsection:collapse_dynamics_newtonian}

\begin{figure}
  \centerline{\includegraphics[width = 85 mm]{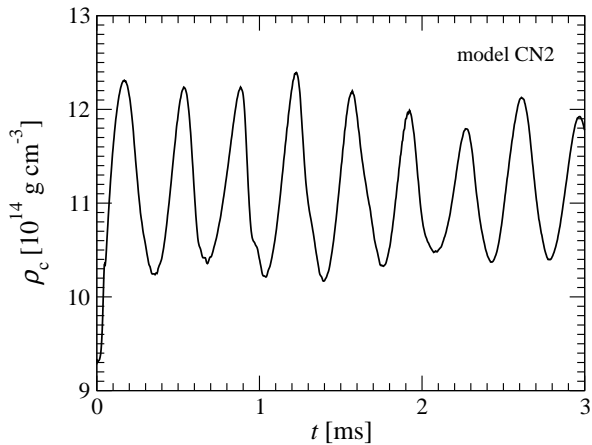}}
  \caption{Time evolution of the central rest-mass density $ \rho_\mathrm{c} $
    for the Newtonian collapse model CN2 (which is identical to model
    R of LCCS).}
  \label{fig:central_density_evolution_sample_newtonian}
\end{figure}

The EoS of the deconfined quark matter in the stellar core generally 
gives a smaller pressure contribution than that of the hadronic 
EoS\footnote{In principle, the quark matter could give as
  large a pressure contribution as the hadronic matter \citep[see,
  e.g.][for a recent discussion]{alford_07_a}. However, in our study
we do not consider those cases.} and so the phase transition in the 
NS core leads to an instability of the progenitor NS (which is in 
equilibrium before the transition) and the entire NS starts to 
contract. Depending on the parameters used and the phase transition 
timescale, the infall phase typically lasts between $ 0.3 $ and $ 0.5 
\mathrm{\ ms} $. As the pressure in the core rises with increasing 
density, the infall decelerates and the contraction of the inner core 
is stopped, while the outer regions are still falling in. In the case 
of a rotating NS, the deceleration of the core can be augmented by the 
increase of centrifugal forces due to angular momentum conservation in 
the contraction phase.

Because of its inertia, the core overshoots its new equilibrium 
configuration, rebounds, expands again and then re-collapses. It 
typically experiences many such distinct sequences of infall, bounce, 
and re-expansion in the form of pronounced, mainly radial ring-down 
pulsations until it finally reaches a new equilibrium state. 
Fig.~\ref{fig:central_density_evolution_sample_newtonian} shows the 
time evolution of the central rest-mass density $ \rho_\mathrm{c} $ 
for the Newtonian model CN2, where this oscillatory behaviour can be 
clearly seen.

\begin{figure}
  \centerline{\includegraphics[width = 85 mm]{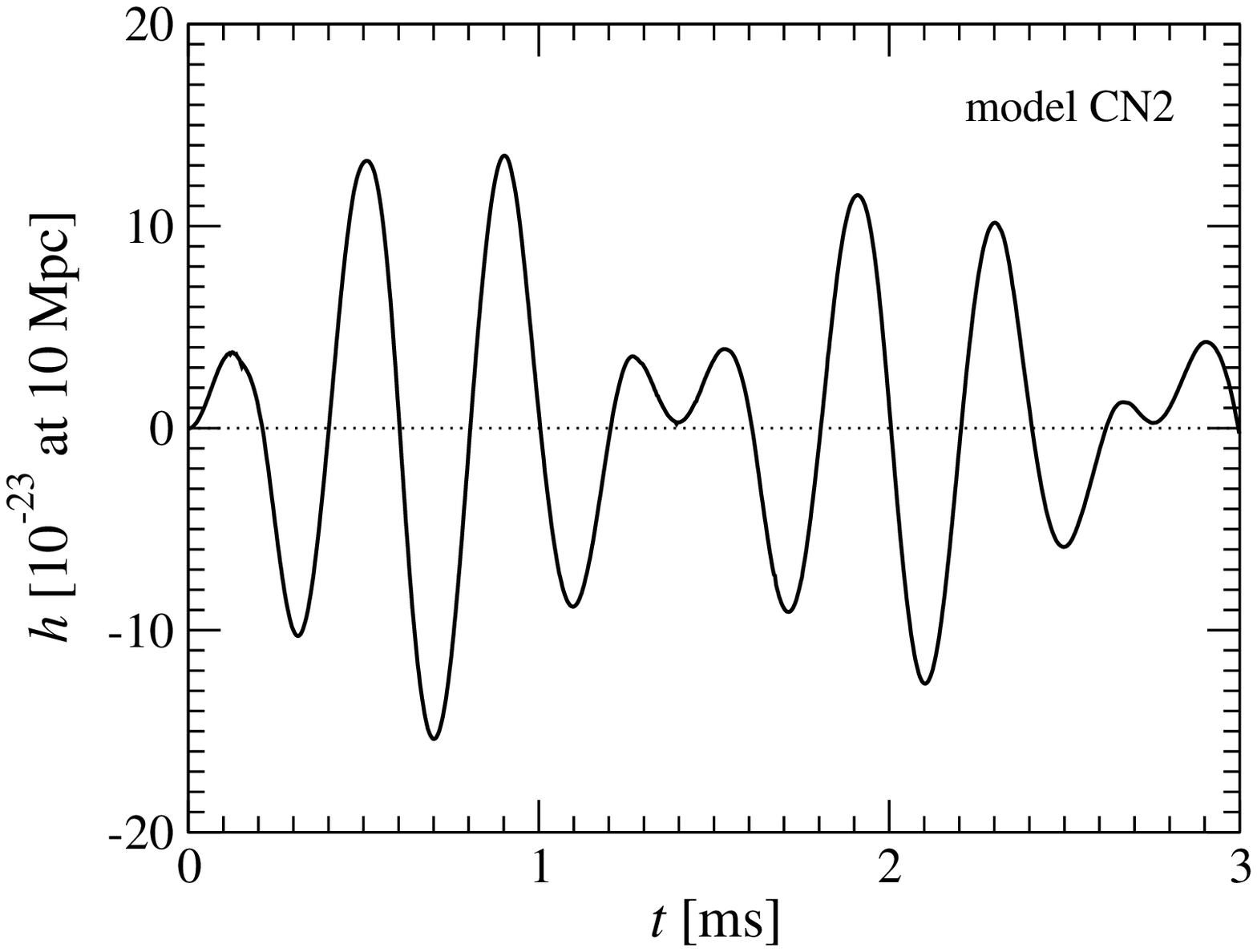}}
  \caption{Time evolution of the GW strain $ h $ at a distance of
    $ 10 \mathrm{\ Mpc} $ for the Newtonian collapse model CN2.}
  \label{fig:waveform_sample_newtonian}
\end{figure}

\begin{figure}
  \centerline{\includegraphics[width = 85 mm]{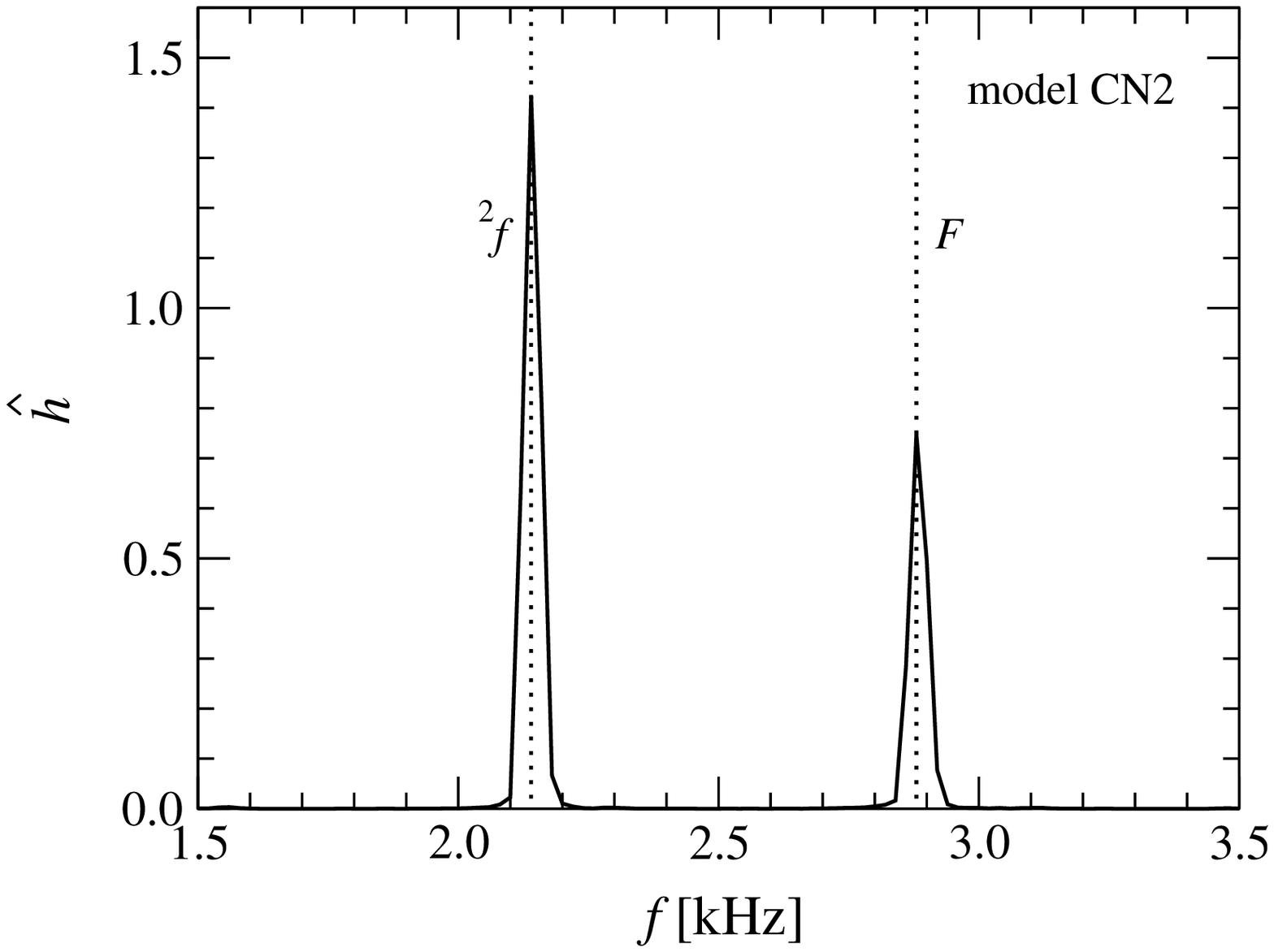}}
  \caption{Power spectrum $\hat{h}$ (in arbitrary units) of the GW
    strain $ h $ for the Newtonian collapse model CN2. The narrow
    peaks of the $ F $-mode and $ {}^{2\!}f $-mode clearly dominate
    the spectrum.}
  \label{fig:waveform_spectrum_sample_newtonian}
\end{figure}

If the initial NS is non-rotating, then the newly-formed HQS pulsates 
only radially, and only $ l = 0 $ modes are present in the oscillation 
spectrum (unless significant convection develops). In this case, no 
GWs are emitted. However, if a rotating initial model undergoes a 
collapse and ring-down, then GWs of considerable amplitude can be 
emitted, as shown in the waveform plot in 
Fig.~\ref{fig:waveform_sample_newtonian}, again for model CN2.

Comparing a simple collapse model with a purely ideal-gas EoS to a 
regular collapse model with a quark contribution to the EoS, LCCS 
demonstrated that the phase-transition-induced collapse of a rotating 
NS to a HQS predominantly excites two quasi-normal pulsation modes: 
the fundamental $ l = 0 $ quasi-radial $ F $-mode and the fundamental 
$ l = 2 $ quadrupolar ${}^{2\!}f$-mode. These stand out in the power 
spectrum of many fluid or metric quantities and, in particular, in 
that of the gravitational radiation waveform as presented in 
Fig.~\ref{fig:waveform_spectrum_sample_newtonian}. All of our collapse 
models exhibit the predominance of these two modes as a generic 
feature\footnote{We perform the mode identification by perturbing 
  equilibrium models that have similar structure to those of the 
  collapse remnants using specific $ l = 0 $ and $ l = 2 $ trial 
  eigenfunctions and analysing the response to these perturbations. 
  This method is described in detail in LCCS.}. For the more slowly 
rotating models, the contribution from higher order modes at higher 
frequencies can become comparable to that of the two fundamental $ l = 
0, 2 $ modes, but these higher frequency modes have damping times that 
are significantly shorter than those of the fundamental modes and 
their impact on the waveform (and other quantities) dies out quickly.

\begin{table*}
  \centering
  \caption{Comparison of various quantities as calculated in our 
    simulations for the Newtonian collapse models CN1, CN2 and CN3 
    (top row for each model) with the results published by LCCS 
    (bottom row). For each quantity we also give the relative 
    difference between our results and theirs. Note that we multiply 
    their values for $ \langle \hat{h}^2 \rangle^{1/2} $ by $ \sqrt{2} 
    $ to undo the angular averaging and obtain $ |h|_\mathrm{max} $. 
    The values of $ \rho_\mathrm{c,b} $ for models CN2 and CN3 are 
    extracted from Figs.~6 and~10 in their article, respectively, 
    while for model CN1 they present no data from which to read off 
    the central rest-mass density at the bounce.}
  \label{tab:comparison}
  \begin{tabular}{@{}lc@{\quad}rc@{\quad}rc@{\quad}rc@{\quad}rc@{\quad}rc@{\quad}r@{}}
    \hline \\ [-1 em]
    Model &
    \multicolumn{2}{c}{$ \rho_\mathrm{c,b} $} &
    \multicolumn{2}{c}{$ |h|_\mathrm{max} $} &
    \multicolumn{2}{c}{$ f_F $} &
    \multicolumn{2}{c}{$ f_{{}^{2\!}f} $} &
    \multicolumn{2}{c}{$ \tau_F $} &
    \multicolumn{2}{c}{$ \tau_{{}^{\,2\!}f} $} \\
    &
    \multicolumn{2}{c}{[$ 10^{14} \mathrm{\ g\ cm}^{-3} $]} &
    \multicolumn{2}{c}{[$ 10^{-23} $ at $ 10 \mathrm{\ Mpc} $]} &
    \multicolumn{2}{c}{[kHz]} &
    \multicolumn{2}{c}{[kHz]} &
    \multicolumn{2}{c}{[ms]} &
    \multicolumn{2}{c}{[ms]} \\ [0.2 em]
    \hline \\ [-1 em]
    \lowentry{CN1} &
    \quad  10.37 & \lowentry{---~~} &
    \quad \z4.41 & \lowentry{4\%} \quad\, &
            2.68 & \lowentry{1\%} &
            1.98 & \lowentry{1\%} &
           72.78 & \lowentry{600\%} &
           80.23 & \lowentry{670\%} \\
    &
    \quad    --- & &
    \quad \z4.24 & &
            2.66 & &
            2.00 & &
           10.39 & &
           10.42 \\ [0.5 em]
    \lowentry{CN2} &
    \quad  12.31 & \lowentry{$ < 1\% $} &
    \quad  15.38 & \lowentry{$ < 1\% $} \quad\, &
            2.88 & \lowentry{2\%} &
            2.14 & \lowentry{3\%} &
          \z8.56 & \lowentry{234\%} &
           20.10 & \lowentry{325\%} \\
    &
    \quad  12.30 & &
    \quad  15.41 & &
            2.82 & &
            2.08 & &
          \z2.56 & &
          \z4.73 \\ [0.5 em]
    \lowentry{CN3} &
    \quad  14.19 & \lowentry{$ < 1\% $} &
    \quad  21.22 & \lowentry{2\%} \quad\, &
            3.18 & \lowentry{2\%} &
            2.34 & \lowentry{4\%} &
          \z1.45 & \lowentry{245\%} &
           14.51 & \lowentry{474\%} \\
    &
    \quad  14.14 & &
    \quad  21.64 & &
            3.12 & &
            2.25 & &
          \z0.42 & &
          \z2.53 \\
    \hline
  \end{tabular}
\end{table*}

Comparing our values of selected quantities describing the collapse 
dynamics with the corresponding ones in LCCS shows that our code is
able to accurately reproduce the original results (see
Table~\ref{tab:comparison}), despite being based on a different
formulation of the hydrodynamic equations and utilising a different
coordinate system. Furthermore,
Figs.~\ref{fig:central_density_evolution_sample_newtonian},
\ref{fig:waveform_sample_newtonian}
and~\ref{fig:waveform_spectrum_sample_newtonian}, which correspond to
the data plotted in Figs.~6, 7 (centre panel) and~12 (dashed line) of
LCCS, also exemplify the excellent agreement both qualitatively and
quantitatively. Making this comparison and demonstrating the good
agreement obtained at a Newtonian level is important for removing
possible doubts about the analysis which we present in the next
subsection concerning the onset and development of convective
instabilities.


\subsection{The role of convection in generating differential rotation}
\label{subsection:convection_newtonian}

A conspicuous difference from the results of LCCS that we observe in 
the simulations of Newtonian models performed with our code is the 
significantly \emph{smaller} damping of the post-bounce oscillations. 
This is not only apparent from the much larger values that we find for 
the damping times $ \tau_F $ and $ \tau_{{}^{\,2\!}f} $ (see 
Table~\ref{tab:comparison}) but can also be noticed by comparing our 
Figs.~\ref{fig:central_density_evolution_sample_newtonian} 
and~\ref{fig:waveform_sample_newtonian} to the corresponding Figs.~6 
and~7 (centre panel) in LCCS. This has important consequences for the 
physical interpretation given in LCCS for the damping of the 
quasi-radial post-bounce pulsations: they suggested that the dominant 
damping mechanism is conversion of the kinetic energy of these 
pulsations into differential rotation. According to their discussion, 
another less significant part of that kinetic energy is lost when 
matter at the boundary of the HQS is ejected by shock waves, while yet 
another small amount of damping is due to numerical dissipation.

The 3D Newtonian code of LCCS used a coarser grid spacing than in our
2D GR calculations and this (together with some other possible
numerical effects) would have led to a higher level of numerical
dissipation. If the damping of post-bounce pulsations which they saw
was indeed mostly caused by \emph{physical} processes such as the
transformation of kinetic energy into differential rotation or mass
shedding at the boundary, then we would not have seen much smaller
damping in our simulations for the same models. We therefore conclude
that numerical dissipation actually did play a major role for the
damping seen in the simulations of LCCS (although physical processes
certainly also played a role).

\begin{figure}
  \centerline{\includegraphics[width = 85 mm]{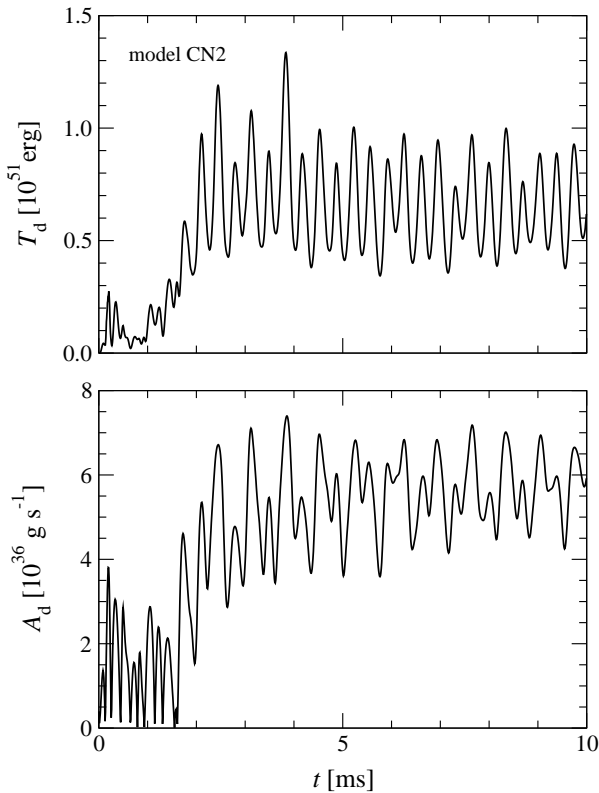}}
  \caption{Time evolution of the differential rotation measures $
    T_\mathrm{d} $ (top panel) and $ A_\mathrm{d} $ (bottom panel) for
    the Newtonian collapse model CN2.}
  \label{fig:differential_rotation_measure_newtonian}
\end{figure}

Furthermore, if the observed exponential damping, which removes energy 
from the pulsations at a constant relative rate, were directly feeding 
the generation of differential rotation, there should be a simple 
correlation between the evolution time $ t $ and the increase in the 
quantity used as a measure of differential rotation by LCCS,
\begin{equation}
  T_\mathrm{d} = \frac{1}{2} \int \rho \, r^2 \sin^2 \theta \,
  (\Omega - \bar{\Omega})^2 \, dV,
  \label{eq:differential_rotation_measure_energy_newtonian}
\end{equation}
where $ \Omega = r^{-1} \sin^{-1} \! \theta \; v_\phi $ is the local 
angular velocity with $ v_\phi $ being the rotation velocity of the 
fluid, and $ \bar{\Omega} $ is the volume averaged angular velocity of 
the HQS\footnote{Note that in contrast to LCCS, we averaged $ 
  \bar{\Omega} $ at each time level and not only once at the time of
  bounce, because when we did the latter, we obtained a very oscillatory
  behaviour for $ T_\mathrm{d} $; we are not clear why LCCS did not
  observe this.}. Note that $ T_\mathrm{d} $ does not follow the 
additive property for energies and so is not a fully satisfactory 
measure of the kinetic energy associated with differential rotation. 
We instead prefer to use as our measure of differential rotation, a 
quantity which is a volume-averaged and density-weighted measure of 
the relevant $ r\phi $ component of the Newtonian shear tensor,
\begin{equation}
  A_\mathrm{d} = \int \rho
  \left| \frac{d v_\phi}{dr} - \frac{v_\phi}{r} \right| \, dV.
  \label{eq:differential_rotation_measure_alternative_newtonian}
\end{equation}
The two 
quantities~(\ref{eq:differential_rotation_measure_energy_newtonian}) 
and (\ref{eq:differential_rotation_measure_alternative_newtonian}) are 
related and they have similar time evolutions, as shown for model CN2 
in Fig.~\ref{fig:differential_rotation_measure_newtonian} (and as 
found for all of the models considered), but we prefer to use $ 
A_\mathrm{d} $ because of it having a clearer physical meaning.

Overall, the evolution of the two measures of differential rotation 
has the following general behaviour: after an initial peak associated 
with the initial collapse and the main bounce (with its height 
correlating with the intensity of the bounce), the average 
differential rotation stays roughly the same for around a millisecond 
(corresponding to several dynamical timescales during which a number 
of post-bounce oscillations occur) following which it grows 
considerably up to a maximum value, then decreases a little and 
finally oscillates around an almost constant state. This behaviour is 
not in accordance with having continuous conversion of pulsational 
kinetic energy into differential rotation as the main damping 
mechanism. Rather, it suggests that some other process is responsible 
for creating the observed differential rotation.

The initial peak of $ T_\mathrm{d} $ and $ A_\mathrm{d} $ can be 
readily explained by the fact that any initially uniform rotation 
profile will become non-uniform during the collapse as a result of the 
non-homology of the collapse. On the other hand, the intermittent 
behaviour after the initial peak and the saturation at a constant 
value can be interpreted straightforwardly in terms of differential 
rotation caused by large-scale convection developing in the HQS 
several dynamical timescales after the initial collapse when it is 
still pulsating but is already close to a new quasi-equilibrium state.

\begin{figure}
  \centerline{\includegraphics[width = 85 mm]{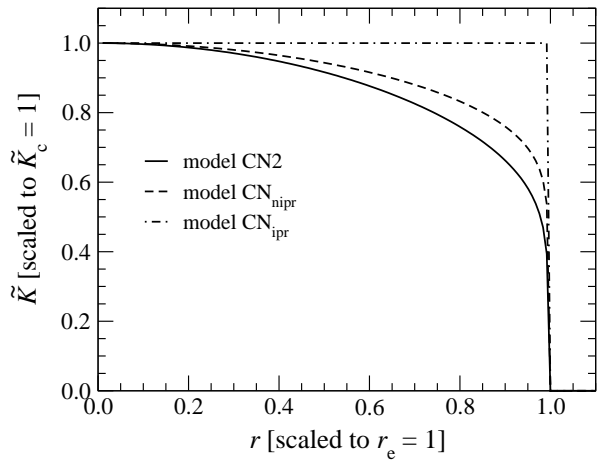}}
  \caption{Radial profiles of the specific entropy measure $ \tilde{K} 
    $ (top panel) and of the $ r $-component of its gradient (bottom 
    panel), 
    both calculated in the equatorial plane at the initial time for the 
    Newtonian
    models CN2 (solid line), \cnn\ (dashed line) and \cni\ (dash-dotted
    line). Both the entropy and the radius are scaled to give
    $ \tilde{K}_\mathrm{c} = 1 $ and $ r_\mathrm{e} = 1 $.}
  \label{fig:negative_entropy_profile}
\end{figure}

In rotating stellar models significant convection can occur if the 
Solberg--H{\o}iland stability criteria are violated \citep[see, 
e.g.][and references therein]{cerda_07_a}. For our simple EoS, these 
translate into the condition that convection can develop if there is a 
sufficiently strong negative radial gradient of specific entropy 
(depending on the rotation rate of the HQS). We find that the method 
of pressure reduction employed by LCCS, which involves uniformly 
lowering $ \gamma $ in the ideal-gas EoS~(\ref{eq:ideal_gas_eos}) for 
the hadronic phase without adjusting the value for the internal 
specific energy $ \epsilon $ at the initial time, indeed results in a 
very large negative specific-entropy gradient for their choices of $ 
\gamma $. This is shown for model CN2 in 
Fig.~\ref{fig:negative_entropy_profile} (solid line), where we plot 
the radial profile at $ t = 0 $ of the (density dependent) measure of 
specific entropy
\begin{equation}
  \tilde{K} = \frac{\rho \epsilon (\gamma - 1)}{\rho^\gamma}
  \label{eq:entropy_measure}
\end{equation}
in the equatorial plane, assuming an ideal-gas EoS for the entire 
star with a $ \gamma $ that is reduced from its initial value of $ 2 $ 
to $ 1.85 $. Note that for a polytrope $ \tilde{K} $ is identical to
the polytropic constant $ K $.

\begin{figure*}
  \centerline{\includegraphics[width = 180 mm]{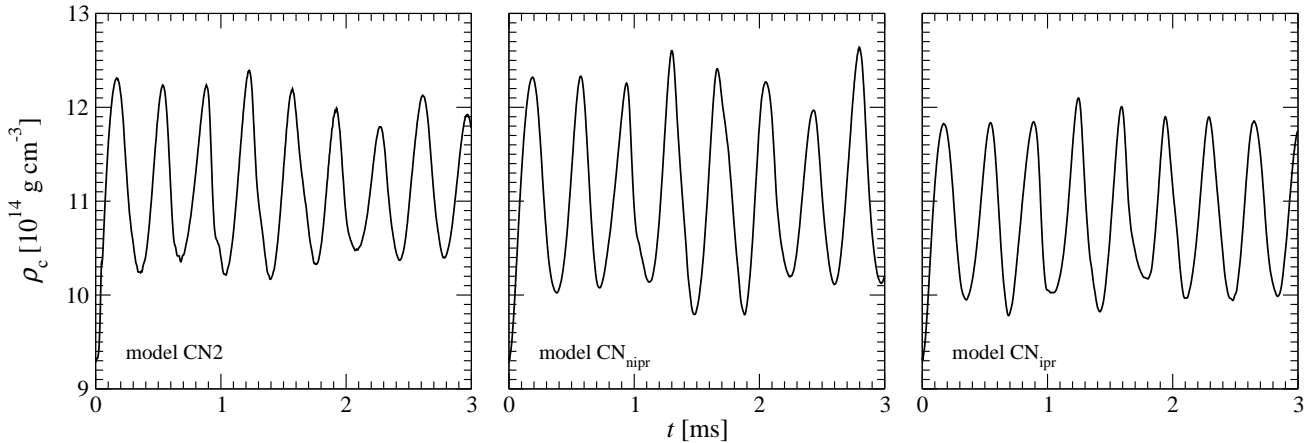}}
  \caption{Time evolution of the central rest-mass density $ \rho_\mathrm{c} $
    for the Newtonian models CN2 (left panel), \cnn\ (centre panel)
    and \cni\ (right panel).}
  \label{fig:central_density_evolution_comparison}
\end{figure*}

\begin{figure*}
  \centerline{\includegraphics[width = 180 mm]{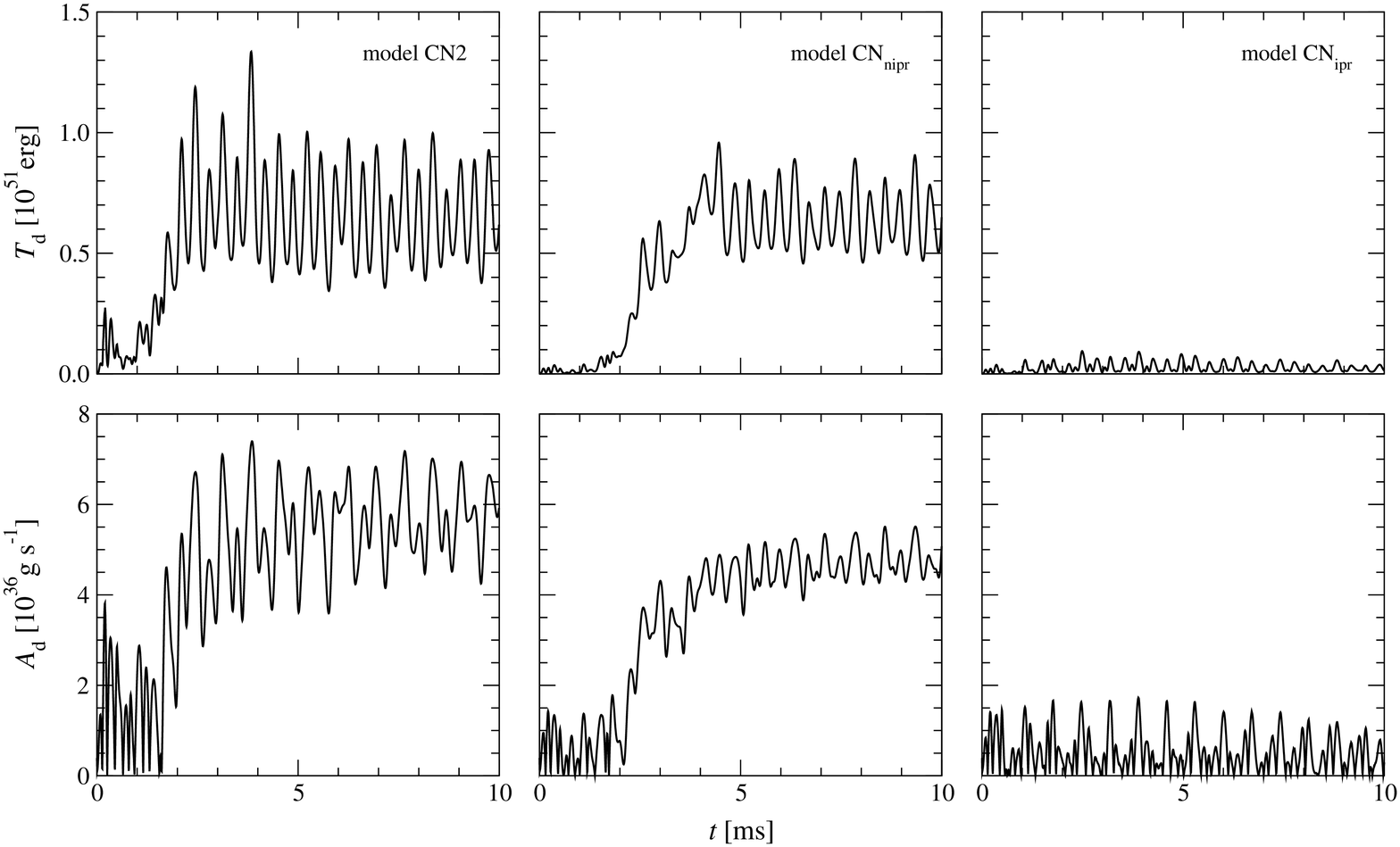}}
  \caption{Time evolution of the differential rotation measures
    $ T_\mathrm{d} $ (top row) and $ A_\mathrm{d} $ (bottom row) for
    the Newtonian collapse models CN2 (left
    panels), \cnn\ (centre panels) and \cni\ (right panels).}
  \label{fig:differential_rotation_measure_comparison}
\end{figure*}

\begin{figure*}
  \centerline{\includegraphics[width = 180 mm]{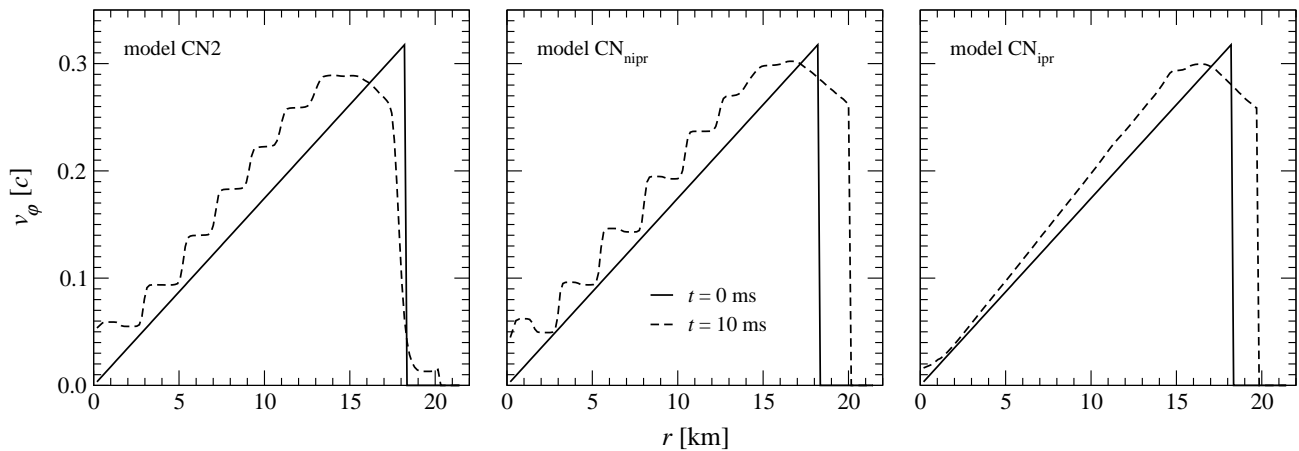}}
  \caption{Radial profile of the rotation velocity $ v_\phi $ in
    the equatorial plane for the the Newtonian collapse models CN2
    (left panel), \cnn\ (centre panel) and \cni\ (right panel) at the
    initial time (solid lines) and at $ t = 10 \mathrm{\ ms} $ (dashed
    lines). For models that develop considerable convection, the
    initially uniform rotation profile is locally destroyed.}
  \label{fig:impact_of_convection_on_rotation}
\end{figure*}

In order to assess unambiguously the occurrence of 
artificially-produced convection and its impact on the development of 
differential rotation, we set up two simple Newtonian collapsing test 
models, which are both based on the initial model N and utilise a 
purely ideal-gas EoS~(\ref{eq:ideal_gas_eos}), so as to simplify the 
discussion by removing the influence of quark matter on the results. 
In model \cnn\ (with non-isentropic pressure reduction) we reduce the 
pressure by lowering the adiabatic index $ \gamma $ from its initial 
value to $ 1.9 $ (without then adjusting $ \epsilon $); this creates a 
strong initial negative specific-entropy gradient that is comparable 
with the one in model CN2 (see 
Fig.~\ref{fig:negative_entropy_profile}). In model \cni\ (for 
isentropic pressure reduction) $ \gamma $ remains at its pre-collapse 
value of $ 2 $ during the evolution, while the pressure reduction is 
now realized by a lowering the polytropic constant $ K $ by 10\%, 
which keeps the specific entropy uniform throughout the entire NS.

Note that the parameters in the EoS for these two test models have 
been chosen in such a way that the collapse dynamics (represented by 
the maximum density reached at the first bounce and the amplitude of 
the post-bounce pulsations) is comparable with that of model CN2, as 
can be seen from Fig.~\ref{fig:central_density_evolution_comparison}. 
Consequently, the gravitational radiation waveforms of these models 
also have amplitudes and waveforms similar to those of model CN2.

However, because of the different behaviour regarding convective 
instability, the dynamics of the three models CN2, \cnn\ and \cni\ 
fall into two very distinct classes, depending on whether the initial 
pressure reduction creates a strong negative specific-entropy profile 
or leaves the specific entropy constant. In all models, analysis of 
the meridional velocity fields shows no noticeable convection being 
present at early times (e.g.\ at the time of the main bounce, around $ 
t = 0.2 \mathrm{\ ms} $). In the isentropic collapse model \cni\, 
convection continues to remain unimportant also at later times and, 
accordingly, both $ T_\mathrm{d} $ and $ A_\mathrm{d} $ remain very 
small at all times (see the right-hand panel of 
Fig.~\ref{fig:differential_rotation_measure_comparison}). In stark 
contrast, the non-isentropic models CN2 and \cnn\ develop several 
convection vortices in the bulk of the collapsed star at $ t \sim 1 
\mathrm{\ ms} $ (corresponding to a few dynamical timescales), which 
is when the two differential rotation measures start to increase. This 
convection grows rapidly and reaches saturation almost 
instantaneously, with typical maximum convection velocities of $ \sim 
0.03\,c $. During the entire period when $ T_\mathrm{d} $ and $ 
A_\mathrm{d} $ are increasing, the convection remains practically 
constant and redistributes angular momentum and entropy locally within 
each vortex. By times $ t \gtrsim 5 \mathrm{\ ms} $ this has led to 
the specific entropy being almost constant within the spatial scale of 
each convection vortex. Convection then subsides again, and $ 
T_\mathrm{d} $ and $ A_\mathrm{d} $ remain approximately constant from 
then on. It is clear, therefore, that the distinct phases in the time 
evolution of the two quantities reflect very accurately the distinct 
phases of convection, which acts as the mechanism that redistributes 
angular momentum and thus creates differential rotation.

\begin{figure*}
  \centerline{\includegraphics[width = 180 mm]{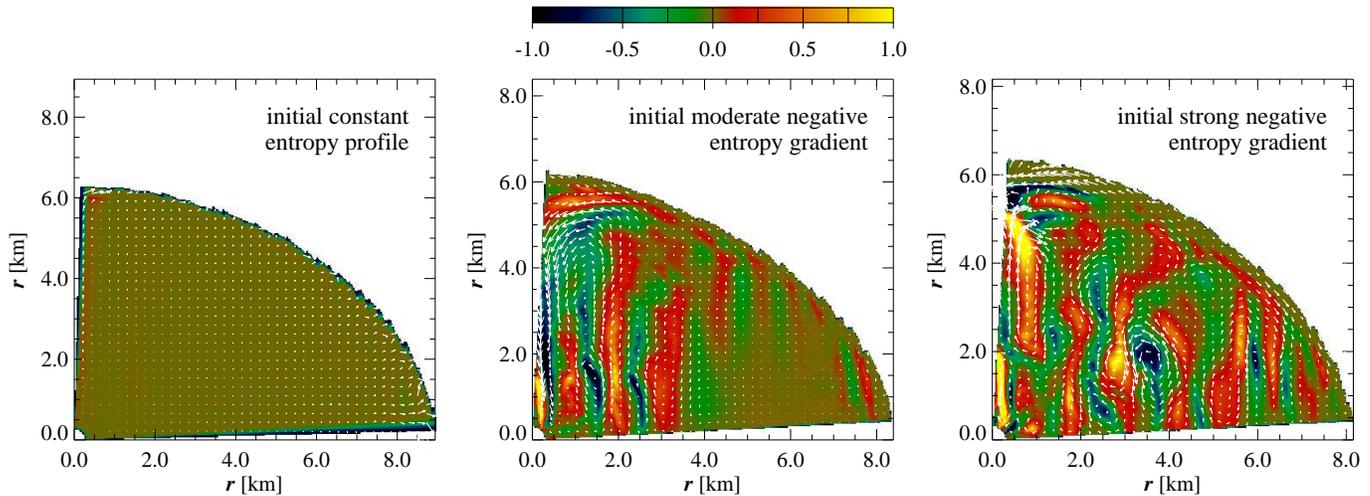}}
  \caption{The velocity field $ v_{r,\theta} $ in the meridional plane
    (with the rotation axis being in the vertical direction) and the
    colour-coded magnitude of the $\varphi$-component of the vorticity
    are shown for equilibrium models whose initial specific-entropy 
    profile is, respectively, constant (left panel), comparable to 
    model CN2 (right panel) and roughly halfway between those two 
    cases (centre panel). The evolution time is at $ t = 3.5 \mathrm{\ 
    ms} $ when convection (if present) has saturated. The length-scale 
    of the vectors is the same for all three models, with the maximum 
    length corresponding to $ \sim 0.03\,c $. The vorticity is 
    measured in arbitrary units.}
  \label{fig:convection}
\end{figure*}

The impact of angular momentum redistribution by convection on the 
initially uniform rotation profile can be seen from 
Fig.~\ref{fig:impact_of_convection_on_rotation} which shows plots of 
the rotation velocity $ v_\phi $ in the equatorial plane for the 
different models. For the strongly convective models CN2 (left panel) 
and \cnn\ (centre panel), the $ v_\phi $ curves are driven away from 
the initial uniform rotation (straight-line) profile to reach a 
step-like profile at late times, whereas the essentially 
non-convective model \cni\ (right panel) maintains an approximately 
uniform rotation profile\footnote{This rather good preservation of
  uniform rotation is a consequence of the infall being nearly 
  homologous here.}. 

\begin{figure*}
  \centerline{\includegraphics[width = 180 mm]{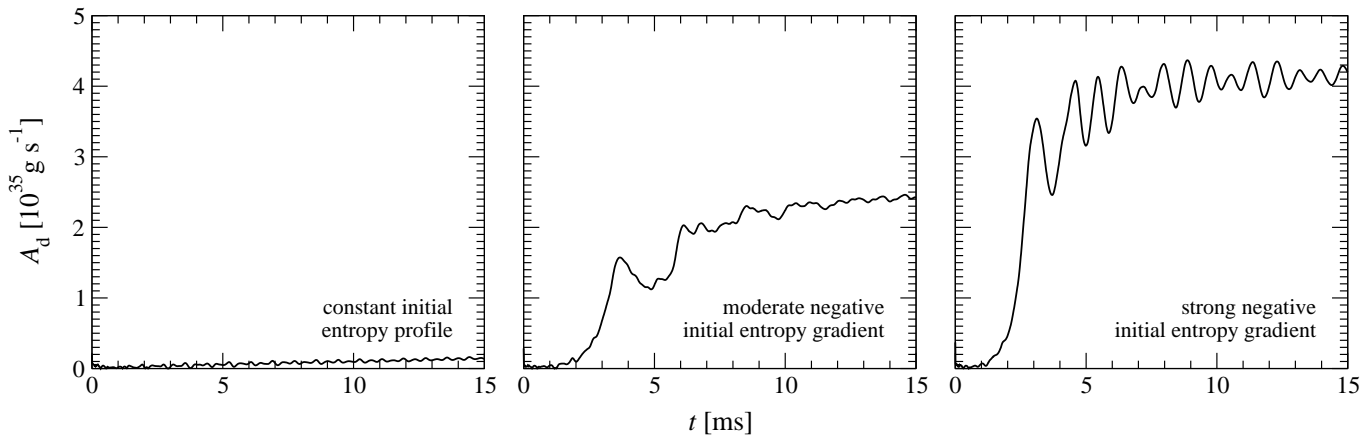}}
  \caption{Time evolution of the differential rotation measure
    $ A_\mathrm{d} $ for equilibrium models whose initial
    specific-entropy profile is, respectively, constant (left panel),
    comparable to model CN2
    (right panel) and roughly halfway between these two cases (centre
    panel). For the isentropic model, $ A_\mathrm{d} $ remains very
    close to zero at all times and is barely visible in the plot.}
  \label{fig:differential_rotation_measure_equilibrium}
\end{figure*}

It is worth stressing that the occurrence of convection here, and thus 
the creation of the differential rotation, is essentially independent 
of the presence and strength of post-bounce pulsations. Indeed, 
setting aside some small spurious convection close to the stellar 
boundary (caused by interaction of matter with the artificial 
low-density atmosphere), the strong \emph{bulk} convection is caused 
by having a negative specific-entropy gradient and can be switched on 
and off at will, depending on whether the initial pressure reduction 
destroys isentropy or not.

In order to demonstrate this connection further, we constructed
initial equilibrium models with a \emph{local} polytropic EoS,
\begin{equation}
  P = K (\rho) \;\rho^\gamma,
  \label{eq:local_polytrope}
\end{equation}
with $ \gamma = 2 $, where the polytropic ``constant'' $ K (\rho) $ 
depends on the rest-mass density and thus varies inside the NS. By 
adjusting $ K (\rho) $ we can thus create models with negative (as 
well as positive) specific-entropy gradients of arbitrary strengths. When these 
initial models are evolved with an ideal-gas 
EoS~(\ref{eq:ideal_gas_eos}), they remain essentially in equilibrium, 
pulsating with only very small amplitudes. However, during the 
evolution they develop convection, and subsequently differential 
rotation, with a strength that directly corresponds to the strength of 
the negative specific-entropy gradient imposed initially. This is 
illustrated in Fig.~\ref{fig:convection} where we show plots of the 
meridional velocity field superimposed on the magnitude of the 
$\varphi$-component of the vorticity. For this figure we selected 
three such equilibrium models for which the initial specific-entropy 
profile varies between constant specific entropy and a gradient that 
is comparable to model CN2\footnote{Note that these equilibrium
  models have central densities comparable with that of CN2 but they 
  are more compact.}. Clearly, convection is practically nonexistent 
in the isentropic model (left panel), whereas the model with moderate 
initial non-isentropy (centre panel) develops considerable convection 
in those parts of the NS that are not too far from the rotation axis. 
Finally, for the model with the strong negative initial 
specific-entropy gradient (right panel), conspicuous convection 
vortices occur throughout the NS. Plots of the vorticity for 
collapsing models show similar patterns but there the meridional 
velocity field is also affected by the contribution from the large 
quasi-radial pulsations.

Also for these equilibrium models the time evolution of $ A_\mathrm{d} 
$ exhibits the expected behaviour, as shown in 
Fig.~\ref{fig:differential_rotation_measure_equilibrium} (we here no 
longer plot $ T_\mathrm{d} $ which, however, exhibits a very similar 
time evolution to that of $ A_\mathrm{d} $). Note that the 
quasi-periodic modulation of $ A_\mathrm{d} $ is not caused by 
pulsations of the star, as their amplitudes are too small to be 
visible in $ A_\mathrm{d} $ and they have higher frequencies. Instead, 
these oscillations (which can also be seen in a power spectrum of $ 
A_\mathrm{d} $ for the collapse models CN2 and \cnn) are due to 
temporal variations in the vortices, with their timescale being 
determined by the typical convection velocity and the average vortex 
size.

In summary, we find that the differential rotation reported by LCCS is 
almost exclusively caused by the transient convection that occurs if a 
negative specific-entropy gradient is generated by the initial pressure 
reduction. Therefore, we are convinced that the conclusion drawn by 
LCCS about the link between the damping of the large amplitude 
post-bounce pulsations and the creation of differential rotation, 
although seemingly plausible, is not correct. They observed that the 
kinetic energy stored in the pulsations is approximately equal to the 
maximum value of $ A_\mathrm{d} $ for both model CN2 and model CN3 
(their models R and G1.75; see their Fig.~8) but this is an 
unfortunate coincidence. Our models \cnn\ and \cni\ demonstrate 
that convection and thus the maximum value of $ A_\mathrm{d} $ can 
vary enormously for roughly constant pulsation amplitude.


\section{General relativistic collapse models}
\label{section:gr_collapse_models}


\subsection{Collapse dynamics and gravitational radiation waveform}
\label{subsection:collapse_dynamics}

We next present our results for the GR models, employing a quark 
contribution to the EoS that is slightly different from that used for 
the Newtonian models (see Section~\ref{subsection:quark_star_model}). 
Overall, we have performed simulations for 23 different models, and in 
Table~\ref{tab:collapse_models} we summarize the most important 
quantities which characterize the dynamics of each model.

\begin{figure}
  \centerline{\includegraphics[width = 85 mm]{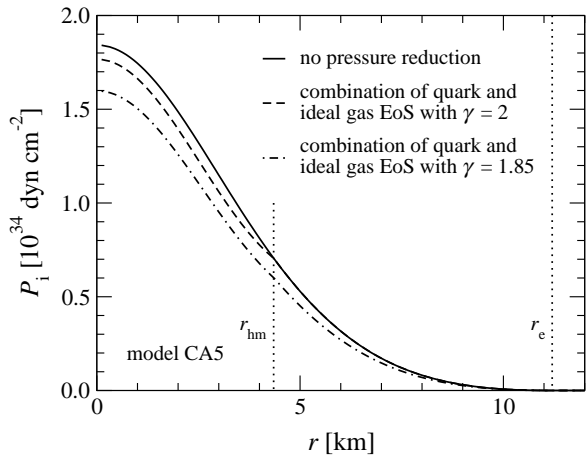}}
  \caption{Radial profile of the initial pressure $ P_\mathrm{i} $ in
    the equatorial plane for the GR collapse model CA5 without any
    pressure reduction (solid line), with the regular EoS treatment
    (combination of quark and ideal-gas EoS with $ \gamma = 2 $;
    dashed line) and with the EoS description used by LCCS
    (combination of quark and ideal-gas EoS with $ \gamma = 1.85 $;
    dash-dotted line). In the latter case $ P $ is reduced throughout
    the star and not just where $ \rho > \rho_\mathrm{hm} $. The
    vertical dotted lines mark the radius
    $ r_\mathrm{hm} = 4.35 \mathrm{\ km} $ where
    $ \rho = \rho_\mathrm{hm} $ and the equatorial radius
    $ r_\mathrm{e} = 11.2 \mathrm{\ km} $ of the initial NS.}
  \label{fig:pressure_reduction}
\end{figure}

As discussed previously, our prescription for the triggering of the 
phase transition and of the subsequent collapse differs from the one 
proposed by LCCS in that we do not change the adiabatic index $ \gamma 
$, leaving the EoS in the hadronic phase unchanged. Although still 
very idealized, we believe that this represents a more consistent 
description of the physics of the phase transition. As a consequence 
of this prescription, the phase transition and collapse in our GR 
models is solely caused by the lower pressure exerted by matter which 
is transformed to the quark phase. The difference between our approach 
and that of LCCS is exemplified in Fig.~\ref{fig:pressure_reduction}, 
where we show the different prescriptions for the initial 
pressure reduction when applied to the representative model CA5.

With our prescription, only a small (central) part of the NS loses 
pressure support, and thus the NS will not, in general, collapse to a 
black hole even if the initial model is close to the stability limit. 
Rather, it reaches a new stable equilibrium state in the form of an 
HQS. For instance, in model CA5 only $ 0.49 \, M_\odot $ out of a 
total mass of $ 1.75 \, M_\odot $ is subject to the pressure 
reduction.  Nevertheless, the change of the central rest-mass density 
during the contraction from its initial value $ \rho_\mathrm{c,i} $ to 
$ \rho_\mathrm{c,b} $ at bounce reaches values of up to $ \sim 50\% $ 
for some models (see Tables~\ref{tab:initial_models} 
and~\ref{tab:collapse_models}), which is comparable to what was 
obtained by LCCS in their Newtonian models with a larger overall 
pressure reduction. There are at least two different reasons behind 
this large and \emph{local} increase of $ \rho $: firstly, the 
stronger gravitational force that the NS experiences in a GR framework 
naturally amplifies the strength of the collapse. Secondly, our 
initial equilibrium models (in particular the ones with high $ 
\rho_\mathrm{c,i} $) are already close to the limit beyond which the $ 
F $-mode becomes unstable. For these models, therefore, even a 
moderate perturbation can lead to a strong local contraction and 
trigger post-bounce oscillations of significant amplitude\footnote{For
  a non-rotating HQS with the EoS of Eq.~(\ref{eq:mixed_eos}) the
  unstable branch starts at $ \rho_\mathrm{c} = 29.6 \times 10^{14}
  \mathrm{\ g\ cm}^{-3} $.}.

\begin{figure}
  \centerline{\includegraphics[width = 85 mm]{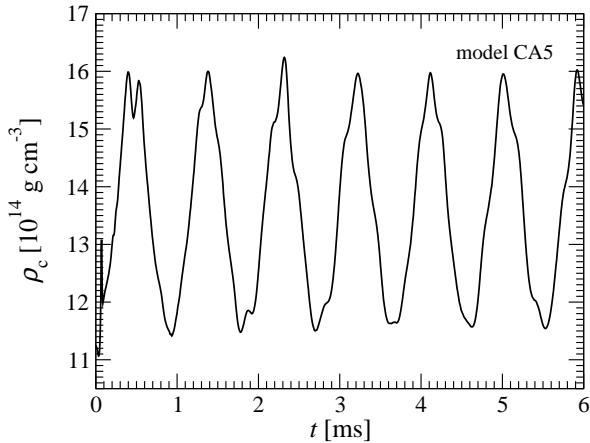}}
  \caption{Time evolution of the central rest-mass density $ \rho_\mathrm{c} $
    for the GR collapse model CA5.}
  \label{fig:central_density_evolution_sample}
\end{figure}

\begin{figure}
  \centerline{\includegraphics[width = 85 mm]{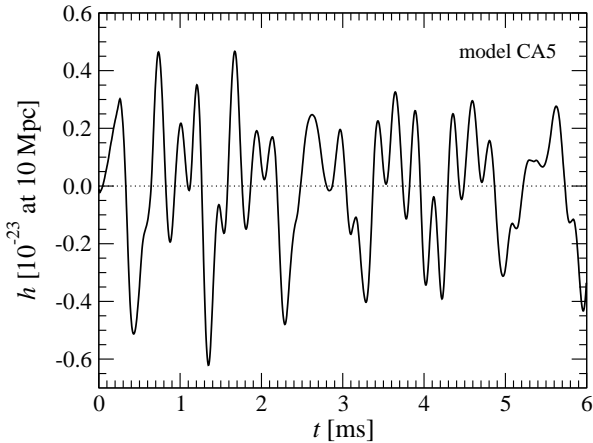}}
  \caption{Time evolution of the GW strain $ h $ at a distance of $ 10
    \mathrm{\ Mpc} $ for the GR collapse model CA5.}
  \label{fig:waveform_sample}
\end{figure}

Despite this conceptually important difference in the way that the 
phase transition (and hence the resulting collapse) is triggered, no 
major \emph{qualitative} differences appear when comparing results 
from the GR simulations with those from the Newtonian simulations of 
LCCS. This is shown for the representative model CA5 in 
Figs.~\ref{fig:central_density_evolution_sample} 
and~\ref{fig:waveform_sample}, where we plot the time evolution of the 
central rest-mass density $ \rho_\mathrm{c} $ and the GW strain $ h $, 
respectively. We point out, however, that in our GR models of
sequences CA, CB, CC, and CD we observe a small spike in the time
evolution of the central rest mass density $ \rho_\mathrm{c} $ at
about $ 0.1 \mathrm{\ ms} $ (see
Fig.~\ref{fig:central_density_evolution_sample}). This spike is caused
by a density compression wave triggered by the non-uniform pressure
reduction in those models, which at the start of the evolution leads
to a noticeable gradient in the first radial derivative of the
pressure at the interface between the mixed and pure hadronic matter
phases.

\begin{figure}
  \centerline{\includegraphics[width = 85 mm]{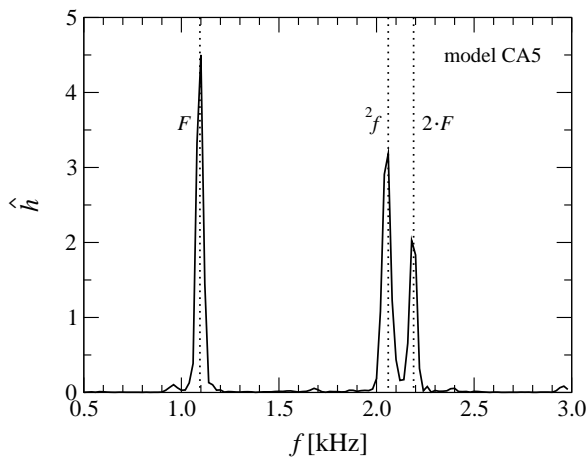}}
  \caption{Power spectrum  $\hat{h}$ (in arbitrary units) of the GW
    strain $ h $ for the GR collapse model CA5. The narrow peaks of
    the $ F $- and $ {}^{2\!}f $-modes (along with a prominent
    nonlinear self-coupling of the $ F $-mode) clearly dominate the
    spectrum.}
  \label{fig:waveform_spectrum_sample}
\end{figure}

As in the Newtonian case, the waveform is mainly composed of the
fundamental $ l = 0 $ quasi-radial $ F $-mode and of the fundamental
$ l = 2 $ quadrupolar $ {}^{2\!}f $-mode (see the Fourier spectrum of
the GW signal in Fig.~\ref{fig:waveform_spectrum_sample}). However, in
contrast to the Newtonian models, the $ F $-mode is now at a lower
frequency than the $ {}^{2\!}f $-mode (with the $ F $-mode now having
frequencies between $ 0.87 $ and $ 1.19 \mathrm{\ kHz} $ for our
selection of models, and  the $ {}^{2\!}f $-mode having frequencies
between $ 1.76 $ and $ 2.15 \mathrm{\ kHz} $). This difference is a
consequence of the different density profile in the GR case and is in
agreement with previous investigations of pulsating $ \gamma = 2 $
equilibrium polytropes \citep[see][and references
therein]{dimmelmeier_06_a}. The prominent peak next to that for the
$ {}^{2\!}f $-mode is a nonlinear self-coupling of the $ F $-mode at
twice the original frequency, which (like several other such nonlinear
modes) is strongly excited due to the violent nature of the collapse.
Using the fitting procedure described in
Appendix~\ref{appendix:damping_times}, we have extracted from the
waveforms the damping times for these two modes, obtaining values
between $ \tau_F = 8 $ and $ 687 \mathrm{\ ms} $, and
$ \tau_{{}^{\,2\!}f} = 18 $ and $ 130 \mathrm{\ ms} $, respectively.
Because of the much smaller numerical dissipation of our code, the
damping times computed for the GR models are considerably longer than
those for the Newtonian models calculated by LCCS.

Another important quantitative difference with respect to the 
Newtonian models appears in the maximum GW strain $ |h|_\mathrm{max} $ 
that is significantly smaller here for comparable overall rotation 
rates. The deeper gravitational potential well in GR gives a shorter 
contraction timescale and higher densities at bounce, which tend to 
amplify the GW signal, while this is counteracted by the more compact 
collapsed remnant in GR having a smaller quadrupole moment \citep[for 
  a detailed discussion of these two competing effects, 
  see][]{dimmelmeier_02_b}. However, the main reason for $ 
|h|_\mathrm{max} $ here being smaller than for the Newtonian models is 
that within our scenario for the destabilisation, a smaller proportion 
of the total matter content of the NS is directly involved in the 
collapse and in undergoing large density variations. For a source at $ 
10 \mathrm{\ Mpc} $, $ |h|_\mathrm{max} $ ranges between $ 0.08 $ and 
$ 1.72 \times 10^{-23} $ for all of the models considered here, while 
the energy $ E_\mathrm{gw} $ emitted in GWs during the first $ 50 
\mathrm{\ ms} $ of the evolution ranges between $ 10^{-6} $ and $ 
10^{-4} \, M_\odot c^2 $. A more detailed discussion about the 
detectability of these sources is presented in 
Section~\ref{section:gravitational_wave_detectability}.


\subsection{Influence of rotation, total rest mass and composition of
  the mixed phase}
\label{subsection:influence_of_parameters}

We next investigate the impact on the collapse dynamics of varying the 
values of the main model parameters: the initial rotation period $ p 
$, the rest mass $ M_0 $, and the exponent $ \delta $ used in the 
mixed phase.

\noindent\textit{Rotation.} As in any gravitational collapse, the 
influence of rotation is twofold. On the one hand, the rotational 
flattening of the collapsing fluid produces an increase of the 
quadrupole moment, which could potentially lead to stronger GW 
emission. On the other hand, centrifugal forces also tend to slow down 
the collapse and, in cases where they are strong enough, they can 
considerably reduce the time variation of the quadrupole moment and 
hence the GW amplitude.

\begin{figure}
  \centerline{\includegraphics[width = 85 mm]{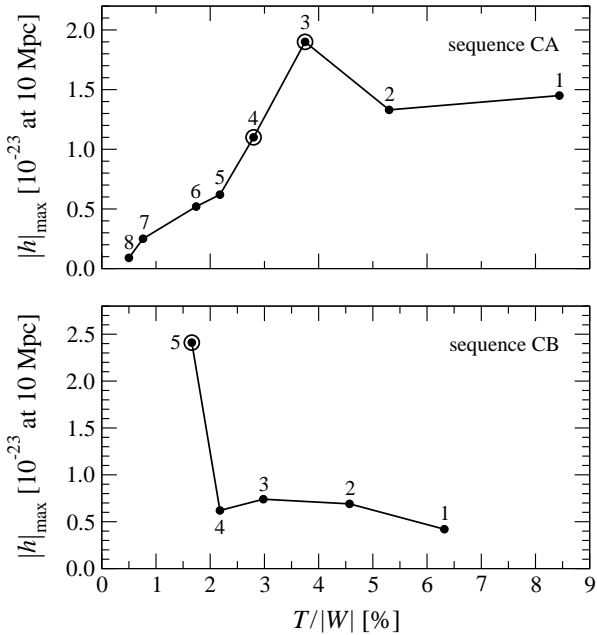}}
  \caption{Dependence of $ |h|_\mathrm{max} $ on the initial rotation
    strength $ T / |W| $ for the collapse model sequences CA (top
    panel) and CB (bottom panel). Each model is denoted with a filled
    circle, and models with enhanced GW emission due to mode
    resonance (see Section~\ref{subsection:mode_resonance}) are
    shown with an additional large circle.}
  \label{fig:maximum_strain}
\end{figure}

For the sequence CA, in which the initial central rest-mass density $ 
\rho_\mathrm{c,b} $ is kept constant, the dependence of the maximum GW 
strain $ |h|_\mathrm{max} $ on the rotation rate is rather 
straightforward to interpret (see the top panel of 
Fig.~\ref{fig:maximum_strain} and Table~\ref{tab:collapse_models}). 
Except for models CA3 and CA4, whose expected GW emission is altered 
by resonance effects (as discussed in detail in 
Section~\ref{subsection:mode_resonance}), $ |h|_\mathrm{max} $ 
increases monotonically with increasing rotation (which we here 
measure in terms of the ratio $ T / |W| $, as this quantity turns out 
to remain almost constant throughout the evolution). Since in this 
sequence the values of the central rest-mass density at bounce are all 
close to $ 16 \times 10^{14} \mathrm{\ g\ cm}^{-3} $, we conclude that 
centrifugal forces do not significantly retard the collapse for these 
models and so there is no significant associated weakening effect for 
the GW emission.

The effect of rotation is also investigated in the constant rest mass
sequence CB, for which we observe a slight initial increase of $
|h|_\mathrm{max} $ with increasing rotation (see the bottom panel of
Fig.~\ref{fig:maximum_strain}) that is then reversed for very rapid 
rotation (model CB5, whose GW emission is again enhanced by mode 
resonance, is an exceptional case). We attribute this behaviour to the 
fact that in this sequence the central rest-mass density, both in the 
initial model and at bounce, drops significantly as rotation grows 
along the sequence, resulting in a much less violent collapse. This 
property can be seen in the central over-density at bounce, which 
amounts to $ \rho_\mathrm{c,b} / \rho_\mathrm{c,i} - 1 = 39\% $ for 
model CB5, but reaches only $ 13\% $ for model CB1. The interplay of 
growing rotation and decreasing collapse strength then explains the 
behaviour of the GW peak amplitude seen for sequence CB.

\begin{figure}
  \centerline{\includegraphics[width = 85 mm]{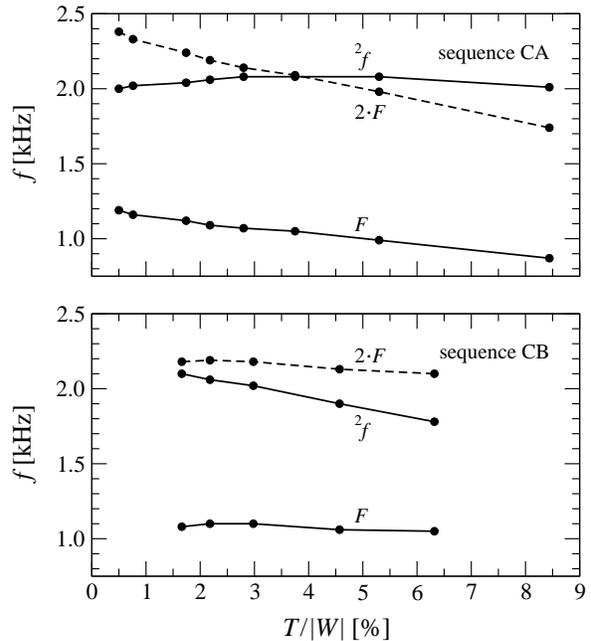}}
  \caption{Dependence of the mode frequencies $ f_F $,
    $ f_{{}^{2\!}f} $ and $ f_{2 \cdot\! F} $ on the initial rotation
    strength $ T / |W| $ for the collapse model sequences CA (top
    panel) and CB (bottom panel). Each model is marked by a filled
    circle.}
  \label{fig:frequencies}
\end{figure}

The influence of rotation on the frequencies of the $ F $-mode and the 
$ {}^{2\!}f $-mode can be compared directly with the results of 
\citet{dimmelmeier_06_a}, who obtained relations between the 
frequencies of these two modes and the strength of rotation for 
pulsating equilibrium models of NSs with a $ \gamma = 2 $ polytropic 
EoS, both for sequences with constant central rest-mass density and 
with constant rest mass. In agreement with that work, for our 
dynamical collapse models, $ f_F $ decreases noticeably with more 
rapid rotation for sequence CA and CB, as shown in 
Fig.~\ref{fig:frequencies} (see also Table~\ref{tab:collapse_models}). 
Also, for $ f_{{}^{2\!}f} $ our models reproduce the initial increase 
and subsequent decrease with growing rotation for the constant initial 
central rest-mass density sequence CA \citep[sequence BU 
in][]{dimmelmeier_06_a} as well as the monotonic decrease with 
increasing rotation for the constant rest mass sequence CB (their 
sequence AU). Our results therefore demonstrate that studies of linear 
pulsations of equilibrium models can be used also in the more general 
context of stellar gravitational collapse to predict the dependence of 
the fundamental mode frequencies on rotation. In our models, the 
creation of differential rotation by non-homologous contraction during 
the collapse phase is rather small. However, even if the deviation 
from uniform rotation were stronger, the study of 
\citet{dimmelmeier_06_a} suggests that the influence on the mode 
frequencies would still be weak.

In Fig.~\ref{fig:frequencies} we also show the behaviour of the 
nonlinear self-coupling of the $ F $-mode at twice the original 
frequency (the $ 2 \cdot\! F $-mode) which is strongly excited in all 
of our models (see also Fig.~\ref{fig:waveform_spectrum_sample})
including the Newtonian ones. Such nonlinear couplings of linear
quasi-normal modes were also observed in the study of initially
linearly perturbed equilibrium models by \citet{dimmelmeier_06_a},
although at much lower excitation levels. In contrast to the initial
low-level perturbations used in their models, in our case the strong
collapse and rebound at bounce manages to channel a large amount of
energy into this particular mode. A detailed discussion of the impact
of exciting the $ 2 \cdot\! F $-mode for GW emission is presented in
Section~\ref{subsection:mode_resonance}.

\noindent\textit{Rest mass.} For their Newtonian models, LCCS reported 
that along a sequence with constant rotation period $ p $, the maximum 
GW strain $ |h|_\mathrm{max} $ first increases with $ M_0 $ and then 
decreases again. In contrast, we find that for our corresponding 
sequence CC, $ |h|_\mathrm{max} $ increases monotonically with $ M_0 $ 
(see Table~\ref{tab:collapse_models})\footnote{We have found the same
  qualitative behaviour also for another sequence with constant period 
  $ p = 1.4 \mathrm{\ ms} $ and varying rest mass, for which we do not 
  present the data here.}. This different behaviour is probably due to 
a fundamental difference between our GR models and their Newtonian 
counterparts. Besides the obvious differences in the structure of the 
HQS and hence in the mode frequencies, one should bear in mind that 
the rest mass of the GR models has an upper limit corresponding to the 
point of transition to the unstable branch of equilibrium solutions. 
This is particularly evident for sequence CC, where model CC4 does not 
reach a stable configuration after the collapse but instead produces a 
black hole as deduced from the appearance of an apparent horizon. As a 
result, the eventual decrease of $ |h|_\mathrm{max} $ with increasing 
$ M_0 $ in the Newtonian case may not occur in GR simply because it 
might require rest masses above the upper limit.

\noindent\textit{Composition of the mixed phase.} The impact on the 
collapse dynamics of varying the parameter $ \delta $ (and hence 
varying the pressure reduction due to the presence of the quarks in 
the mixed phase) is quite obvious. Both for the collapse models CD1 to 
CD5, which are based on the initial model B2, and for the models CD6 
to CD9, which use the initial model B5, the rest-mass density at 
bounce $ \rho_\mathrm{c,b} $ and the amplitude of the post-bounce 
oscillations grow when the pressure reduction is enlarged by 
increasing the exponent $ \delta $ (see 
Table~\ref{tab:collapse_models}). Since within each of the two 
subsequences the initial rotation rate is constant, the growth of $ 
\rho_\mathrm{c,b} $, and hence of the compactness, with a greater 
pressure reduction is directly reflected in stronger GW emission and 
higher values for $ |h|_\mathrm{max} $. Also in this case, an 
exception arises with model CD7 which is identical to the already 
mentioned model CB5. Interestingly, when $ \delta = 4 $, the pressure 
reduction in the mixed phase is so great that the corresponding model 
CD9 collapses directly to a black hole without experiencing a bounce.

Finally, we note that the $ F $-mode frequencies in the sequence of 
models CD1 to CD5 first decrease with increasing $ \delta $ and then 
increase again. This may be the result of a near cancellation of
opposite effects: while a larger pressure reduction produces a higher
overall density of the post-collapse HQS (potentially resulting in
higher values of $ f_F $), it also leads to more rapid rotation due
the comparatively greater compactness (which should lower $ f_F $). We
also note that the large amplitude quasi-radial oscillations in the
last model(s) of each subsequence of sequence CD are so strongly
damped shortly after the collapse, that making a precise determination
of the frequency $ f_F $ is difficult.


\subsection{Damping of the stellar pulsations}
\label{subsection:damping}

\begin{figure*}
  \centerline{\includegraphics[width = 180 mm]{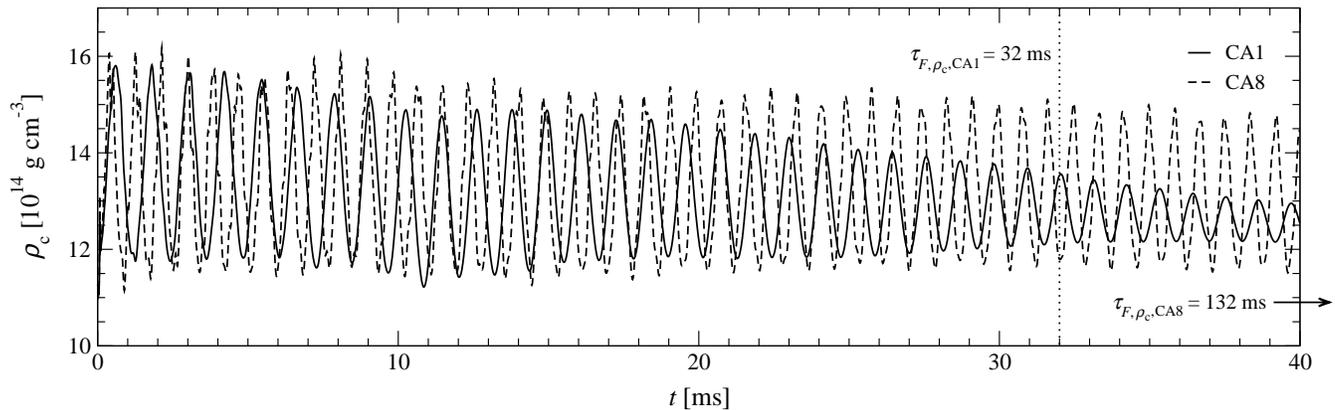}}
  \caption{Time evolution of the central rest-mass density $ 
    \rho_\mathrm{c} $ for models CA1 and CA8, the most and the least 
    rapidly rotating models of sequence CA. The dotted line marks the 
    pulsation damping time $ \tau_{F,\rho_\mathrm{c}} $ for model 
    CA1.}
  \label{fig:central_density_evolution_damping}
\end{figure*}

Both from the values for the damping time $ \tau_F $ of the 
fundamental quasi-radial $ F $-mode in the GW signal given in 
Table~\ref{tab:collapse_models} and from 
Fig.~\ref{fig:central_density_evolution_damping}, where we plot the 
time evolution of $ \rho_\mathrm{c} $ for the most and least rapidly 
rotating collapse models of sequence CA, it is evident that there can 
be significant damping of the post-bounce pulsations in the HQS, for 
some values of the model parameters. In Table~\ref{tab:damping} we 
also report the values for the damping time $ \tau_{F,\rho_\mathrm{c}} 
$ extracted from the time evolution of the central rest-mass density $ 
\rho_\mathrm{c} $: these values range from $ 8 $ to $ 200 \mathrm{\ 
ms} $ for the models considered here. With the exception of those 
models for which the GW emission is dominated by mode resonance, the 
two estimated timescales $ \tau_F $ and $\tau_{F,\rho_\mathrm{c}} $ 
are very similar.

In Section~\ref{subsection:comparison}, we have argued that the 
kinetic energy stored in the quasi-radial post-bounce pulsations is 
not responsible for generating differential rotation, and therefore 
the observed damping cannot be attributed to this mechanism as 
previously suggested. Since, also, the numerical dissipation of our 
code is much too small to be the main cause for the strong damping 
observed in some models (and anyway affects all models in the same 
way) it is necessary to offer an alternative physical explanation for 
why the pulsations are strongly damped for some models, with $ 
\tau_{F,\rho_\mathrm{c}} $ being a few ms (and thus comparable with 
the dynamical timescale), while for other models the damping is much 
slower, with $ \tau_{F,\rho_\mathrm{c}} \sim 100 \mathrm{\ ms} $ (and 
thus orders of magnitude longer than both the dynamical timescale and 
the pulsation period of the star; in this case, the damping seems to 
be limited essentially to that caused by numerical viscosity). Since 
the gravitational radiation back-reaction on the system is also 
negligible (and is not taken into account by our conformally flat 
approximation of the metric equations anyway), the strong damping seen 
can only be due to hydrodynamic effects.

\begin{figure}
  \centerline{\includegraphics[width = 85 mm]{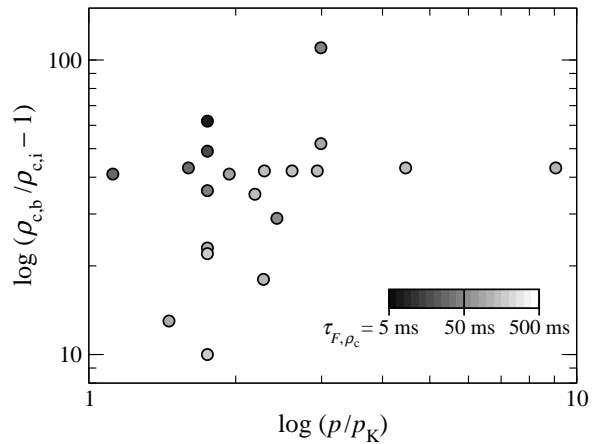}}
  \caption{Location of all GR collapse models in the $ (p / 
    p_\mathrm{K}) $--$ (\rho_\mathrm{c,b} / \rho_\mathrm{c,i} - 1) $
    plane. The damping time $ \tau_{F,\rho_\mathrm{c}} $ of the
    $ F $-mode, as calculated from the time evolution of the central 
    rest-mass density, is encoded in the grey scale, with dark/light 
    grey corresponding to strong/weak damping. Only models with very 
    rapid rotation (close to the Kepler limit) \emph{and} violent 
    post-bounce oscillations due to a very dynamical collapse (upper 
    left corner of the plot) have short damping times.}
  \label{fig:damping}
\end{figure}

A careful analysis of our model parameters reveals that damping is 
significant only for those models that both rotate very rapidly 
\emph{and} experience strong pulsations. This can be seen in 
Fig.~\ref{fig:damping}, where the value of $ \tau_{F,\rho_\mathrm{c}} 
$ for each model is given by the grey scale. Only models located in 
the upper left part of the plane spanned by the rotation period ratio 
$ p / p_\mathrm{K} $ and the over-density at bounce $ \rho_\mathrm{c,b} 
/ \rho_\mathrm{c,i} - 1 $, have the very short damping times.

\begin{table}
  \centering
  \caption{Relevant quantities for mass-shedding-induced damping: $ p 
    $ is the rotation period of the NS, $ p_\mathrm{K} $ is the 
    rotation period of a free particle on a circular orbit at the 
    equator (Kepler limit), $ p / p_\mathrm{K} $ is the ratio of these 
    two periods, $ \rho_\mathrm{c,b} / \rho_\mathrm{c,i} - 1 $ is the
    central over-density at bounce compared to the initial value, and
    $ \tau_{F,\rho_\mathrm{c}} $ and $ \tau_F $ are the $ F $-mode
    damping times extracted from the time evolution of 
    $ \rho_\mathrm{c} $ and from the GW signal, respectively. Note
    that we omit here those models that collapse to a black hole. For
    completeness, we also include the Newtonian collapse models CN1,
    CN2 and CN3.}
  \label{tab:damping}
  \begin{tabular}{@{}lcccccc@{}}
    \hline \\ [-1 em]
    Model &
    $ p $ &
    $ p_\mathrm{K} $ &
    $ p / p_\mathrm{K} $ &
    $ \rho_\mathrm{c,b} / \rho_\mathrm{c,i} - 1 $ &
    $ \tau_{F,\rho_\mathrm{c}} $ &
    $ \tau_F $ \\
    &
    [ms] &
    [ms] & &
    [\%] &
    [ms] &
    [ms] \\
    \hline \\ [-1 em]
    CA1 & 1.00 & 0.89 & 1.12 & \z41 & \z32 & \z40 \\
    CA2 & 1.20 & 0.75 & 1.60 & \z43 & \z36 & \z49 \\
    CA3 & 1.40 & 0.72 & 1.94 & \z41 & \z93 &  --- \\
    CA4 & 1.60 & 0.70 & 2.29 & \z42 &  155 &  319 \\
    CA5 & 1.80 & 0.69 & 2.61 & \z42 &  160 &  418 \\
    CA6 & 2.00 & 0.68 & 2.94 & \z42 &  150 &  270 \\
    CA7 & 2.99 & 0.67 & 4.46 & \z43 &  152 &  711 \\
    CA8 & 5.98 & 0.66 & 9.06 & \z43 &  132 &  --- \\ [0.5 em]
    CB1 & 1.30 & 0.89 & 1.46 & \z13 &  113 & \z99 \\
    CB2 & 1.40 & 0.80 & 1.75 & \z23 &  128 &  133 \\
    CB3 & 1.60 & 0.73 & 2.19 & \z35 &  156 &  196 \\
    CB4 & 1.80 & 0.69 & 2.61 & \z42 &  160 &  418 \\
    CB5 & 2.00 & 0.67 & 2.99 & \z52 &  103 &  687 \\ [0.5 em]
    CC1 & 1.80 & 0.79 & 2.28 & \z18 &  136 &  143 \\
    CC2 & 1.80 & 0.74 & 2.43 & \z29 & \z62 & \z71 \\
    CC3 & 1.80 & 0.69 & 2.61 & \z42 &  160 &  418 \\ [0.5 em]
    CD1 & 1.40 & 0.80 & 1.75 & \z10 &  200 &  150 \\
    CD2 & 1.40 & 0.80 & 1.75 & \z23 &  128 &  133 \\
    CD3 & 1.40 & 0.80 & 1.75 & \z36 & \z49 & \z54 \\
    CD4 & 1.40 & 0.80 & 1.75 & \z49 & \z19 & \z19 \\
    CD5 & 1.40 & 0.80 & 1.75 & \z62 & \zz8 & \zz8 \\
    CD6 & 1.40 & 0.80 & 1.75 & \z22 &  191 &  248 \\
    CD7 & 2.00 & 0.67 & 2.99 & \z52 &  103 &  687 \\
    CD8 & 2.00 & 0.67 & 2.99 &  110 & \z53 &  --- \\ [0.5 em]
    CN1 & 1.20 & 0.86 & 1.40 & \z11 & \z91 & \z73 \\
    CN2 & 1.20 & 0.86 & 1.40 & \z32 & \z16 & \zz9 \\
    CN3 & 1.20 & 0.86 & 1.40 & \z52 & \zz1 & \zz1 \\
    \hline
  \end{tabular}
\end{table}

For models rotating close to the mass-shedding limit (the Kepler 
limit), the effective gravity near to the equator is significantly 
weakened, vanishing at the mass-shedding point. As one goes to models 
having larger amplitude quasi-radial post-bounce pulsations, an 
increasing amount of matter is ejected from low latitudes on the 
stellar surface during each oscillation. This matter goes into the 
initially artificial very-low-density atmosphere and creates an 
expanding envelope of weakly-bound (or even unbound) material around 
the HQS. This mass shedding causes strong damping of the pulsations, 
as pulsational kinetic energy is converted into gravitational 
potential energy of the ejected matter. Since polar perturbations 
(mainly radial ones) are coupled to axial perturbations in rotating 
stars, the damping rapidly affects all modes.

This mass-shedding-induced damping of oscillations in rapidly rotating 
NSs near to the Kepler limit was first observed and discussed by 
\citet{stergioulas_04_a} for weakly pulsating equilibrium polytropic 
NS models in uniform rotation, treated within the Cowling 
approximation (neglecting the dynamics of the spacetime). In a 
subsequent study, \citet{dimmelmeier_06_a} found that if the Cowling 
approximation is abandoned and the spacetime is dynamically evolved, 
the effect of mass-shedding is drastically reduced, giving a 
consequent decrease in the damping of the pulsations. However, both of 
these studies treated small-amplitude pulsations of equilibrium models 
(working within the linear regime), whereas in our collapse models the 
pulsations can have very large amplitudes.

Considering the example of model CD8: this has a large initial
pulsation amplitude, with relative variations in $ \rho_\mathrm{c} $ 
of around 55\% (which is about half of the over-density at bounce) but 
it is rotating too slowly to be close to the Kepler limit
(with $ p / p_\mathrm{K} \simeq 3 $) and does not show
strong damping. The five models with the shortest
damping timescales, however, ($ \tau_{F,\rho_\mathrm{c}} < t_\mathrm{f} = 50
\mathrm{\ ms} $) all have both quite large initial pulsation amplitudes $
> 18\% $ and are also rapidly rotating, with rotation-period ratios 
$ p / p_\mathrm{K} < 1.75 $. This is consistent with equatorial 
mass-shedding being the predominant mechanism for the strong damping 
of the post-bounce pulsations seen for models that both rotate close 
to the Kepler limit \emph{and} pulsate with large amplitude. For 
models that are not affected by equatorial mass shedding, other 
damping mechanisms predominate (such as the conversion of kinetic 
energy into internal energy by shocks, nonlinear coupling of modes, 
numerical or physical dissipation) but all of these operate on 
timescales much longer than the dynamical timescale\footnote{As far as
  viscosity is concerned, in the quark phase the shear viscous damping
  timescale is comparable to that of normal nuclear matter, which is of the order 
  of $ 10^8 \mathrm{\ s} $ for a typical NS, while the bulk viscous 
  damping timescale is of the order of seconds, assuming a rotation
  period $ p = 1 \mathrm{\ ms} $ and an s-quark mass $ m_s = 100
  \mathrm{\ MeV} $ \citep{madsen_00_a}.}.


\subsection{The role of convection}
\label{subsection:convection}

\begin{figure}
  \centerline{\includegraphics[width = 85 mm]{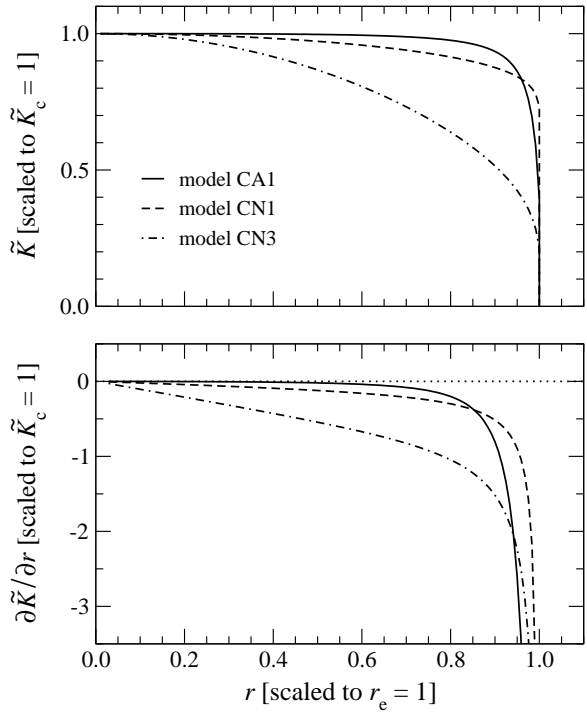}}
  \caption{Radial profile of the specific-entropy measure $ \tilde{K}
    $ (top panel) and its radial derivative (bottom panel) in the 
    equatorial plane at the initial time for the GR model CA1 (solid 
    line), and the corresponding profiles for the Newtonian models CN1 
    (dashed line) and CN3 (dash-dotted line). The specific-entropy measure and 
    the radius are scaled to give $ \tilde{K}_\mathrm{c} = 1 $ and $ 
    r_\mathrm{e} = 1 $.}
  \label{fig:entropy_profile}
\end{figure}

\begin{figure*}
  \centerline{\includegraphics[width = 180 mm]{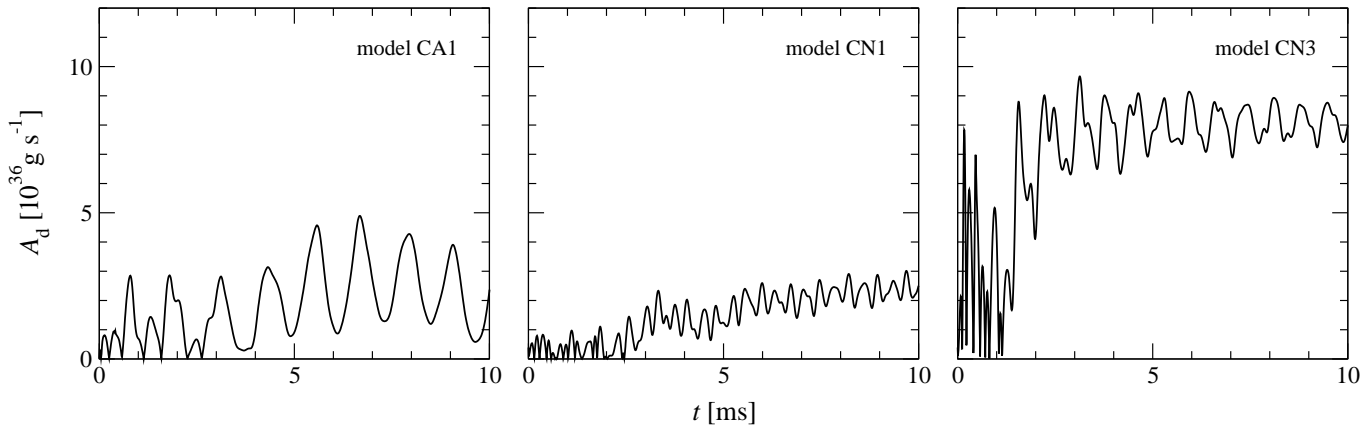}}
  \caption{Time evolution of the differential rotation measure $
    A_\mathrm{d} $ for the GR collapse model CA1 (left panel) and the
    Newtonian collapse models CN1 (centre panel) and CN3 (right 
    panel).}
  \label{fig:differential_rotation_gr}
\end{figure*}

As discussed above, the damping of post-bounce pulsations and the
development of differential rotation seen by LCCS seems to 
have been mainly a manifestation of the convective motion artificially 
produced in their simulations by the way in which they induced the 
collapse. Real physical convection might well be induced by energy 
input coming from the phase transition but this has no connection with 
the artificially-induced convection. Calculating the real convection 
is beyond the scope of the present simplified treatments and remains a 
topic for future work. While we have eliminated from our calculations 
the main source of artificial convection present in the LCCS 
simulations, some artificially-induced convection still remains. There 
are at least two different origins for this which apply for two 
different classes of initial models.

The first origin is related to the CFC approximation itself: when the 
data from the initial-model solver are mapped to the evolution code and 
the initial-value problem is subsequently re-solved to satisfy the 
constraints~(\ref{eq:cfc_metric_equations}), small errors due to the 
CFC approximation create spurious departures from constant entropy. 
This side effect of the CFC approximation is clearly larger for 
rapidly rotating and very compact models such as CA1, and becomes 
very small for slowly rotating models, vanishing in the non-rotating 
limit. Fig.~\ref{fig:entropy_profile} shows the initial radial profile 
of the specific-entropy measure $ \tilde{K} $ [cf.\ 
Eq.~(\ref{eq:entropy_measure})] for model CA1, together with those for 
the Newtonian models CN1 and CN3 discussed previously. It can be seen 
that the CFC approximation produces a negative specific-entropy gradient for 
model CA1 at the start of the evolution. While this gradient is very 
small in the bulk of the star (for $ r / r_\mathrm{e} \lesssim 0.8 $), 
near to the surface it becomes comparable to (or even greater than) 
that for model CN1, for which $ \gamma $ was initially decreased from 
$ 2 $ to $ 1.95 $. However, it is still much smaller than that 
occurring when reducing $ \gamma $ to $ 1.75 $ as in model CN3.

The CFC-induced violation of the Solberg--H{\o}iland criteria drives
this model (and other rapidly rotating ones) to develop significant
convection, which we measure in terms of the general relativistic
equivalent of the averaged shear [cf.\
Eq.~(\ref{eq:differential_rotation_measure_alternative_newtonian})],
\begin{equation}
  A_\mathrm{d} = \int \sqrt{\gamma} \, \rho h W^2
  \left| \frac{d v_\phi}{dr} - \frac{v_\phi}{r} \right| \, dV,
  \label{eq:differential_rotation_measure_alternative_gr}
\end{equation}
where $ v_\phi = \sqrt{v_3 v^3} $ is the rotation velocity of the 
  fluid\footnote{As already noted for the Newtonian models,
  the Newtonian equivalent of $ A_\mathrm{d} $ has a qualitatively 
  very similar behaviour to that of $ T_\mathrm{d} $ but only the first 
  of these has a clearly-defined physical meaning.}. 
Fig.~\ref{fig:differential_rotation_gr} shows the time evolution of $ 
A_\mathrm{d} $ for model CA1 (whose convection is, in fact, among the
strongest for the GR models; left-hand panel), as well as for the 
Newtonian models CN1 (centre panel) and CN3 (whose convection is among
the strongest for the Newtonian models; right-hand panel). It can be 
seen that both the development of the convection with time for model 
CA1 and its maximum saturation level are comparable with those for the 
Newtonian model CN3, while the convection in model CN1 is stronger and 
occurs on a much shorter timescale, which is natural since the initial 
non-isentropy is larger in that case.

The second origin of spurious convection appears in slowly rotating 
models; this is not due to the CFC approximation since the errors 
coming from that are essentially negligible in these cases. Instead, 
convection is produced here by the small violation of isentropy caused 
by interaction of the matter just inside the surface of the HQS with 
the surrounding artificial atmosphere. This perturbation slowly 
propagates inwards and finally affects the entire star at late times. 
Since in uniformly rotating configurations, centrifugal forces tend to 
have a stabilising effect within the Solberg--H{\o}iland criteria, it 
is easy to understand why this process occurs only for the slowly 
rotating models. An example of this mechanism in action is given by 
model CA8, which is the slowest rotator in our set of GR models: here 
$ A_\mathrm{d} $ reaches a maximum that is close to the saturation 
level of models CA1 and CN1 but is still much smaller than the value 
for the strongly convective model CN3.

In models that are neither very rapidly nor very slowly rotating 
(e.g.\ model CC1), convection is almost absent and $ A_\mathrm{d} $ 
remains at very low values, which can be explained as resulting from 
small deviations away from homology during the initial contraction and 
the subsequent post-bounce oscillations. We repeat, however, that 
consistent treatment of other possible real sources of convection 
remains a major topic for future work.


\subsection{Enhanced emission of gravitational waves via mode resonance}
\label{subsection:mode_resonance}

\begin{figure*}
  \centerline{\includegraphics[width = 180 mm]{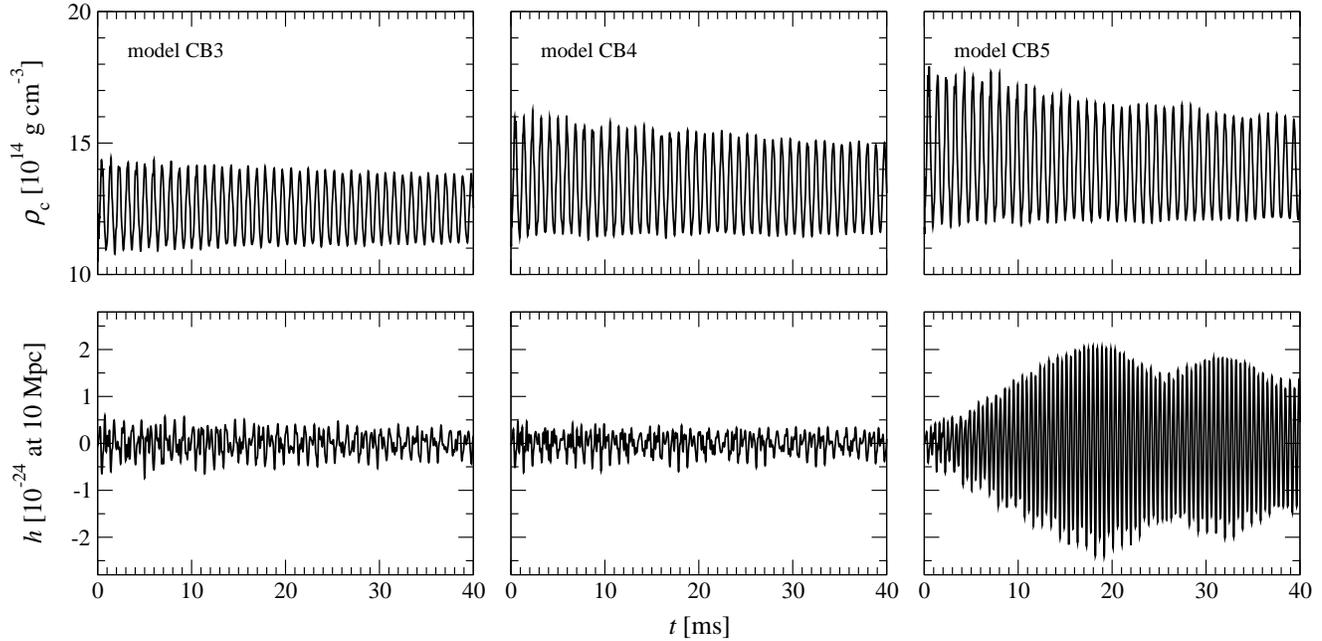}}
  \caption{Time evolution of the central rest-mass density $ \rho_\mathrm{c} $
    (top panels) and the GW strain $ h $ at a distance of $ 10 
    \mathrm{\ Mpc} $ (bottom panels) for the GR collapse models CB3 
    (left panels), CB4 (centre panels) and CB5 (right panels). The 
    enhanced GW emission for model CN5 due to mode resonance is 
    clearly visible.}
  \label{fig:resonance_time_evolution}
\end{figure*}

\begin{figure*}
  \centerline{\includegraphics[width = 180 mm]{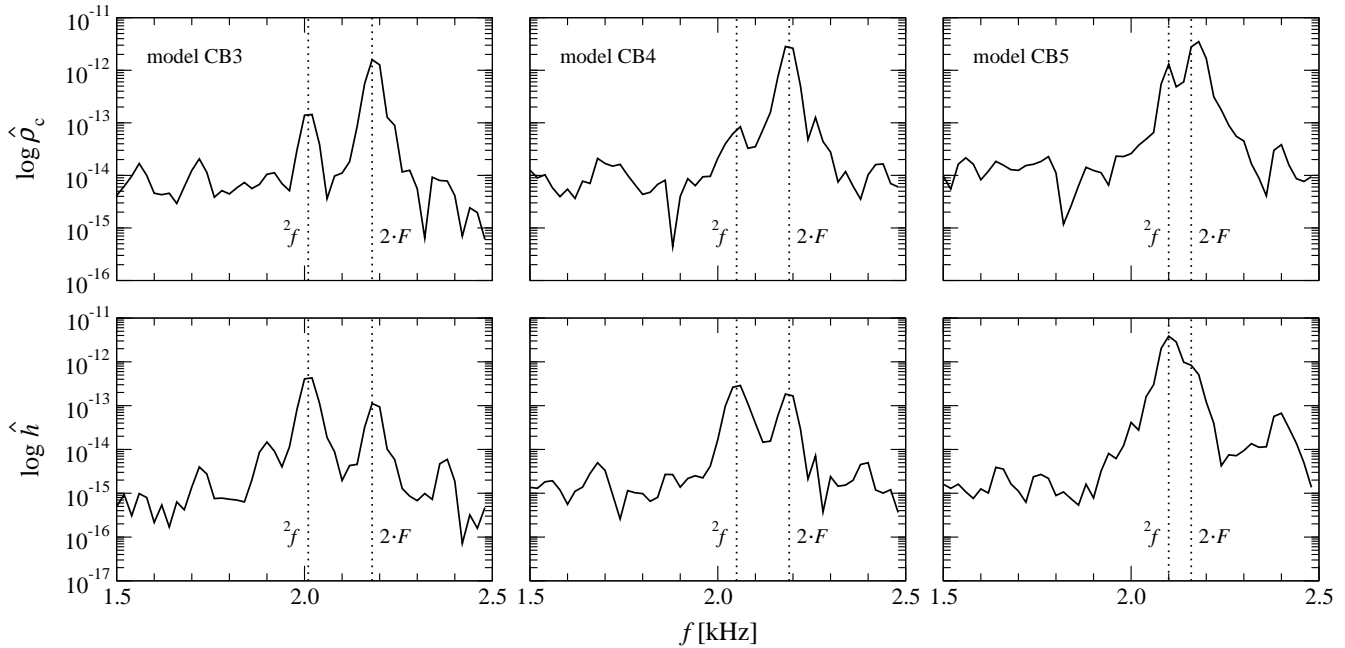}}
  \caption{Logarithmic power spectrum (in arbitrary units) of the
    central rest-mass density $ \rho_\mathrm{c} $ (top panels) and the
    GW strain $ h $ (bottom panels) for the same models as in
    Fig.~\ref{fig:resonance_time_evolution}. The resonance between the
    $ {}^{2\!}f $-mode and the $ 2 \cdot\! F $-mode sets in as their
    frequencies (marked by dotted lines) approach each other.}
  \label{fig:resonance_spectrum}
\end{figure*}

In Section~\ref{subsection:influence_of_parameters} we have discussed 
the influence of the initial rotation speed on the frequencies of the 
post-bounce oscillations and also commented that for models with the 
same rest mass, those with higher rotation generally have higher GW 
strain due to their increased quadrupole moments. This is indeed what 
is seen, for instance, in the sequence CB as reported in 
Table~\ref{tab:collapse_models}. This general behaviour, however, has 
a notable exception in the case of the comparatively slowly rotating 
model CB5, which has by far the largest GW strain for any of those in 
the sequence CB, with the energy emitted in GWs in the first $ 50 
\mathrm{\ ms} $ being at least an order of magnitude larger than for 
any of the other models in this sequence.

In the top panels of Fig.~\ref{fig:resonance_time_evolution} we plot 
the time evolution of the central rest-mass density $ \rho_\mathrm{c} 
$ for models CB3, CB4 and CB5. It can be seen that the collapse 
dynamics of model CB5 are not qualitatively different from those of 
the more rapidly rotating members of the same sequence. As rotation 
decreases from CB3 to CB5, the strength of the contraction increases, 
leading to higher central over densities $ \rho_\mathrm{c,b} $ at the 
bounce and stronger post-bounce oscillations. This, however, cannot account 
satisfactorily for the very large maximum GW strain amplitude $ 
|h|_\mathrm{max} $ of model CB5, nor for the growth of $ h $ during 
the first $ 20 \mathrm{\ ms} $ \emph{after} the bounce in this case 
(see bottom panels of Fig.~\ref{fig:resonance_time_evolution}).

Rather, the time evolution of $ h $ in model CB5 (the delayed growth, 
the saturation and the subsequent decay) suggests the presence of a 
resonance effect among several modes, at least one of which should be 
an efficient emitter of GWs. One of the obstacles to having strong GW 
emission is that the modes which are most strongly excited during the 
bounce are (quasi-)radial modes which are not efficient GW emitters; 
indeed they would not emit at all if rotation were not present to 
introduce non-radial contributions. However, it was suggested many 
years ago \citep[see, e.g.][]{eardley_83_a} that the pulsational 
energy from a quasi-radial mode, which contains a significant amount 
of kinetic energy but radiates GWs only weakly, could be transferred 
to a much more strongly radiating quadrupolar mode by means of 
resonance effects and even parametric instabilities. 
\citet{dimmelmeier_06_a}, as well as \citet{passamonti_05_a} and 
\citet{passamonti_07_a} have discussed this possibility in the context 
of nonlinear coupling of quasi-normal modes for nearly-equilibrium 
models of NSs.

In order to investigate whether this mechanism might be responsible 
for the enhanced GW emission observed for model CB5, we performed a 
mode analysis for models CB3, CB4 and CB5. The power spectrum of the 
time evolution of the central rest-mass density $ \rho_\mathrm{c} $, 
shown in the top panels of Fig.~\ref{fig:resonance_spectrum}, 
indicates that there is a lot of energy in the $ 2 \cdot\! F $-mode 
which has a peak about an order of magnitude higher than the 
corresponding one for the $ {}^{2\!}f $-mode, located at slightly 
lower frequencies. As rotation decreases from model CB3 to model CB5, 
the frequencies of these two modes get closer until the two peaks 
almost merge for model CB5, with the relative difference between the 
two frequencies decreasing to about $ 4\% $ (see the bottom panel of 
Fig.~\ref{fig:frequencies} and also 
Table~\ref{tab:resonance})\footnote{We have also performed simulations
  for models with other parameter values in the close vicinity of those 
  for model CB5 and found that CB5 actually exhibits almost the 
  maximum possible resonance between the $ 2 \cdot\! F $-mode and the 
  $ {}^{2\!}f $-mode.}. Under these conditions of resonance, the $ 2 
\cdot\! F $-mode is able to transfer a considerable amount of energy 
into the $ {}^{2\!}f $-mode, as can be clearly seen in the power 
spectrum of the waveform amplitude shown in the lower panels of 
Fig.~\ref{fig:resonance_spectrum}. If the spectra for model CB5 from
Fig.~\ref{fig:frequencies} are produced for several time windows, this
energy transfer between the two modes becomes visible. While the peaks
of both $ \rho_\mathrm{c} $ and $ h $ corresponding to the
$ {}^{2\!} f $-mode first gradually grow in the early phases of the
evolution (showing the initial amplification at the expense of the
$ 2  \cdot\! F $-mode) and start descreasing only at later times,
the peaks of the $ 2 \cdot\! F $-mode always descrease.

The strong dependence of the maximum GW strain $ |h|_\mathrm{max} $ on
the rotation rate, as seen in Table~\ref{tab:resonance} as well as in
the bottom panel of Fig.~\ref{fig:maximum_strain}, indicates that the
resonant behaviour  becomes important only when the frequencies of the
$ {}^{2\!}f $-mode and the $ 2 \cdot\! F $-mode are very close to each
other. This small difference between the two frequencies is also
responsible for the clear beating pattern seen in the waveform for
model CB5 (see the bottom right panel in
Fig.~\ref{fig:resonance_time_evolution}).
Fig.~\ref{fig:maximum_strain} also shows that while the pulsation
energy contained in the $ F $-mode (represented by the spectrum of
$ \rho_\mathrm{c} $ in the top panel) is always larger than the
corresponding one in the $ 2 \cdot\! F $-mode, when more resonance
between the $ 2 \cdot\! F $-mode and the $ {}^{2\!} f $-mode is at
work, the $ 2 \cdot\! F $-mode becomes a very efficient emitter of GWs
as well (see the spectrum of $ h $ in the lower panel), surpassing the
$ F $-mode here. This is most likely a consequence of the altering of
the $ 2 \cdot\! F $-mode's previously quasi-radial eigenfunction by
the interaction with the quadrupolar $ {}^{2\!} f $-mode \citep[see,
e.g.][]{dimmelmeier_06_a}.

\begin{table}
  \centering
  \caption{The frequencies $ f_{{}^{2\!}f} $ and $ f_{2 \cdot\! F} $ 
    of the fundamental quadrupolar $ {}^{2\!}f $-mode and the 
    self-coupling of the $ F $-mode, for the collapse models of 
    sequence CB. $ \Delta f_\mathrm{rel} = f_{2 \cdot\! F} / 
    f_{{}^{2\!}f} - 1 $ is the relative difference between the 
    frequencies of these two modes.}
  \label{tab:resonance}
  \begin{tabular}{@{}lccccc@{}}
    \hline \\ [-1 em]
    Model &
    $ f_{{}^{2\!}f} $ &
    $ f_{2 \cdot\! F} $ &
    $ \Delta f_\mathrm{rel} $ &
    $ |h|_\mathrm{max} $ &
    $ E_\mathrm{gw} $ \\
    &
    \highentry{[kHz]} &
    \highentry{[kHz]} &
    \highentry{[\%]} &
    $ \displaystyle \left[ \!\!\!
      \begin{array}{c}
        10^{-23} \\ [-0.2 em]
        \mathrm{\ at\ 10\ Mpc}
      \end{array}
      \! \right] $ &
    \highentry{[$ 10^{-4} M_\odot c^2 $]} \\ [0.7 em]
    \hline \\ [-1 em]
    CB1 & 1.78 & 2.10 &  18 & 0.38 & 0.01 \\
    CB2 & 1.90 & 2.13 &  12 & 0.57 & 0.05 \\
    CB3 & 2.02 & 2.18 & \z8 & 0.10 & 0.09 \\
    CB4 & 2.06 & 2.19 & \z6 & 0.10 & 0.11 \\
    CB5 & 2.10 & 2.18 & \z4 & 1.72 & 1.16 \\
    \hline
  \end{tabular}
\end{table}

Applying this analysis to the complete set of investigated models, we
find that the same resonance between the $ {}^{2\!}f $-mode and the $
2 \cdot\! F $-mode is also at work in models CA3 and CA4 (this is clearly
visible in the top panel of Fig.~\ref{fig:maximum_strain}) as well as
in model CD8. Also in these cases, the waveforms show an initial growth
in amplitude over several cycles and then a strong beating at
later times. In the case of model CD8, the growth in $ h
$ due to the mode resonance becomes prominent long after the bounce, 
at $ t \gtrsim 20 \mathrm{\ ms} $, not reaching its maximum $
|h|_\mathrm{max} $ until $ t \simeq 48 \mathrm{\ ms} $, when the
quasi-radial oscillations have already been damped to around 38\% of 
their initial amplitude. In contrast, for models whose GW emission is 
not influenced by this resonance, $ |h|_\mathrm{max} $ is reached at 
the time of bounce or very close to it. In addition, here the spectral
power of both the $ 2 \cdot\! F $-mode and the $ {}^{2\!}f $-mode
decreases at all times if successive spectra are obtained with the
technique of shifted time windows.


\subsection{Detectability prospects for the gravitational wave
  emission}
\label{section:gravitational_wave_detectability}

Although the maximum GW strain $ |h|_\mathrm{max} $ in the waveforms 
of our GR models is about an order of magnitude smaller than that 
computed by LCCS for Newtonian models, the long quasi-periodic GW 
emission that is possible for phase-transition-induced collapse may 
still make this scenario a plausible source for GW detectors. Assuming 
that the strong post-bounce oscillations are not damped by any other 
physical mechanisms apart from dissipation of kinetic energy by 
shocks, the damping times that we obtain suggest that the effective 
total emission time for GWs can be much longer than the time for which 
we have followed the evolution of our models ($ t_\mathrm{f} = 50 
\mathrm{\ ms} $), extending to hundreds of oscillations.

\begin{figure*}
  \centerline{\includegraphics[width = 180 mm]{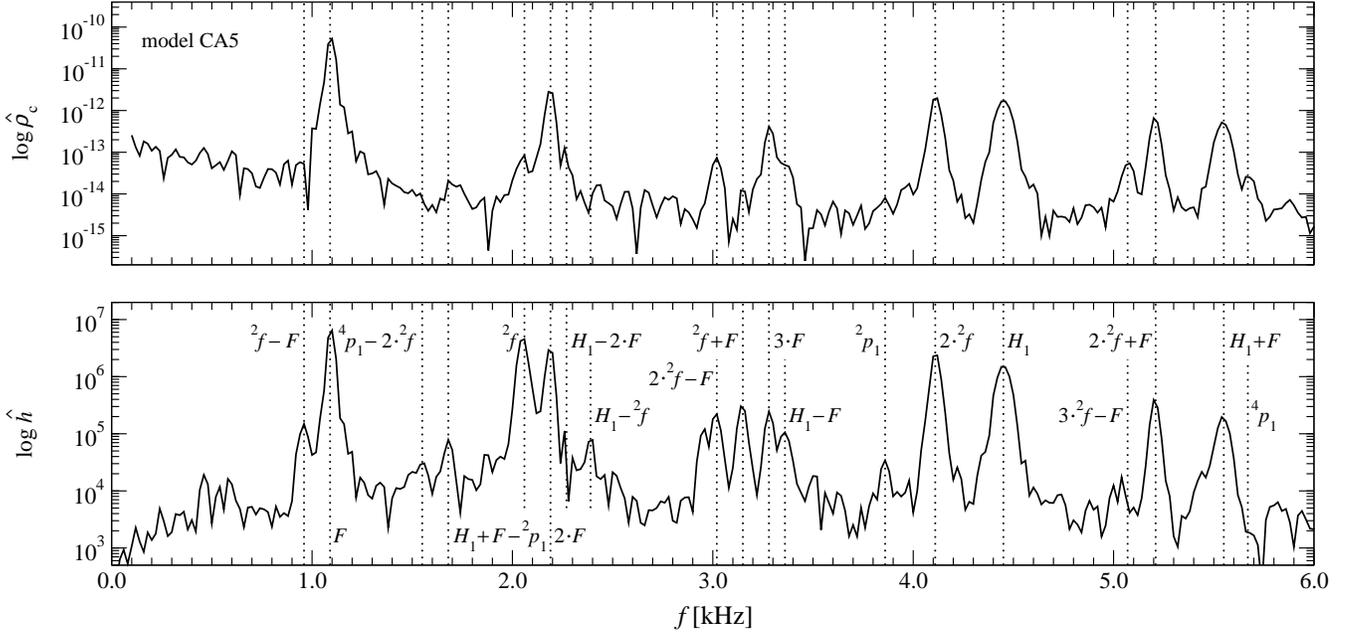}}
  \caption{Logarithmic power spectrum (in arbitrary units) of the
    central rest-mass density $ \rho_\mathrm{c} $ (top panel) and the
    GW strain $ h $ (bottom panel) for the GR collapse model CA5. The
    linear quasi-normal modes and the nonlinear (self-)couplings are 
    marked with dotted lines and labelled. $ H_1 $ denotes the first
    overtone of the $ F $-mode, while $ {}^2 p_1 $ and $ {}^4 p_1 $
    are the first overtones of the fundamental $ l = 2 $ and $ l = 4 $
    modes, respectively.}
  \label{fig:waveform_spectrum_sample_logarithmic}
\end{figure*}

The main GW emission modes are the $ F $-mode and the $ {}^{2\!}f 
$-mode, which have comparable energies in the power spectra, except 
where the resonance between the $ {}^{2\!}f $-mode and the $ 2 \cdot\! 
F $-mode is important, when the combined $ {}^{2\!}f $-mode/$ 2 \cdot 
F $-mode dominates the GW spectrum. Model CA5, for which we show in 
Fig.~\ref{fig:waveform_spectrum_sample_logarithmic} the power spectrum 
of the time evolution of $ \rho_\mathrm{c} $ (top panel) and of the GW 
signal (bottom panel), is an example where the frequencies of $ 
{}^{2\!}f $-mode and $ 2 \cdot F $-mode are already close and the two 
modes start to merge.

Exploiting the fact that, for most models, the waveform can be very 
well approximated as a combination of two essentially monochromatic 
damped sinusoids, it is straightforward to construct a 
phenomenological waveform expressed in terms of six parameters: the 
mode frequencies $ f_F $ and $ f_{{}^{2\!}f} $, the damping times
$ \tau_F $ and $ \tau_{{}^{\,2\!}f} $, and the initial phases 
and relative amplitude of the two modes (all of which can be found 
from Table~\ref{tab:collapse_models}). This can then be applied as a
template in matched filtering data-analysis algorithms so as to search
for the waveforms in the data stream more effectively.

In order to assess the prospects for detection by current and planned 
interferometric detectors, we next calculate characteristic quantities 
for the GW signal following \citet{thorne_87_a}. Making a Fourier 
transform of the dimensionless GW strain $ h $,
\begin{equation}
  \hat{h} =
  \int_{-\infty}^\infty \!\! e^{2 \pi i f t} h \, dt\,,
  \label{eq:waveform_fourier_tranform}
\end{equation}
we can compute the (detector dependent) integrated characteristic
frequency
\begin{equation}
  f_\mathrm{c} =
  \left( \int_0^\infty \!
  \frac{\langle \hat{h}^2 \rangle}{S_h}
  f \, df \right)
  \left( \int_0^\infty \!
  \frac{\langle \hat{h}^2 \rangle}{S_h}
  df \right)^{-1}\!\!\!\!\!\!\!,
  \label{eq:characteristic_frequency}
\end{equation}
and the dimensionless integrated characteristic strain
\begin{equation}
  h_\mathrm{c} =
  \left( 3 \int_0^\infty \!
  \frac{S_{h\mathrm{\,c}}}{S_h} \langle \hat{h}^2 \rangle
  f \, df \right)^{1/2}\!\!\!\!\!\!\!\!,
  \label{eq:characteristic_amplitude}
\end{equation}
where $ S_h $ is the power spectral density of the detector and $ 
S_{h\mathrm{\,c}} = S_h (f_\mathrm{c}) $. We approximate the average $ 
\langle \hat{h}^2 \rangle $ over randomly distributed angles by $ 
\hat{h}^2 $, assuming optimal orientation of the detector. From 
Eqs.~(\ref{eq:characteristic_frequency}, 
\ref{eq:characteristic_amplitude}) the signal-to-noise ratio (SNR) can 
be calculated as $ h_\mathrm{c} / [h_\mathrm{rms} (f_\mathrm{c})] $, 
where $ h_\mathrm{rms} = \sqrt{f S_h} $ is the value of the rms strain 
noise for the detector (which gives the theoretical sensitivity
window).

\begin{table*}
  \centering
  \caption{Detection prospects for the GWs: $ f_\mathrm{c} $ is the
    characteristic frequency, $ h_\mathrm{c} $ is the integrated
    characteristic GW signal strain and $ \mathrm{SNR} $ is the
    signal-to-noise ratio, each given for the current LIGO and VIRGO
    detectors, and for the future advanced LIGO detector. The values 
    given for all of the quantities assume a total emission time of $ 
    t_\mathrm{f} = 50 \mathrm{\ ms} $ and are dependent on the rms 
    strain noise $ h_\mathrm{rms} $ of the detector. Note that models 
    CB6, CC4 and CD9 collapse to form a black hole as a result of the 
    phase transition and so are omitted here.}
  \label{tab:gravitational_wave_emission}
  \begin{tabular}{@{}lc@{~~}c@{~~}c@{~~}c@{~~}c@{~~}c@{~~}c@{~~}c@{~~}c@{~~}}
    \hline \\ [-1 em]
    Model &
    $ f_\mathrm{c,LIGO} $ &
    $ f_\mathrm{c,VIRGO} $ &
    $ f_\mathrm{c,adv.\,LIGO} $ &
    $ h_\mathrm{c,LIGO} $ &
    $ h_\mathrm{c,VIRGO} $ &
    $ h_\mathrm{c,adv.\,LIGO} $ &
    $ \mathrm{SNR}_\mathrm{LIGO} $ &
    $ \mathrm{SNR}_\mathrm{VIRGO} $ &
    $ \mathrm{SNR}_\mathrm{adv.\,LIGO} $ \\
    &
    \highentry{[kHz]} &
    \highentry{[kHz]} &
    \highentry{[kHz]} &
    $ \displaystyle \left[ \!\!\!
      \begin{array}{c}
        10^{-20} \\ [-0.2 em]
        \mathrm{\ at\ 10\ kpc}
      \end{array}
    \! \right] $ &
    $ \displaystyle \left[ \!\!\!
      \begin{array}{c}
        10^{-20} \\ [-0.2 em]
        \mathrm{\ at\ 10\ kpc}
      \end{array}
    \! \right] $ &
    $ \displaystyle \left[ \!\!\!
      \begin{array}{c}
        10^{-20} \\ [-0.2 em]
        \mathrm{\ at\ 10\ kpc}
      \end{array}
    \! \right] $ &
    \highentry{[at 10 kpc]} &
    \highentry{[at 10 kpc]} &
    \highentry{[at 10 kpc]} \\ [0.7 em]
    \hline \\ [-1 em]
    CA1 & 0.954 & 0.978 & 1.029 & \z2.55 & \z2.59 & \z2.68 & \z7.5 &  12.0 &  179 \\
    CA2 & 1.755 & 1.798 & 1.864 & \z5.65 & \z5.75 & \z5.92 & \z6.8 &  11.8 &  201 \\
    CA3 & 1.845 & 1.876 & 1.923 & \z8.52 & \z8.63 & \z8.80 & \z9.4 &  16.6 &  287 \\
    CA4 & 1.619 & 1.660 & 1.735 & \z3.56 & \z3.62 & \z3.75 & \z4.8 & \z8.3 &  139 \\
    CA5 & 1.349 & 1.382 & 1.450 & \z1.78 & \z1.81 & \z1.87 & \z3.1 & \z5.3 & \z86 \\
    CA6 & 1.415 & 1.452 & 1.527 & \z1.55 & \z1.58 & \z1.63 & \z2.5 & \z4.4 & \z71 \\
    CA7 & 1.449 & 1.485 & 1.572 & \z0.72 & \z0.73 & \z0.76 & \z1.1 & \z2.0 & \z32 \\
    CA8 & 1.609 & 1.672 & 1.788 & \z0.29 & \z0.30 & \z0.31 & \z0.4 & \z0.7 & \z11 \\ [0.5 em]
    CB1 & 1.349 & 1.374 & 1.417 & \z1.31 & \z1.33 & \z1.36 & \z2.3 & \z4.0 & \z64 \\
    CB2 & 1.304 & 1.335 & 1.388 & \z2.12 & \z2.16 & \z2.22 & \z3.9 & \z6.7 &  108 \\
    CB3 & 1.322 & 1.353 & 1.411 & \z2.09 & \z2.12 & \z2.18 & \z3.8 & \z6.4 &  104 \\
    CB4 & 1.349 & 1.382 & 1.450 & \z1.78 & \z1.81 & \z1.87 & \z3.1 & \z5.3 & \z86 \\
    CB5 & 2.073 & 2.078 & 2.086 &  13.18 &  13.27 &  13.24 &  12.3 &  21.9 &  389 \\ [0.5 em]
    CC1 & 1.370 & 1.394 & 1.435 & \z1.06 & \z1.08 & \z1.10 & \z1.8 & \z3.1 & \z51 \\
    CC2 & 1.370 & 1.394 & 1.442 & \z1.48 & \z1.50 & \z1.54 & \z2.6 & \z4.4 & \z71 \\
    CC3 & 1.349 & 1.382 & 1.450 & \z1.78 & \z1.81 & \z1.87 & \z3.1 & \z5.3 & \z86 \\ [0.5 em]
    CD1 & 1.346 & 1.370 & 1.407 & \z1.23 & \z1.25 & \z1.27 & \z2.2 & \z3.7 & \z61 \\
    CD2 & 1.304 & 1.335 & 1.388 & \z2.12 & \z2.16 & \z2.22 & \z3.9 & \z6.7 &  108 \\
    CD3 & 1.313 & 1.345 & 1.406 & \z2.79 & \z2.84 & \z2.93 & \z5.1 & \z8.7 &  140 \\
    CD4 & 1.555 & 1.585 & 1.636 & \z4.54 & \z4.60 & \z4.71 & \z6.5 &  11.2 &  188 \\
    CD5 & 1.661 & 1.682 & 1.718 & \z5.09 & \z5.14 & \z5.23 & \z6.6 &  11.6 &  197 \\
    CD6 & 1.350 & 1.383 & 1.449 & \z0.82 & \z8.36 & \z0.86 & \z1.5 & \z2.5 & \z40 \\
    CD7 & 2.073 & 2.078 & 2.086 &  13.18 &  13.21 &  13.24 &  12.3 &  22.0 &  389 \\
    CD8 & 1.722 & 1.780 & 1.862 & \z3.88 & \z3.99 & \z4.13 & \z4.8 & \z8.3 &  141 \\
    \hline
  \end{tabular}
\end{table*}

In Table~\ref{tab:gravitational_wave_emission} we summarize the values 
of $ f_\mathrm{c} $, $ h_\mathrm{c} $ and the SNR for all of the 
models (except those which collapse to a black hole) for the currently 
operating LIGO and VIRGO detectors and for the future advanced LIGO 
detector. For all of the detectors we consider a source inside our own 
Galaxy at a reference distance of $ 10 \mathrm{\ kpc} $. The 
proportion of NSs that undergo phase-transition-induced collapse at
some stage in their lifetimes 
is not well-known. The phenomenon could occur at, or soon after, the
formation stage (giving an event rate roughly proportional to that for
core collapse supernovae) or it could come at a later stage when an
old NS is spun up and has its mass increased by accretion from a
binary partner in an LMXB \citep[an interesting case but with an event
rate which is thought to be very much lower, probably $\sim
10^{-5} \mathrm{\ yr}^{-1}$ for the Milky Way; see][]{pfahl_03_a}. Another
point is that a phase-transition-induced collapse occurring for a NS
which is not rapidly rotating would not be so interesting for our
purposes. Even under the most extremely optimistic assumption that the
phase-transition-induced collapse rate equals that for core-collapse
supernovae, that would only give a rate of up to $ 1 $ per $ 20
\mathrm{\ yr} $ for our Galaxy, which is prohibitively low. On the
other hand, if such an event did occur in our Galaxy, for current
interferometric detectors of the LIGO class and assuming an emission
time $ t_\mathrm{f} = 50 \mathrm{\ ms} $, all of our models except one
have a SNR above $ 1 $. For the strongest emitting model CB5, where
mode resonance significantly enhances the GW emission, the SNR even
exceeds $ 10 $. With the advanced LIGO detector, the SNR lies
comfortably above $ 10 $ for all models and reaches almost $ 400 $ in
model CB5.

For substantially increasing the event rate, it would be necessary for 
the detector to be sensitive to signals coming from distances out to 
the Virgo cluster, at $ \sim 20 \mathrm{\ Mpc} $, (for which the supernova 
rate would rise to more than 1 per year). However, at this distance
the SNR for our models drops to well below $ 1 $ even for advanced LIGO. 
Therefore, as for GW signals from supernova core collapse \citep[see 
the discussion in][]{dimmelmeier_07_a}, the second generation 
detectors will improve the SNR of a local event, but will not increase 
the event rate much on account of the inhomogeneous galaxy 
distribution in the local region of the Universe. Only third 
generation detectors will have the required sensitivity in the kHz 
range to achieve a robust SNR at the distance of the Virgo cluster.

\begin{figure*}
  \centerline{\includegraphics[width = 180 mm]{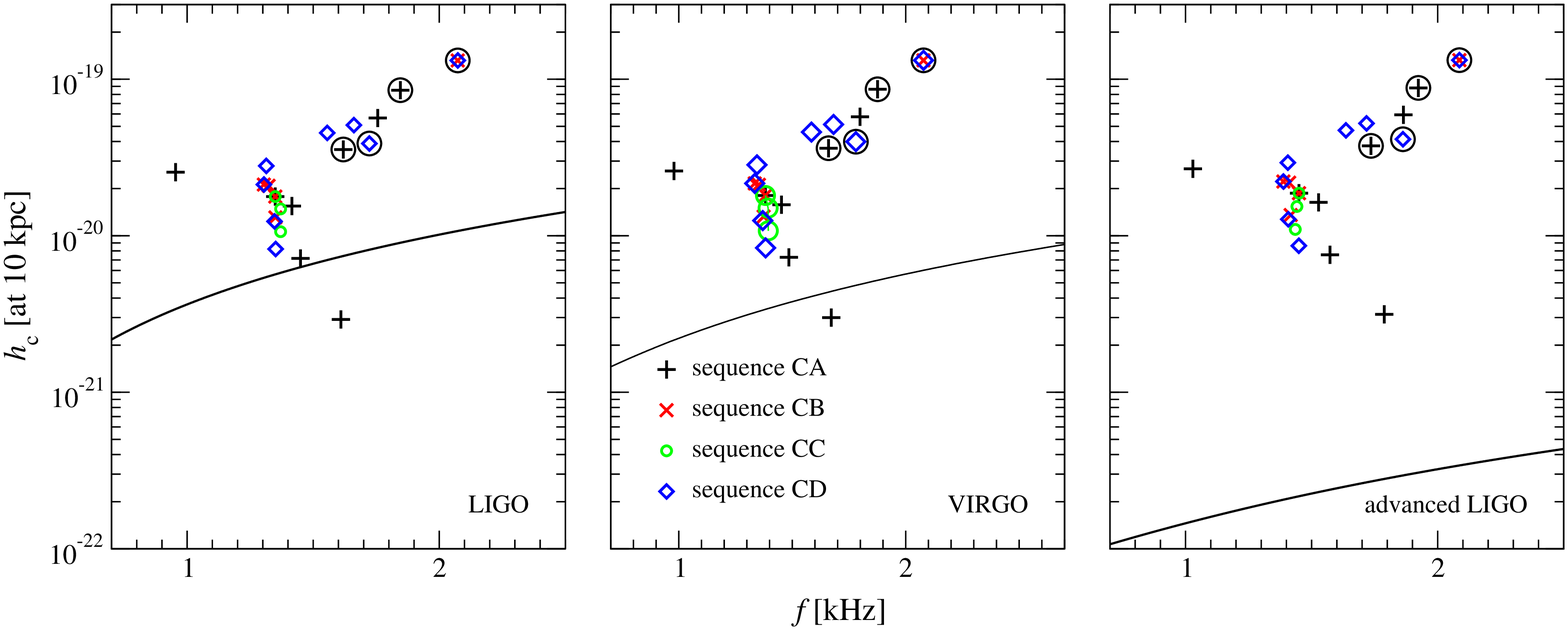}}
  \caption{Location of the GW signals from all of our GR models 
    in the $ h_\mathrm{c} $--$ f_\mathrm{c} $ plane shown relative to 
    the sensitivity curves (giving the rms strain noise 
    $h_\mathrm{rms}$) for the current LIGO detector (left panel) and
    VIRGO detector (centre panel) and for the advanced LIGO detector 
    (right panel). The sources are taken to be at a distance of $ 10 
    \mathrm{\ kpc} $. Note that some of the models belong to more than 
    one sequence. Models for which mode resonance boosts the GW emission 
    are additionally marked with circles.}
  \label{fig:detector_sensitivity}
\end{figure*}

Note that for most of the models the \emph{integrated} characteristic 
frequency $ f_\mathrm{c} $ given in 
Table~\ref{tab:gravitational_wave_emission} is not very close to 
either of the two main GW emission frequencies $ f_F $ and $ 
f_{{}^{2\!}f} $. This is because $ f_\mathrm{c} $ reflects the 
frequency dependence of the sensitivity of the detector, because 
nonlinear mode couplings and higher order linear modes also contribute 
to the GW signal (although with lower relative amplitudes; see 
Fig.~\ref{fig:waveform_spectrum_sample_logarithmic}) and, most 
importantly, because for many models the GW power spectrum of the 
signal exhibits nearly equally strong peaks in the $ F $-mode and the 
$ {}^{2\!}f $-mode.

The detector dependence of $ f_\mathrm{c} $ and $ h_\mathrm{c} $ is 
also illustrated in Fig.~\ref{fig:detector_sensitivity}, where the 
locations of the GW signals for all of the models are plotted 
relative to the rms strain noise $ h_\mathrm{rms} $ of the current 
LIGO and VIRGO detectors (left and centre panels, respectively) and 
the future advanced LIGO detector (right panel), all for a distance to 
the source of $ 10 \mathrm{\ kpc} $. As a general feature, we note 
that the upper parts of these sensitivity diagrams are occupied by 
those models whose GW signal strength is enhanced by mode resonance. 
When mode resonance is not important, the largest characteristic 
amplitudes occur for those models which are most rapidly rotating, 
unless strong post-bounce damping is in action as in model CA1.

Both in Table~\ref{tab:gravitational_wave_emission} and in 
Fig.~\ref{fig:detector_sensitivity} the GW characteristics have been 
evaluated for a total emission time of $ t_\mathrm{f} = 50 \mathrm{\ 
ms} $. In Appendix~\ref{appendix:damping_times} we describe how to
obtain the characteristic GW strain $ h_\mathrm{c} $ for an 
\emph{arbitrary} emission time using simple scaling laws.


\section{Conclusion}
\label{section:conclusion}

In this paper, we have described numerical simulations of the 
phase-transition-induced collapse of rotating neutron stars to become 
hybrid quark stars. The simulations were made using a code which 
solves the general relativistic hydrodynamic equations in axisymmetry 
and within the conformally flat approximation. The initial stellar 
models were taken as being rapidly rotating $ \gamma = 2 $ polytropes, 
while during the evolution we used an equation of state composed of 
three parts, depending on density: a normal hadronic-matter phase, a 
mixed phase of normal matter together with deconfined quark matter, 
and a pure quark-matter phase. The hadronic component of the equation 
of state was described with an ideal-gas type of equation of state, 
while for the deconfined quark matter phase we used the MIT bag model.

To validate our code, we first repeated several of the Newtonian 
simulations performed previously by \citet{lin_06_a}. We found that 
the differential rotation which develops in these models during the 
post-bounce phase is almost entirely due to strong transient 
convection which arises because the way in which they treated the 
onset of the phase transition leads to an artificial negative
specific-entropy gradient. We argue that their conclusion about there
being a causal link between the large amplitude post-bounce pulsations
and the creation of differential rotation was a misinterpretation of
the results. We also suggest that a significant part of the damping of
the pulsations which they observed was a consequence of the numerical
dissipation present in their calculations, rather than being related to
differential rotation, although a part of this damping was related to
the other physical processes which we have discussed here and which
were present in both calculations. 

Having clarified the dynamics of the collapse in the Newtonian 
framework, we then investigated the corresponding situation within 
general relativity, using a modified prescription for triggering the 
phase transition and initiating the collapse. Here we change the 
equation of state only in the regions where the phase transition has 
taken place and leave it unchanged elsewhere. We recognize that our 
procedure for describing the phase transition remains extremely 
idealized but we believe that it represents a step forward.

Despite this difference in the way in which the phase transition is 
triggered, we have not found any major qualitative differences in the 
waveforms produced when comparing the relativistic simulations with 
the earlier Newtonian ones. Also in the general relativistic 
simulations, the waveforms produced are mainly composed of the 
fundamental $ l = 0 $ quasi-radial $ F $-mode and the fundamental $ l 
= 2 $ quadrupolar $ {}^{2\!}f $-mode. However, in contrast to the 
Newtonian models, the $ F $-mode is at a lower frequency than the $ 
{}^{2\!}f $-mode as a consequence of the different density profiles. 
In addition to these modes, a nonlinear self-coupling of the $ F 
$-mode at twice the original frequency, the $ 2 \cdot F $-mode, is 
strongly excited due to the violent nature of the collapse. Although 
qualitatively similar to their Newtonian counterparts, the 
relativistic models exhibit quantitative differences in their 
dynamics. In order to investigate these, we have considered a set of 
23 different models organized in several sequences. In each of these 
sequences, only one of the characteristic parameters of the models was 
allowed to vary.

The main trends observed were as follows. For the sequence with 
constant initial central rest-mass density, the maximum gravitational 
wave strain $ |h|_\mathrm{max} $ increases monotonically with the 
rotation rate (except for some models where the waveform is strongly 
altered by mode resonances). For the constant rest mass sequence, on 
the other hand, we observe first an increase of $ |h|_\mathrm{max} $ 
with the rotation rate and then a decrease for very rapid rotation. 
For the sequence with constant rotation period but varying rest mass, 
we see $ |h|_\mathrm{max} $ increasing monotonically with the rest 
mass, which is a different behaviour from that seen for the Newtonian 
models, where $ |h|_\mathrm{max} $ first increases and then decreases 
again as the rest mass is increased. The reason for this difference 
may simply be that having a decreasing part of the curve would require 
progenitor neutron stars with rest masses beyond the upper limit for 
general-relativistic models. Finally, changing the equation of state 
in the mixed phase has straightforward consequences: the central 
rest-mass density at bounce, the amplitude of the post-bounce 
oscillations, and the maximum gravitational wave strain all increase 
as the overall pressure in the mixed phase is reduced.

Other points to arise include the following. Firstly, the influence of 
rotation on the frequencies of the $ F $ and $ {}^{2\!}f $-modes 
agrees well with what was found by \citet{dimmelmeier_06_a} for 
pulsations of equilibrium neutron stars, suggesting that studies of 
linear pulsations of equilibrium models can (at least in some cases) 
correctly predict the properties of the eigenfrequencies also when the 
pulsations are excited in a dynamical situation. Secondly, in some models 
the emission of gravitational radiation is considerably enhanced by 
the nonlinear resonance between the $ 2 \cdot F $-mode and the $ 
{}^{2\!}f $-mode. When the frequencies of these two modes are 
sufficiently close, the weakly radiating quasi-radial $ 2 \cdot F 
$-mode transfers some of its kinetic energy to the strongly radiating 
quadrupolar $ {}^{2\!}f $-mode, leading to a considerable increase in 
$ |h|_\mathrm{max} $. Thirdly, we have proposed a simple explanation 
for the strong damping of the post-bounce pulsations seen for a 
sub-set of our models. Our analysis reveals that that these models are 
all both rotating close to the Kepler limit and also undergoing 
large-amplitude post-bounce pulsations, resulting in significant mass 
shedding from the stellar surface close to the equator. As already 
discussed by \citet{stergioulas_04_a}, this ejection of loosely bound 
matter is very efficient in the damping quasi-radial pulsations.

In order to assess the prospects for the detection of 
phase-transition-induced collapse events by gravitational wave 
interferometers, special attention has been paid to making an accurate 
calculation of the gravitational wave emission resulting from this 
scenario. We find that the dimensionless gravitational wave strain $ h 
$ from a source at a distance of $ 10 \mathrm{\ Mpc} $ ranges between 
$ 0.1 $ to $ 2.4 \times 10^{-23} $ for all of the models considered 
and that the total energy emitted in gravitational waves during first 
$ 50 \mathrm{\ ms} $ of evolution is between $ 10^{-6} $ and $ 10^{-4} 
M_\odot c^2 $, corresponding to $ 2 \times 10^{48} $ and $ 2 \times 
10^{50} \mathrm{\ erg}$, respectively. These numbers are considerably
smaller than those presented by \citet{lin_06_a} for their Newtonian
calculations and so are disappointing for the prospects of detecting
these sources. The damping times for the post-bounce oscillations, as
computed from the gravitational radiation waveform, range from $ 8 $
to $ 687 \mathrm{\ ms} $ for the $ F $-mode, and from $ 18 $ to $ 130
\mathrm{\ ms} $ for the $ {}^{2\!}f $-mode. For all of the models
considered, we have also calculated the characteristic frequency $
f_\mathrm{c} $, the characteristic strain $ h_\mathrm{c} $, and the
signal-to-noise ratio for current and future detectors. For current
detectors such as LIGO or VIRGO, all of the models (except one) have a
signal-to-noise ratio above $ 1 $ for a source at $ 10 \mathrm{\ kpc}
$. For the advanced LIGO detector, the signal-to-noise ratio rises to
well above $ 10 $ for a source at $ 10 \mathrm{\ kpc} $ for all of the
models and reaches almost $ 400 $ when mode resonance is
active. However, for detecting sources within the Virgo cluster at a
distance of $ 20 \mathrm{\ Mpc} $, which is probably necessary in
order to reach a realistic event rate, third generation detectors
would be needed.

In conclusion, we note that while our study represents an improvement 
over previous ones, it still lacks a number of very important aspects 
which would be necessary for a properly realistic modelling of these 
objects. Firstly, there is the treatment of the phase transformation 
itself which remains extremely crude, containing no detailed picture 
of the transformation of the material, the local heat input or the 
emission of neutrinos or photons. A consistent treatment of radiative 
transfer is likely to be essential for following the cooling phase of 
the newly formed hybrid quark star and could highlight that the 
radiative losses would produce differences in the specific-entropy 
stratification and hence trigger real convection. Also, in our 
discussion, we have been considering the phase transition as occurring 
by means of a detonation; the conclusions would be drastically altered 
if it takes place instead via a slow deflagration. A second aspect 
concerns the treatment of the standard neutron-star matter: while 
using a gamma-law equation of state to model this can be reasonable 
for some simplified calculations, it gives an extremely poor 
approximation to the complex physics actually occurring in real 
neutron stars. Thirdly, we have not been considering the influence of 
magnetic fields which are not only expected to affect the dynamics, 
but could also lead to a dramatic modification of the phase transition 
process itself \citep{lugones_02_a}. It is clear that future studies 
will need to take these aspects into account.


\section*{Acknowledgments}

It is a pleasure to thank Michal Bejger, Alessandro Drago, Kostas
Glampedakis, Nikolaos Stergioulas, Shin Yoshida and J.~Leszek Zdunik
for helpful comments and discussions. We are grateful to Lap-Ming Lin
for providing us with the reformulation of the source term for the
energy equation, and to Marie Anne Bizouard for information about the
scaling laws for the characteristic strain. Part of this work was done
during a visit by E.~A.\ to the AEI in Golm, Germany, and a visit of
H.~D.\ to CAMK in Warsaw, Poland. We gratefully acknowledge the
hospitality of the respective groups. This work was supported by DFG
(SFB/TR~7) and by the DAAD and IKY (IKYDA German--Greek research
travel grant). H.~D.\ is a Marie Curie Intra-European Fellow within
the 6th European Community Framework Programme (IEF 040464). The
computations were performed at SISSA, AEI and MPA.



\appendix


\section{Finite time for the initial phase-transformation}
\label{appendix:conversion_velocity}

In the study by LCCS, it was assumed that the timescale of
the phase transition is much smaller than the hydrodynamic timescale
for a NS, which is roughly $ 0.1 \mathrm{\ ms} $. They then
ignored the finite velocity of the conversion process and instead
induced the phase transition by instantaneously replacing the initial
polytropic EoS by one including the quark matter, which gives a lower 
pressure.

However, since the timescale for the phase transition even to 
two-flavour quark matter can be as long as $ 0.05 \mathrm{\ ms} $ 
within our picture (which is comparable to the hydrodynamic 
timescale), it is far from clear that treating it as instantaneous 
will not significantly affect the subsequent dynamics of the forming 
HQS. To check on this, we performed simulations for a set of 
representative collapse models in which we take the initial phase 
transformation to occur gradually over a finite timescale with the 
values of $ \rho_\mathrm{nm} $ and $ \rho_\mathrm{qm} $ in 
Eqs.~(\ref{eq:mixed_eos}, \ref{eq:alpha}) depending on the time $ t $ 
according to
\begin{equation}
  \begin{array}{rcll}
    \rho_\mathrm{hm} (t) & = & \left\{
      \begin{array}{ll}
        \displaystyle
        \rho_\mathrm{c} - \frac{t}{\tau_\mathrm{conv}}
        (\rho_\mathrm{c} - \rho_\mathrm{hm}) &
        \phantom{\rho_\mathrm{qm}}
        \mathrm{for\ } t < \tau_\mathrm{conv},
        \\
        \rho_\mathrm{hm} &
        \phantom{\rho_\mathrm{qm}}
        \mathrm{for\ } t \geq \tau_\mathrm{conv},
      \end{array}
    \right.
    \\ [2 em]
    \rho_\mathrm{qm} (t) & = & \left\{
      \begin{array}{ll}
        \displaystyle
        \rho_\mathrm{qm} + \rho_\mathrm{c} - \rho_\mathrm{hm} &
        \phantom{\rho_\mathrm{c}}
        \mathrm{for\ } t < \tau_\mathrm{conv},
        \\
        \displaystyle
        \phantom{\rho_\mathrm{qm}}
        - \frac{t}{\tau_\mathrm{conv}}
        (\rho_\mathrm{c} - \rho_\mathrm{hm})
        \\
        \rho_\mathrm{qm} &
        \phantom{\rho_\mathrm{c}}
        \mathrm{for\ } t \geq \tau_\mathrm{conv},
      \end{array}
    \right.
    \label{eq:conversion_time}
  \end{array}
\end{equation}
 where $ \tau_\mathrm{conv} $ is the timescale of the initial phase 
transformation and $ \rho_\mathrm{c} $ is the central rest-mass 
density. This means that $ \rho_\mathrm{hm} (t) $ starts initially at 
$ \rho_\mathrm{c} $ and then evolves linearly with time to reach its 
final value $ \rho_\mathrm{hm} $ at $ t = \tau_\mathrm{conv} $, while 
$ \rho_\mathrm{qm} (t) $ has a similar behaviour shifted by $ 
\rho_\mathrm{qm} - \rho_\mathrm{hm} $. For this test we set the 
timescale equal to the approximate upper limit for a detonation type 
phase transition, $ \tau_\mathrm{conv} = 0.05 \mathrm{\ ms} $, and 
compared the results obtained with those for the regular models 
(corresponding to $ \tau_\mathrm{conv} = 0 \mathrm{\ ms} $).

\begin{figure}
  \centerline{\includegraphics[width = 85 mm]{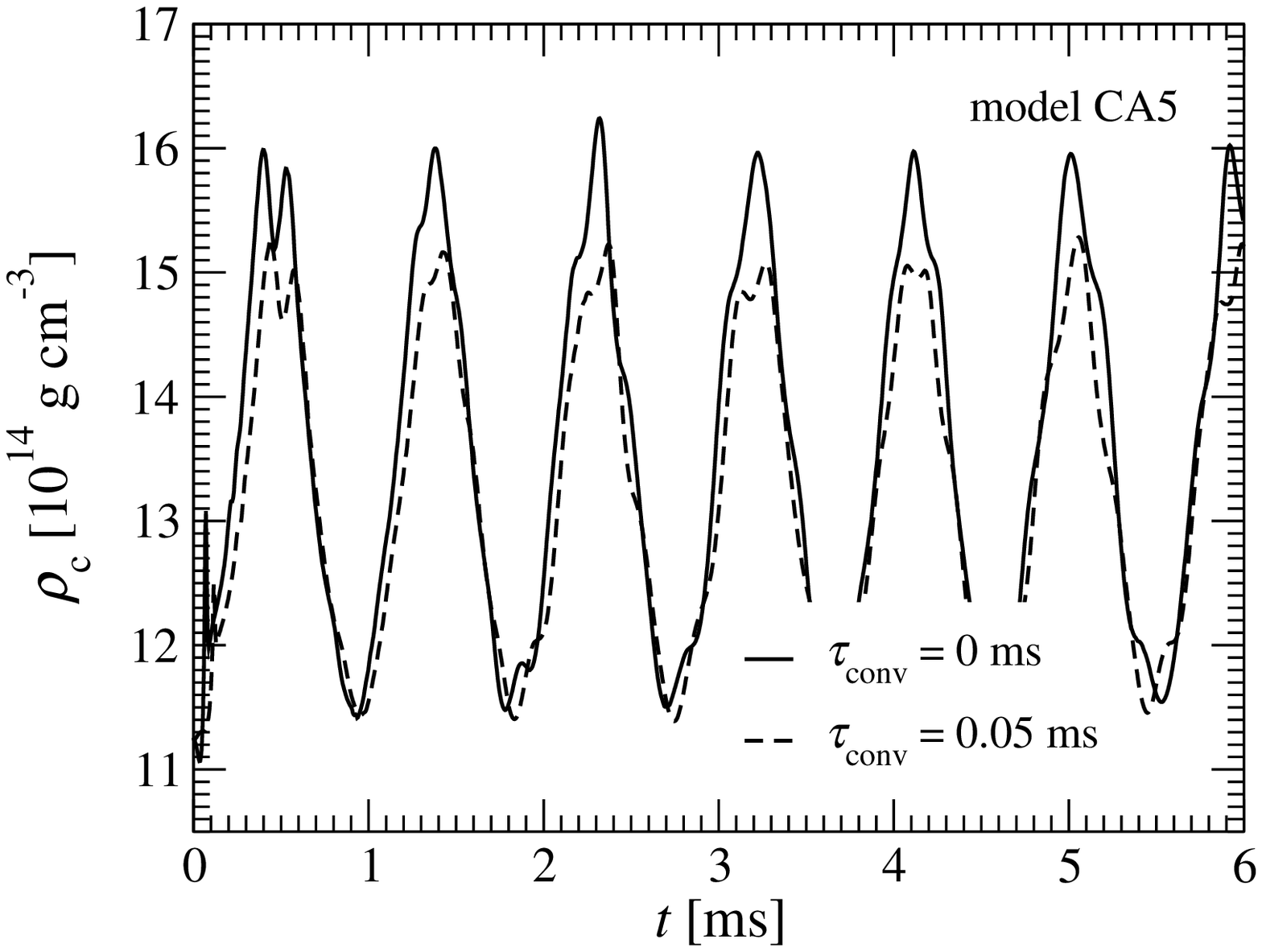}}
  \caption{Time evolution of the central rest-mass density $ \rho_\mathrm{c} $
    for the GR collapse model CA5 with
    $ \tau_\mathrm{conv} = 0 \mathrm{\ ms} $ (solid line) and
    $ \tau_\mathrm{conv} = 0.05 \mathrm{\ ms} $ (dashed line).}
  \label{fig:central_density_evolution_phase_transition}
\end{figure}

\begin{figure}
  \centerline{\includegraphics[width = 85 mm]{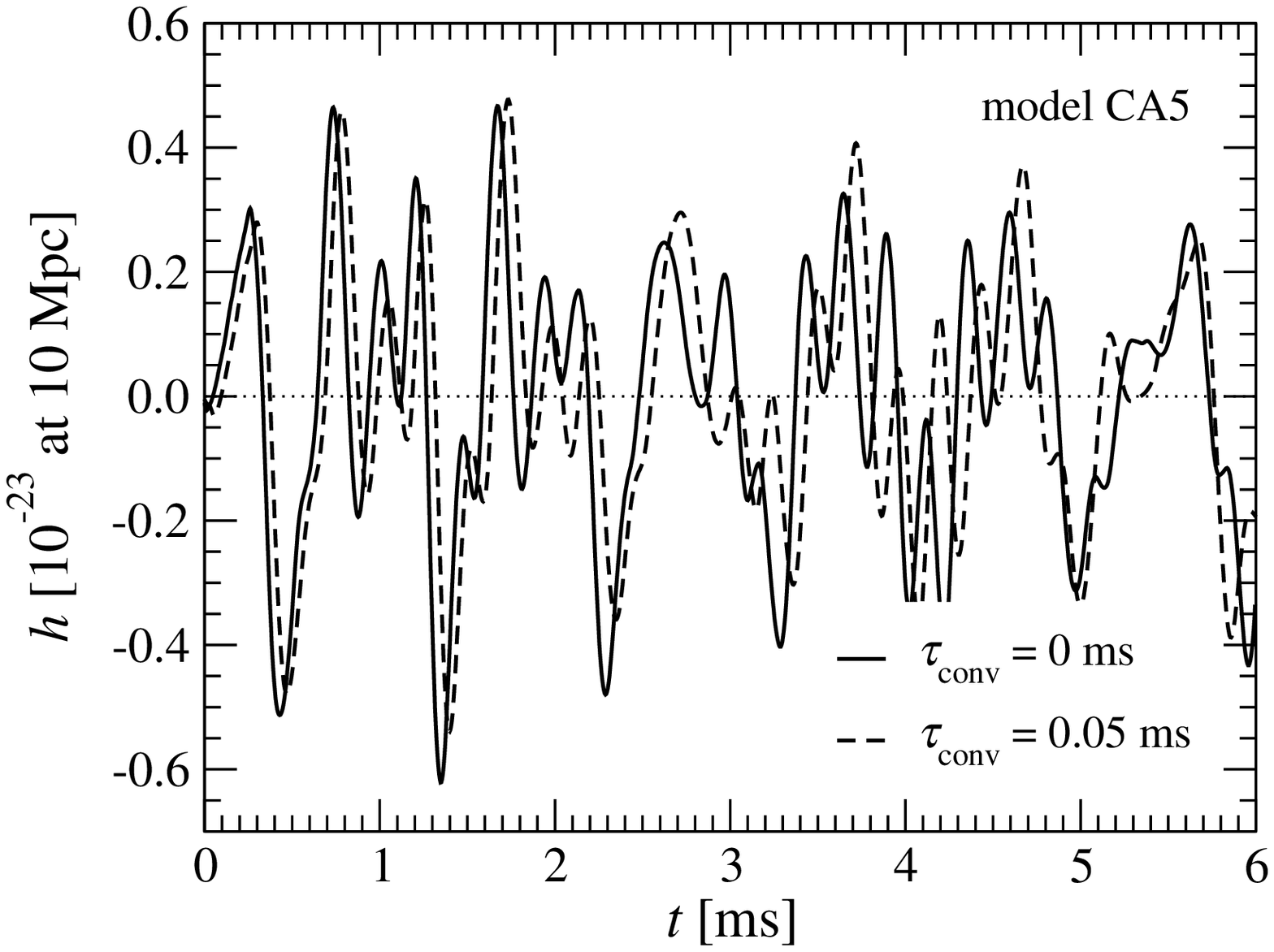}}
  \caption{Time evolution of the GW strain $ h $ at a distance of
    $ 10 \mathrm{\ Mpc} $ for the GR collapse model CA5 with
    $ \tau_\mathrm{conv} = 0 \mathrm{\ ms} $ (solid line) and
    $ \tau_\mathrm{conv} = 0.05 \mathrm{\ ms} $ (dashed line).}
  \label{fig:waveform_phase_transition}
\end{figure}

We expected that having the initial phase transformation occuring 
gradually in this way would produce a time lag of the same order as $ 
\tau_\mathrm{conv} $ in the initial contraction and give a less 
violent bounce occuring at a lower density with post-bounce pulsations 
of smaller amplitude than before. All of this was indeed the case for 
the models which we investigated, as can be seen in 
Fig.~\ref{fig:central_density_evolution_phase_transition}, where we 
show the time evolution of the central rest-mass density $ 
\rho_\mathrm{c} $ for the representative model CA5 with $ 
\tau_\mathrm{conv} = 0 $ and $ 0.05 \mathrm{\ ms} $. The waveform of 
the emitted GWs remains essentially unaltered except for a small 
reduction in the first large amplitude peaks and the expected phase 
shift (see Fig.~\ref{fig:waveform_phase_transition}). As the final HQS 
is less compact in the case of a noninstantaneous phase transition due 
to the less dynamic initial contraction, the frequencies of the 
predominantly excited quasi-normal modes, the $ F $- and $ {}^{2\!}f 
$-mode, are modified only slightly, changing from $ 1.09 $ to $ 1.12
\mathrm{\ kHz} $ and from $ 2.04 $ to $ 2.06 \mathrm{\ kHz} $, giving
relative changes of $ 3\% $ and $ 1\% $, respectively.

\begin{figure}
  \centerline{\includegraphics[width = 85 mm]{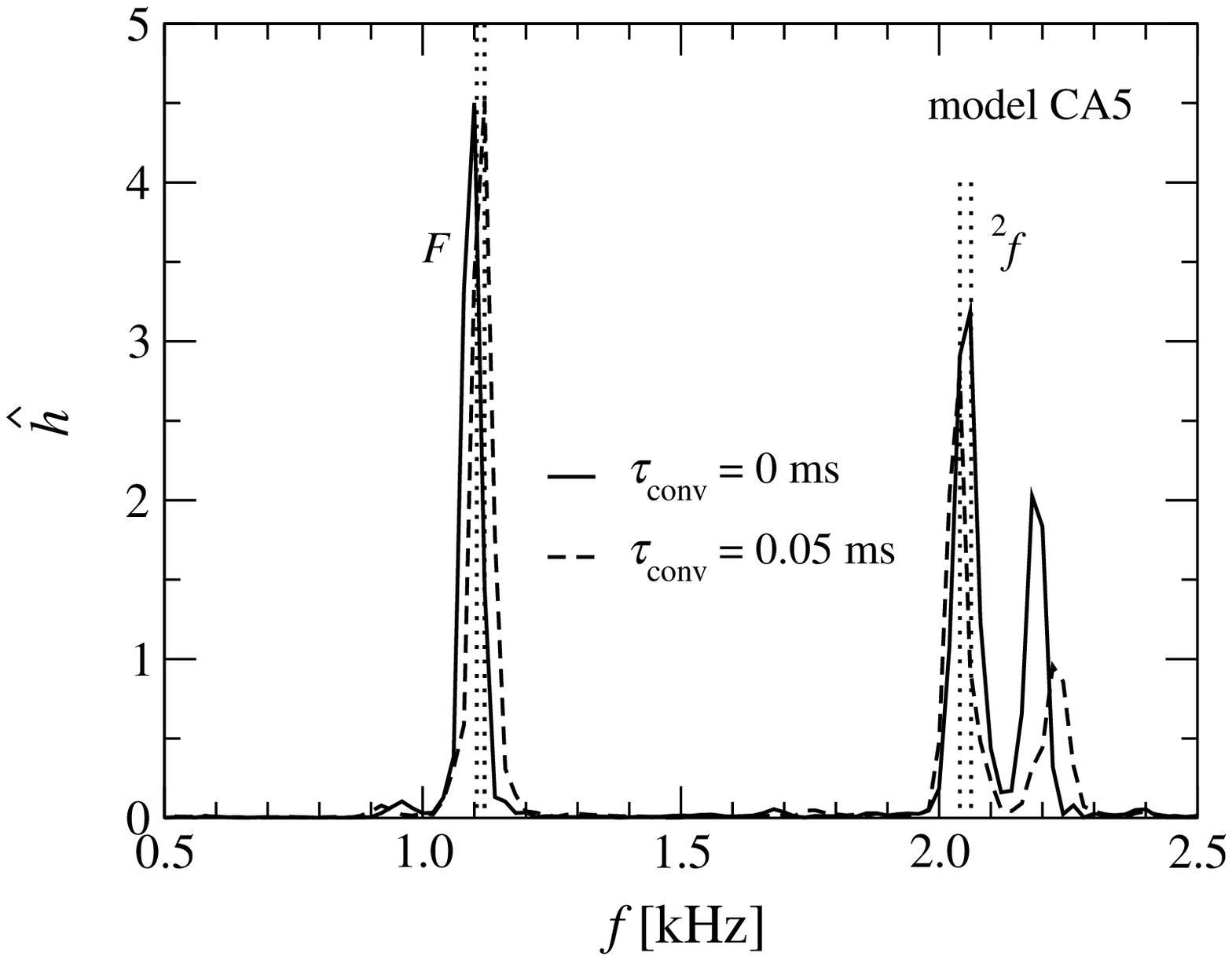}}
  \caption{Power spectrum $ \hat{h} $ (in arbitrary units) of the GW
    strain $ h $ for the GR collapse model CA5 with $
    \tau_\mathrm{conv} = 0 \mathrm{\ ms} $ (solid line) and $
    \tau_\mathrm{conv} = 0.05 \mathrm{\ ms} $ (dashed line).} 
  \label{fig:waveform_spectrum_phase_transition}
\end{figure}

The result that the differences seen when the finite $ 
\tau_\mathrm{conv} $ is introduced are so small is a direct 
consequence of $ \tau_\mathrm{conv} = 0.05 \mathrm{\ ms} $ still being 
about an order of magnitude smaller than the collapse timescale in the 
case of an instantaneous initial phase transformation (although it is 
of the same order as the \emph{dynamical} timescale). The collapse 
timescale can be approximated by the time of bounce, which is $ 0.40 
\mathrm{\ ms} $ for model CA5 and takes similar values for all of the 
other models. Since the choice $ \tau_\mathrm{conv} = 0.05 \mathrm{\ 
ms} $ is an upper limit within our picture, we conclude that the 
taking $ \tau_\mathrm{conv} = 0 \mathrm{\ ms} $ for our 
regular collapse models was reasonable, giving values for 
quantities such as $ \rho_\mathrm{c,b} $, $ |h|_\mathrm{max} $, $ f_F 
$ and $ f_{{}^{2\!}f} $ that are upper, but close, limits for a 
detonation type phase-transition-induced collapse of a NS to a HQS.


\section{Calculation of the damping times of the waveforms}
\label{appendix:damping_times}

In order to determine the damping times for the GW emission we apply
a curve fitting method to the numerically-obtained waveform: the 
waveform is fitted by a series of damped sinusoids,
\begin{equation}
  A_{20}^\mathrm{E2} =
  \sum_{i = 0}^n A_i \, e^{ - t / \tau_i} \, \cos (2 \pi f_i t + \phi_i).
  \label{eq:curve_fit}
\end{equation}
The parameters of this series 
(the damping times $ \tau_i $, the amplitudes $ A_i $ and the phases $ 
\phi_i $) are fixed by finding the best fitting curve. Depending on 
the model, we use $ n = 4\mbox{\,--\,}5 $ terms for the fitting 
procedure.

Note that this method of determining the damping times is more general 
than that presented in LCCS, where the series consisted of only two 
terms, so that only the fundamental $ l = 0 $ and $ l = 2 $ modes are 
taken into account, which leads to some inaccuracies at early 
post-bounce times when higher frequency modes in the GW signal are 
present. Also, they take $ A_1 = A_2 $ and $ \phi_i = 0 $, which may 
not be the case in general. With our approach we are able fit the 
original waveform with the correlation coefficient between the 
numerical data and Eq.~(\ref{eq:curve_fit}) exceeding $ 0.99 $ for all 
models, whereas the method used by LCCS gives a correlation 
coefficient of less than $ 0.9 $ for some models.

Assuming that the GW signal is essentially a linear combination of the 
$ F $-mode and the $ {}^{2\!}f $-mode, approximating each of them as a 
damped sinusoid and using knowledge of the mode frequencies $ f_F $ 
and $ f_{{}^{2\!}f} $, the amplitudes $ A_1 $ and $ A_2 $, and the 
phases $ \phi_1 $ and $ \phi_2 $, one can obtain the value of the 
characteristic GW strain $ h_\mathrm{c} $ for an \emph{arbitrary} 
emission time using simple scaling laws. For a single damped sinusoid,
\begin{equation}
  h = h_0 \, e^{- t / \tau} \sin \, (2 \pi f t - \phi)
\end{equation}
where $ h_0 $ is the amplitude, $ \tau $ is the damping timescale, $ 
f $ is the frequency and $ \phi $ is the phase. If $ h_\mathrm{c} $ is 
known for some emission time $ t_\mathrm{f} $, then its value for a 
multiple $ n $ of the original emission time can be calculated as
\begin{equation}
  h_\mathrm{c} (n \, t_\mathrm{f}) = h_\mathrm{c} (t_\mathrm{f}) \,
  \sqrt{\frac{1 - e^{- 2 n t_\mathrm{f} / \tau}}{1 - e^{- 2 t_\mathrm{f} / \tau}}},
  \label{eq:characterstic_strain_scaling}
\end{equation}
 provided that $ f^{-1} \ll \tau $ (which is fulfilled for most of our 
models) and the power spectral density $ S_h $ of the detector is 
reasonably constant in the vicinity of $ f $. For an undamped sinusoid 
with $ \tau = \infty $, Eq.~(\ref{eq:characterstic_strain_scaling}) 
gives $ h_\mathrm{c} (n \, t_\mathrm{f}) = \sqrt{n} \, h_\mathrm{c} $ 
as expected; in other words, $ h_\mathrm{c} $ scales like the square 
root of the number of cycles in the GW signal. In the limit of 
infinite emission time, but with finite $ \tau $, the exponential 
damping of the signal results in a finite value for the total 
characteristic GW strain,
\begin{equation}
  h_\mathrm{c} (t = \infty) = h_\mathrm{c} (t_\mathrm{f}) \,
  \frac{1}{\sqrt{1 - e^{- 2 t_\mathrm{f} / \tau}}},
\end{equation}

For many of our models, we have $ F $-mode damping times $ \tau_F $ 
which are much longer than $ t_\mathrm{f} $. On these timescales, 
other damping mechanisms such as physical viscosity or GW 
back-reaction (which are not taken into account in our study) could 
become important. We therefore do not give values of the total 
characteristic GW strain $ h_\mathrm{c} (t = \infty) $ for our models.

\label{lastpage}

\end{document}